# Temporal Dynamics of Microbial Communities in Anaerobic Digestion: Influence of Temperature and Feedstock Composition on Reactor Performance and Stability.

Ellen Piercy [a], Xinyang Sun [a], Peter R. Ellis [b], Mark Taylor [c], Miao Guo [a]*.

[a] Department of Engineering, Faculty of Natural, Mathematical & Engineering Sciences, King's College London, Strand Campus, London, WC2R 2LS, UK.

[b] Biopolymers Group, Departments of Biochemistry and Nutritional Sciences, Faculty of Life Sciences & Medicine, King's College London, Franklin-Wilkins Building, London, SE1 9NH, UK.

[c] Fermentation Lead, Marlow Ingredients, Nelson Ave, Billingham, North Yorkshire, TS23 4HA, UK

E-mail: miao.guo@kcl.ac.uk.

### Abstract

Anaerobic digestion (AD) offers a sustainable biotechnology to recover resources from carbon- and nutrient-rich wastewater streams, such as food-processing wastewater. Despite the wide adoption of crude wastewater characterisation, the impact of detailed chemical fingerprinting on AD remains underexplored. This study investigated the influence of fermentation-wastewater composition and operational parameters on AD over time to identify critical parameters influencing microbiome diversity and reactor performance. Eighteen bioreactors were operated under various operational conditions using mycoprotein fermentation wastewater. Detailed chemical analysis fingerprinted the molecules in the fermentation-wastewater throughout the AD process including sugars, sugar alcohols and volatile fatty acids (VFAs). High-throughput sequencing revealed distinct microbiome profiles linked to temperature and reactor configuration, with mesophilic conditions supporting a more diverse and densely connected microbiome. Importantly, significant elevations in *Methanomassiliicoccus* were correlated to high butyric acid concentrations and decreased biogas production, further elucidating the role of this newly discovered methanogen in AD. Reactors from different experimental runs had distinct VFA profiles, which impacted microbial taxonomy and diversity. Dissimilarity analysis demonstrated the importance of individual VFAs, sugars and sugar alcohols on microbiome diversity, highlighting the need for detailed chemical fingerprinting in AD studies of microbial trends. Furthermore, machine learning models predicting reactor performance achieved high accuracy based on operational parameters and microbial taxonomy. Operational parameters were found to have the most substantial influence on chemical oxygen demand removal, whilst *Oscillibacter* and two *Clostridium* species were highlighted as key factors in biogas production. By integrating detailed chemical and biological fingerprinting with explainable machine learning models this research presents a novel approach to advance our understanding of AD microbial ecology, offering insights for industrial applications of sustainable waste-to-energy systems.

# 1. Introduction

Global wastewater is estimated to have reached 400 billion $m^3$/year [1]. Alarmingly, 48% of wastewater is inadequately treated, posing significant environmental and public health risks[2]. Compared with municipal wastewater, food-processing wastewater is often characterised by high organic matter concentrations, which require advanced treatment technologies [3]. Aerobic wastewater treatments are energy-intensive, cost ineffective, and contribute to greenhouse gas emissions [4]. This realisation has prompted a shift towards more sustainable approaches, which recover valuable compounds from wastewater. Anaerobic digestion (AD) offers a biological process that converts organic matter to energy-carrier gas (biogas and nutrient-rich digestate), which has the potential to reduce environmental impacts and operational costs [5,6].

Many studies have focused on lab-scale reactors using synthetic feedstock, such as acetic acid or glucose-based growth mediums (Figure 1a). However, synthetic feedstock is not representative of the fluctuations and contaminants within real-world wastewater. Detailed chemical composition of wastewater feedstock remains largely undefined, especially within studies including detailed biological characterisation. Diverse microbes show different substrate affinities; therefore, it is important to understand how the molecules present in wastewater streams impact microbial diversity (Figure 1b). Moreover, previous studies often focus on singular time points, which do not capture the dynamic nature of the microbiome in response to fluctuating feedstock structure and composition. Understanding the interactions between these parameters and microbial communities over temporal scales is essential for enhancing biogas production, process stability, and overall efficiency.

The findings of a comprehensive literature review comprising 197 publications of AD species growth and experimental conditions are synthesised in Figure 1. On the phylum level, *Firmicutes* has the most diverse substrate utilisation, which could help to explain the wide diversity of *Firmicutes* reported in AD (Figure 1b). The majority of *Firmicutes* and *Bacteroidetes* species utilised sugars, sugar alcohols, polysaccharides and carboxylates. Specific substrate categorisation is detailed in according to Supplementary Table 1.1. Conversely, *Proteobacteria* showed complementary substrate utilisation to *Bacteroidetes*, with higher utilisation of amino acids, proteins, redox substrates (such as hydrogen and nitrates) and pyridines, benzoates and phenols (Figure 1c). These findings provide insights into how different feedstocks influence microbial community identity. However, the results are not comprehensive, and further research is required to relate detailed feedstock chemical characterisation with various microbial enrichments.

Operational parameters play a crucial role in AD performance and stability. In industrial-scale AD plants, microbial communities are subjected to fluctuating conditions, including feedstock structure and composition, temperature, and pH variations [7,8]. These fluctuations could lead to community changes, potentially causing process instability [9]. Most species had a mesophilic optimum temperature (63%); however, many species had a wide range of temperatures that could support growth (Figure 1d). Most bacteria (56%) had an optimum pH range at or close to neutral, although one-quarter of species did not have a specified pH growth range (Figure 1d). Archaea had a higher proportion of alkaliphilic species compared to more acidophilic bacteria (21% versus 11%, respectively, Figure 1d).

Furthermore, machine learning (ML) is a rapidly expanding field that is being applied to human microbiome studies [10,11]. Nevertheless, the application of ML in applied and industrial microbiology remains an emerging field. Current studies have shown promise for predicting AD performance from operational and biological parameters [10]. However, these studies suffer from limited sample sizing (between 17 to 50 samples), potentially limiting statistical power, and also lack detailed feedstock and biological characterisation [11,12]. Furthermore, model interpretability is often limited due to the need to reduce high dimensionality datasets. Applying explainable ML techniques, such as Shapley Additive exPlanations (SHAP) values, to datasets including detailed chemical and biological fingerprinting, allows interpretability of factor impact to improve our understanding of how operational, chemical and biological factors contribute to AD reactor performance.

To address this knowledge gap, our research aimed to understand the resource recovery potential of complex wastewater streams from fermentation processes and investigate the dynamics of the microbiome underpinning AD across different operational parameters and temporal scales. Here, we present the results of experiments exploring the effect of feedstock composition on AD using characterised mycoprotein fermentation wastewater (MFWW) obtained from the food industry.

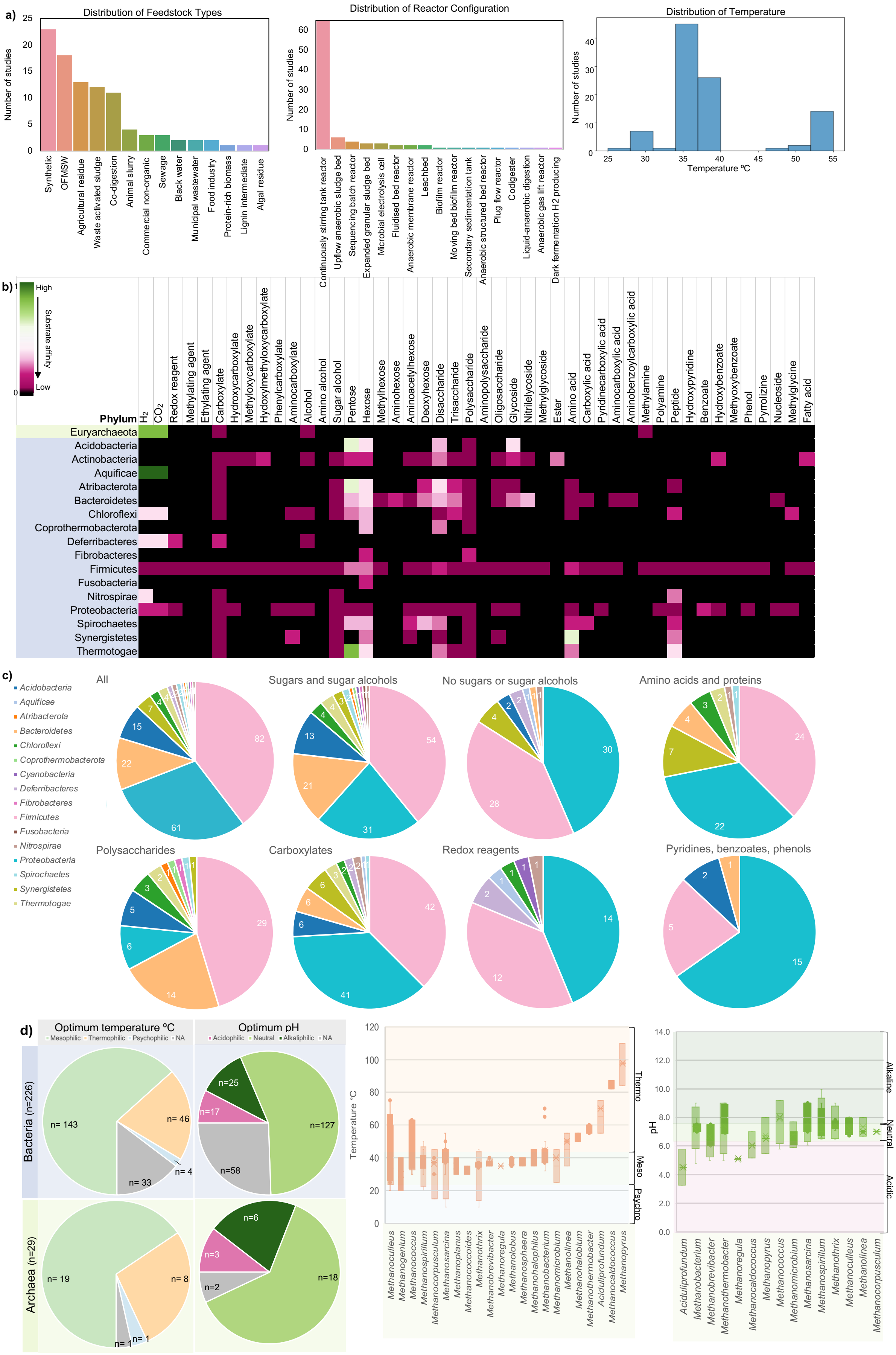

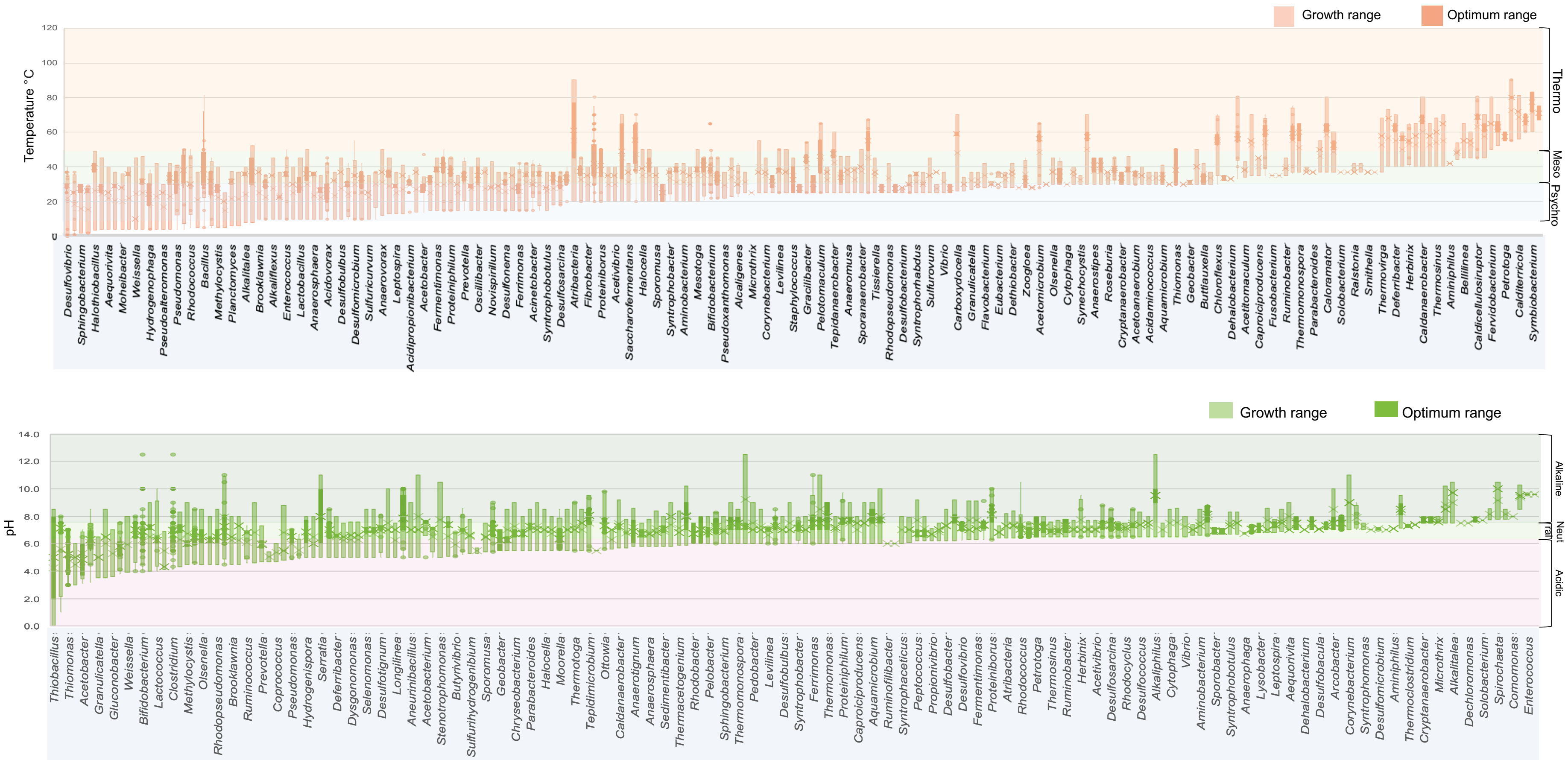

**Figure 1.** Growth conditions reported to support microbial growth for anaerobic digestion from a literature review of 197 papers (supplementary materials and methods 1.1). **(a)** Distribution of feedstock, reactor configuration and operational temperature. **(b)** Substrates (187) reported to support microbial growth were investigated. Affinity was based on average substrate utilisation on a scale of 0 to 1, where 1 (green) indicates that substrate supports the growth of all species within a phyla, and 0 (black) indicates no reported growth on a given substrate. Substrates are categorised by key functional groups (Supplementary Table 1.1). **(c)** Pie charts displaying the number of genera belonging to 16 bacteria phyla reported to utilise different categories of substrate. **(d)** Temperature and pH range reported to support growth. Temperature is categorised as psychrophilic (<25ºC), mesophilic (25ºC> temperature >42ºC) or thermophilic (>42ºC). The pH range is categorised as acidic (≤ 6.5, pink), neutral (6.5< pH <7.5, light green), or alkaliphilic (≥7.5, dark green). NA (grey) indicates information not available; n represents number of genera.

# 2. Materials and methods

MFWW, a by-product of Quorn™ food production, was obtained from Marlow Ingredients (Billingham, UK). Four batches of MFWW were characterised to evaluate batch variation and the effect of the feedstock on AD (Figure 2a). Two experiments were designed to address the research objectives. In the first experimental run, mesophilic and thermophilic reactors were compared to evaluate the effect of temperature. In run 2, single- and two-stage reactors were assessed to determine the impact of reactor configuration (Figure 2b-c). Influent composition was monitored in a time-series sampling schedule (Figure 2d). Reactor chemistry, performance and microbiome biodiversity were analysed to address how MFWW composition affects these parameters across different operational parameters and temporal scales. Predictive ML models were developed to forecast AD reactor performance in terms of chemical oxygen demand (COD) removal, cumulative biogas and specific daily biogas based on biological, chemical and operational parameters.

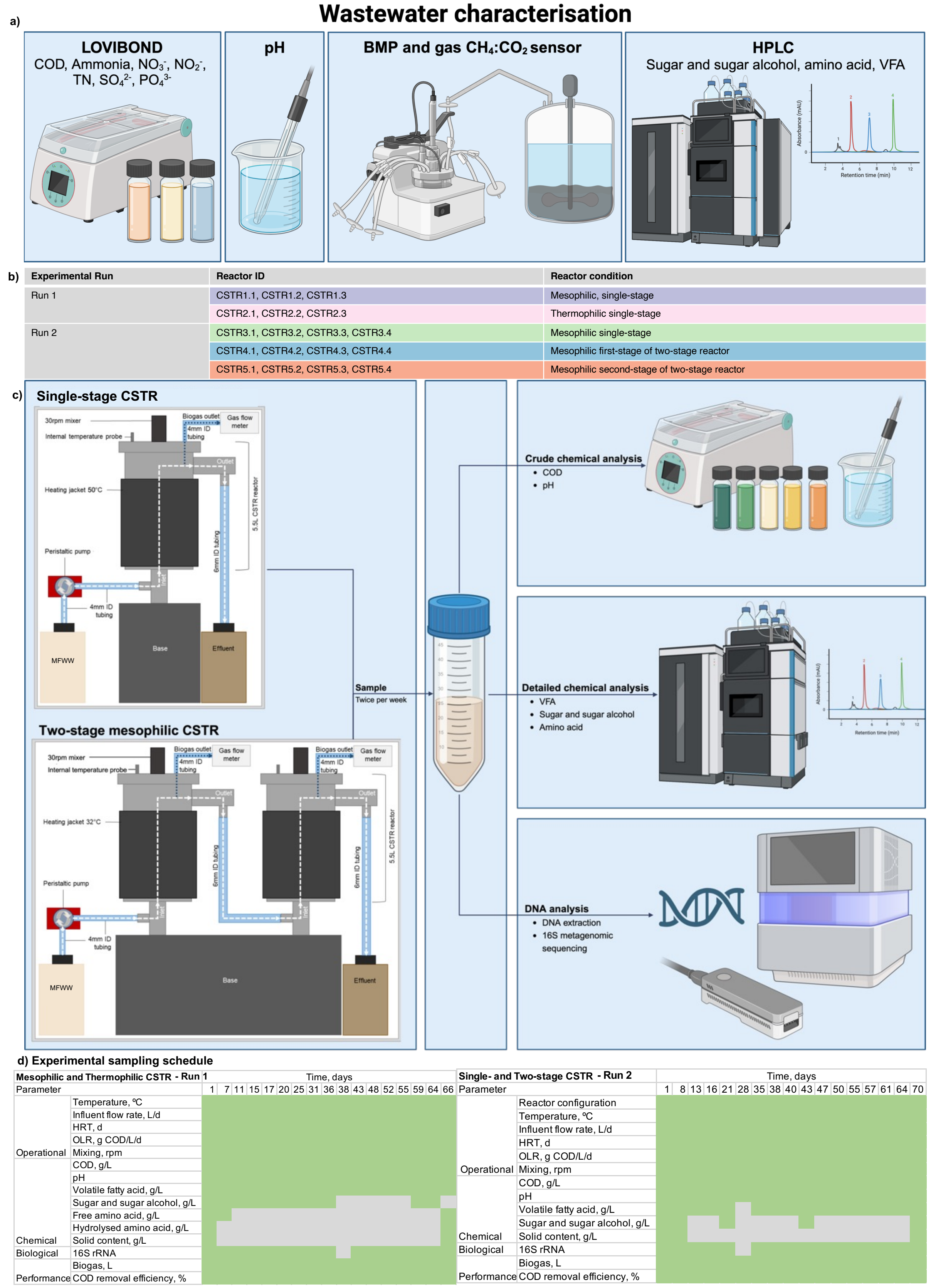

**Figure 2.** Experimental design. **(a)** Mycoprotein fermentation wastewater characterisation. **(b)** Experimental grouping. **(c)** Bioreactor set up, sampling and analysis. **(d)** Sampling schedule. *Abbreviations: chemical oxygen demand (COD), total nitrogen (TN), biomethane potential (BMP), high performance liquid chromatography (HPLC), volatile fatty acid (VFA), and continuously stirred tank reactor (CSTR).

## 2.1. Digester description.

Eighteen 5.5L continuously-stirred tank reactors (CSTR) were operated with a 14.7-day hydraulic retention time and 0.32L/d flow rate with MFWW (Figure 2c). Reactors were operated for a two-month adaptation period with MFWW as feedstock at 0.18L/d. CSTR were grouped by operational conditions. Group 1 represented single-stage mesophilic reactors (37±1ºC), group 2 represented single-stage thermophilic reactors (50±1ºC) and group 3-5 represented mesophilic single-, first- and second-stage reactors, respectively (Figure 2b). Groups 1 and 2 contained biological triplicates, and groups 3, 4 and 5 contained four biological replicates each. 2mL samples were taken in a time-series schedule in triplicate, and operational parameters were monitored throughout (Figure 2d). Samples were evaluated using the methods detailed below. Groups 1-2 and 3-5 were set up as two independent runs using the same batch of inoculum obtained from a commercial food-waste AD plant treating municipal source-separated domestic kitchen waste, industrial food waste and a fraction of abattoir waste (Devon, UK). Volatile suspended solid content was quantified as a measure of active biomass using standard methods (Supplementary Figure 2.1) [13].

## 2.2. Analytical methods.

### 2.2.1. Biogas production.

Real-time biogas production and composition ($CH_4$ and $CO_2$) from each individual reactor was measured in real-time using an Anaero technology gas flow meter and sensor.

### 2.2.2. Chemical oxygen demand (COD).

COD was measured directly from MFWW and reactor samples using Vario HR/COD, MR/COD, LR/COD Lovibond kits (as appropriate), according to standard methods (ISO, 1989; Supplementary materials and methods 2.3). COD removal was calculated using Supplementary Equation S3.

### 2.2.3. pH.

Fresh MFWW and reactor samples were measured directly in triplicate with a pH meter (Mettler Toledo).

### 2.2.4. High performance liquid chromatography.

Analysis of sugars and sugar alcohols (melibiose, glucose, maltitol, glycerol, mannitol and arabitol) and VFAs (acetic, propionic, isobutyric, butyric, isovaleric and isovaleric acid) for MFWW and reactor samples were conducted using a Shimadzu U-HPLC with SPD-M40 photodiode array and RID-20A refractive index detectors (Shimadzu). Prior to analysis samples were filtered through a 0.22µm pore filter (Claristep® Filtration system, Sartorius). Analysis was carried out in triplicate, and the mean and standard deviation were calculated. Calibration curves were constructed by plotting peak area against concentration. The linearity of the line of best fit was calculated using the least square regression method within the LabSolutions software. The limit of detection and quantification were calculated using LabSolutions software to verify quality. Detailed methodologies are provided in Supplementary Materials 2.4.

## 2.3. Metagenomic Sequencing and Bioinformatics.

### 2.3.1. DNA extraction.

Reactor samples were centrifuged at 10,000rpm for 5 minutes to obtain a pellet which was used for the DNA extraction, using the DNeasy® PowerSoil® Pro Kit according to standard protocol (QIAGEN GmbH, Germany). PCR amplification was carried out using 16S rRNA primers (Supplementary Figure 2.5.1c). Amplicon barcoding PCR was conducted using BiomekFX robot and automated software. Sequencing was conducted using Illumina MiSeq v3 kit and the run quality was assessed based on quality score (% bases>Q30), raw cluster density (k/mm$^2$), data output (Gbp), read number and % PhiX aligned (Supplementary Figure 2.5.1).

### 2.3.2. Sequence analysis.

Taxonomic classification was conducted using the Mothur pipeline in Galaxy referenced against the Silva_v4 database with ≥97% sequence identity threshold [14,15]. Additional pre-processing steps were conducted to filter for sequences ≥1200bp with Phred qualities >20 per base (Supplementary Figure 2.5.2).

### 2.3.3. Phylogenetic tree construction.

A Newick tree of genera was constructed from the taxonomic classification data using the anytree library in python and visualised using the international tree of life [16].

### 2.3.4. Network analysis.

The relative abundance of classified microbial taxa for each reactor sample were analysed across all time points, using the Molecular Ecology Network Analysis pipeline (Supplementary Materials and Methods 2.6) [17]. Networks were visualised in Cytoscape at the genus level [18].

### 2.3.5. Alpha diversity.

Shannon, Simpson's and Chao1 diversity indices were calculated using Supplementary Equation S4-6.

### 2.3.6. Beta diversity and dissimilarity index.

The calculation of Bray-Curtis similarity and db-RDA was implemented using the scipy.spatial, skbio and statsmodels library in python (Supplementary Eq. S7) [19]. The dissimilarity matrix served as the response variable, while environmental variables were used as predictors. Detailed methodology is available in Supplementary Materials 2.8.

## 2.4. Machine learning models

Lasso regression, random forest and bagging regression models were evaluated using algorithms defined in Eq.S9-11 (Supplementary Materials 2.9). The models were evaluated based on their performance in predicting reactor performances. Root mean squared error (RMSE), and coefficient of determination ($R^2$) were used to assess performance.

### 2.5. Statistical analysis.

For normally distributed data with two comparisons a pairwise t-test was applied, for three or more comparisons a one-way ANOVA analysis was performed with Tukey *post-hoc* analysis. For non-normally distributed independent data a Kruskal-Wallis H test was conducted with Dunn's *post-hoc* test for significant results. Confidence intervals of 95%, 99% and 99.9% were used. To investigate correlations between syntrophic relationships, a non-parametric Spearman's rank correlation analysis was conducted.

## 3. Results

### 3.1. MFWW represents a complex feedstock suitable for AD.

The composition of four different batches (A-D) of MFWW, compared with a previously reported value (sample R) obtained from the food industry, were measured over the time course of four fermentation cycles to determine resource recovery potential (Figure 3). We quantified physicochemical parameters driving the AD process, including ammonium, protein, pH, COD, and soluble organic compounds in MFWW (Figure 3). MFWW presented as a complex but balanced feedstock with a neutral pH ($6.2\pm0.1$) and high potential for AD (Figure 3a). Inhibitory ammonium was present at low concentrations ($0.03\pm0.01$g/L, Figure 3b). High COD ($11.32\pm4.09$g/L) and a mixed sugar profile, paired with high protein content ($2.19\pm1.31$g/L) facilitating good C/N ratios and suggests high resource recovery potential (Figure 3c-e). Biomethane potential assays showed high biodegradability ($92\pm6$%, Figure 3f).

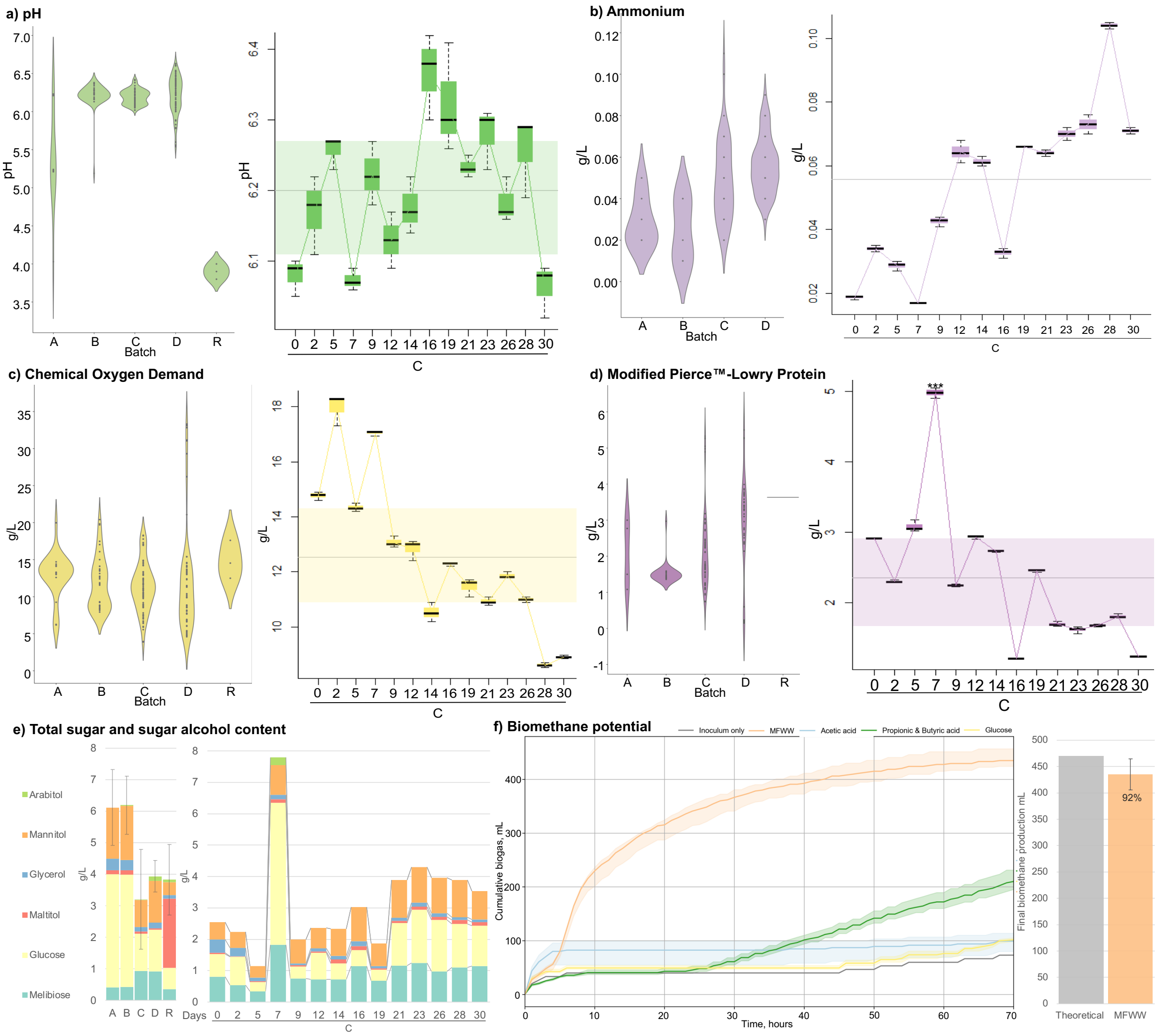

**Figure 3.** Chemical characterisation for four different batches (A-D) of mycoprotein fermentation wastewater (MFWW) compared to previously internally reported values (batch R). Batch C was measured at 14 times points within a 30-day fermentation cycle. Violin plots represent the distribution of results, box plots represent the interquartile range, the central bar represents the mean average, error bars represent the standard deviation and dots represent outliers. Statistical significance compared to R are shown at 95% *, 99% ** and 99.9% *** confidence intervals[20,21]. **(a)** pH. **(b)** Ammonium. **(c)** Chemical oxygen demand. **(d)** Protein. **(e)** Sugars and sugar alcohols. **(f)** Biomethane potential assay results including cumulative biogas of MFWW compared to inoculum only, acetic acid, propionic and butyric acid, and glucose. Final biomethane production compared as a percentage of the theoretical value calculated based on the COD to $CH_4$ mass balance equation. Error bars represent standard deviation.

## 3.2. Bioreactor performance

### 3.2.1. Run 1 reactors had better COD removal but poorer biogas production than Run 2 reactors.

COD removal in samples obtained from run 1 reactors were consistently greater than the 75% regulatory threshold and run 2 reactor samples (p=0.000, Figure 4b) [22]. COD removal efficiency in run 2 reactor samples regularly fell below the regulatory threshold, particularly at the beginning of the experiment up to day 28 and again at day 38, but gradually increased over time (Figure 4b). Conversely, run 2 reactors produced higher cumulative biogas than run 1 ($p=0.000$, Figure 4b). We compared the chemical and biological differences between reactors with high and low biogas production (Figure 4-5).

Mesophilic reactors had better biogas production than thermophilic ($p=0.000$) despite similar COD removal ($p=1.000$, Figure 4b). This result indicates that thermophilic conditions created an unfavourable environment for biogas production, possible due to poor adaptation increased temperature (Figure 4b). This could be indicative of the inoculum coming from a mesophilic source, meaning that thermophilic genera were underrepresented. Another possibility is that the temperature was low enough to support mesophilic genera with thermotolerant traits, but caused thermal stress, leading to underperformance, suggesting a specialised microbiome with little flexibility for adaptation (Figure 4b).

#### 3.2.2. VFA content showed marked differences between Run 1 and 2 reactor samples.

Samples obtained from run 2 had elevated VFA concentration compared to run 1 ($p=0.000$) and were characterised by higher acetic ($p=0.000$), isovaleric ($p=0.020$) and valeric acid ($p=0.000$, Figure 4c). Isobutyric acid constituted a large proportion of the VFA content produced for all run 1 reactors, likely owing to the influence of the influent containing amino acids, such as valine and leucine, which can be converted during AD to isobutyric acid and isovaleric acid, respectively [23-25]. The influent contained relatively high concentrations of isobutyric acid and isovaleric acid (Figure 4a) [26].

CSTR3.1-3.2, had elevated total VFA contents compared to CSTR3.3-3.4 ($p=0.007$ and $p=0.001$, respectively, Figure 4c). CSTR3.1 had elevated isovaleric acid, and CSTR3.2 had significantly elevated valeric acid compared to other group 3 reactors ($p=0.000$, Figure 4c). Samples in run 1 and 2 reactors had a spike in VFA content on day 7 and 13, respectively (Figure 4c). This peak consisted of elevated isovaleric content in samples for most reactors (except CSTR3.1) and increased acetic acid for run 2 reactors (Figure 4c). For second-stage reactors CSTR5.2, had elevated valeric acid ($p=0.000$), whilst CSTR5.1 had lower total VFA content than other group 5 reactors ($p<0.05$, Figure 4c).

The average reactor pH ranged from 8.00 to 8.44 and was slightly higher for groups 1, 2 and 5 than 3 and 4 ($p=0.000$, Figure 4c). Group 2 reactors pH showed high variability, suggesting potential instability (Figure 4c). Reactor pH is a key factor influencing microbial activity, and different pH ranges support different species' growth (Figure 1d). Therefore, it will be interesting to investigate the effect of this pH difference by comparing microbial communities between reactors (Figure 1d, Figure 4c).

**a) Influent composition**

| | pH | COD | TS | VS | TSS | VSS | Ammonium | Nitrite | Nitrate | TN | Protein | Sulphate | Phosphate | Total sugar & sugar alcohol | Melibiose | Maltitol | Glucose | Glycerol | Mannitol | Arabitol | Total VFA | Formic acid | Acetic acid | Propionic acid | Isobutyric acid | Butyric acid | Isovaleric acid | Valeric acid |
|---|---|---|---|---|---|---|---|---|---|---|---|---|---|---|---|---|---|---|---|---|---|---|---|---|---|---|---|---|
| Avg. | 6.17 | 11.32 | 11.77 | 9.59 | 4.00 | 2.00 | 0.03 | 0.00 | 0.01 | 1.32 | 2.19 | 1.03 | 0.89 | 2.56 | 0.60 | 0.06 | 0.95 | 0.17 | 0.80 | 0.17 | 0.20 | 0.08 | 0.02 | 0.00 | 0.03 | 0.09 | 0.00 | 0.00 |
| SD | 0.27 | 4.09 | 1.66 | 1.53 | 0.30 | 0.40 | 0.01 | 0.00 | 0.00 | 0.17 | 1.31 | 0.03 | 0.32 | 1.63 | 0.66 | 0.07 | 0.92 | 0.25 | 0.41 | 0.82 | 0.19 | 0.06 | 0.03 | 0.02 | 0.02 | 0.10 | 0.00 | 0.00 |

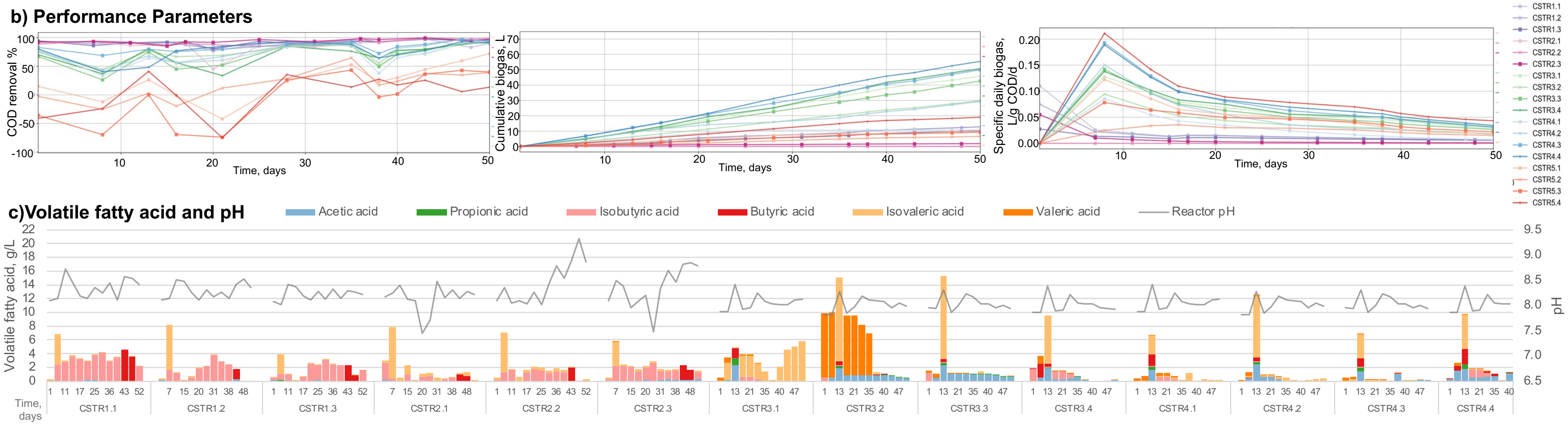

| | CSTR1.1 | CSTR1.2 | CSTR1.3 | CSTR2.1 | CSTR2.2 | CSTR2.3 | CSTR3.1 | CSTR3.2 | CSTR3.3 | CSTR3.4 | CSTR4.1 | CSTR4.2 | CSTR4.3 | CSTR4.4 | CSTR5.1 | CSTR5.2 | CSTR5.3 | CSTR5.4 |
|---|---|---|---|---|---|---|---|---|---|---|---|---|---|---|---|---|---|---|
| Acetic acid | 0.03±0.06 | 0.04±0.08 | 0.03±0.06 | 0.04±0.07 | 0.01±0.03 | 0.13±0.09 | 0.21±0.63 | 0.74±0.44 | 0.90±0.52 | 0.32±0.48 | 0.13±0.45 | 0.35±0.63 | 0.24±0.45 | 0.81±0.47 | 0.22±0.75 | 0.17±0.58 | 0.30±0.68 | 0.18±0.63 |
| Propionic acid | 0.00±0.00 | 0.00±0.00 | 0.02±0.04 | 0.00±0.00 | 0.00±0.00 | 0.00±0.00 | 0.07±0.25 | 0.06±0.08 | 0.05±0.10 | 0.02±0.05 | 0.04±0.12 | 0.03±0.12 | 0.05±0.16 | 0.08±0.18 | 0.05±0.17 | 0.04±0.14 | 0.04±0.14 | 0.05±0.18 |
| Isobutyric acid | 2.40±1.40 | 1.27±1.24 | 1.31±1.06 | 0.46±0.65 | 0.90±0.70 | 1.34±0.84 | 0.12±0.21 | 0.05±0.15 | 0.10±0.32 | 0.37±0.60 | 0.12±0.22 | 0.01±0.02 | 0.01±0.02 | 0.23±0.43 | 0.16±0.24 | 0.01±0.04 | 0.26±0.50 | 0.11±0.32 |
| Butyric acid | 0.63±1.49 | 0.13±0.41 | 0.25±0.64 | 0.13±0.31 | 0.15±0.53 | 0.28±0.67 | 0.11±0.37 | 0.05±0.17 | 0.04±0.13 | 0.20±0.55 | 0.13±0.44 | 0.04±0.15 | 0.12±0.33 | 0.26±0.59 | 0.03±0.08 | 0.29±0.57 | 0.22±0.68 | 0.12±0.39 |
| Isovaleric acid | 0.44±1.14 | 0.55±1.71 | 0.27±0.74 | 0.95±1.91 | 0.74±1.43 | 0.43±0.90 | 3.29±3.28 | 1.17±3.25 | 1.27±3.28 | 0.72±1.89 | 0.51±0.75 | 1.33±2.68 | 0.38±0.90 | 0.73±1.68 | 0.22±0.55 | 1.64±3.45 | 3.58±5.41 | 0.93±1.46 |
| Valeric acid | 0.00±0.00 | 0.00±0.00 | 0.00±0.00 | 0.00±0.00 | 0.00±0.00 | 0.00±0.00 | 0.14±0.24 | 4.15±4.03 | 0.08±0.15 | 0.09±0.30 | 0.17±0.23 | 0.07±0.15 | 0.09±0.14 | 0.06±0.13 | 0.86±0.61 | 15.06±5.16 | 1.44±1.02 | 4.17±2.88 |
| Total VFA | 3.50±1.42 | 1.99±2.11 | 1.87±1.06 | 1.57±1.99 | 1.80±1.67 | 2.19±1.17 | 3.93±3.07 | 6.23±4.65 | 2.43±3.83 | 1.73±2.47 | 1.10±1.67 | 1.84±3.41 | 0.88±1.77 | 2.17±2.46 | 1.54±1.06 | 17.22±2.10 | 5.84±5.37 | 5.55±2.46 |
| pH | 8.34±0.20 | 8.29±0.15 | 8.22±0.12 | 8.14±0.27 | 8.44±0.40 | 8.36±0.38 | 8.06±0.14 | 8.01±0.13 | 8.03±0.13 | 8.00±0.15 | 8.06±0.14 | 8.01±0.13 | 8.03±0.13 | 8.00±0.15 | 8.00±0.40 | 8.25±0.12 | 8.31±0.12 | 8.05±0.59 |

**Figure 4.** Reactor operation, performance and chemistry **(a)** Influent composition including mean (Avg.) ± standard deviation (SD) for pH, chemical oxygen demand (COD), total solids (TS), volatile solids (VS), total suspended solids (TSS), volatile suspended solids (VSS), ammonium, nitrites, nitrates, total nitrogen (TN), protein, sulphate, phosphate, total sugar & sugar alcohol, melibiose, maltitol, glucose, glycerol, mannitol, arabitol, total volatile fatty acid (VFA), formic acid, acetic acid, propionic acid, isobutyric acid, butyric acid, isovaleric acid and valeric acid. All units are g/L (except pH). (**b)** Performance parameters including COD removal, cumulative biogas, and specific daily biogas. (**c**) Reactor pH and VFA content. Bar charts represent the cumulative sum of individual VFA over time (days). Line chart represents pH (grey). Mean values ± standard deviation are presented for VFA concentrations and pH.

## 3.3. Microbiome composition and function

### 3.3.1. Run 1 reactor samples had a distinct taxonomic profile, characterised by increased *Methanosaeta* compared to run 2 samples.

In total, 1,982,343 unique sequences were retained after quality control over 25 times points. Archaea accounted for 1.23±0.72% of the total population. *Proteobacteria, Firmicutes, Bacteroidetes,* and *Actinobacteria* represented the major phyla across all reactors, although run 1 had a distinct taxonomic profile compared to run 2 reactors (Figure 5a). Notably, the relative abundance of *Firmicutes* was significantly elevated in run 1 compared to run 2 reactors, indicating elevated *Firmicutes* is associated with reduced biogas production ($p<0.05$, Figure 5a). Additionally, the relative abundance of *Synergistota* was significantly elevated in group 1 reactors compared to groups 2, 3 and 4 ($p$=0.000, Figure 5a).

### 3.3.2. The *Firmicutes* taxonomic profile was highly dynamic.

Core microbiome construction and times series network analysis revealed that *Firmicutes* was highly dynamic, with little conservation over time for run 2 reactors with high biogas production (Figure 4b, Figure 5e). This result suggests that a dynamic *Firmicutes* community improves reactor performance (Figure 5e). Interestingly, *Firmicutes* had the most diverse substrate utilisation of key AD phyla, which could likely be contributing to their dynamic profile (Figure 1b). By contrast, there was a much higher presence of conserved *Proteobacteria* and *Bacteroidetes* across run 2 reactors that were not shared by run 1 (Figure 5e).

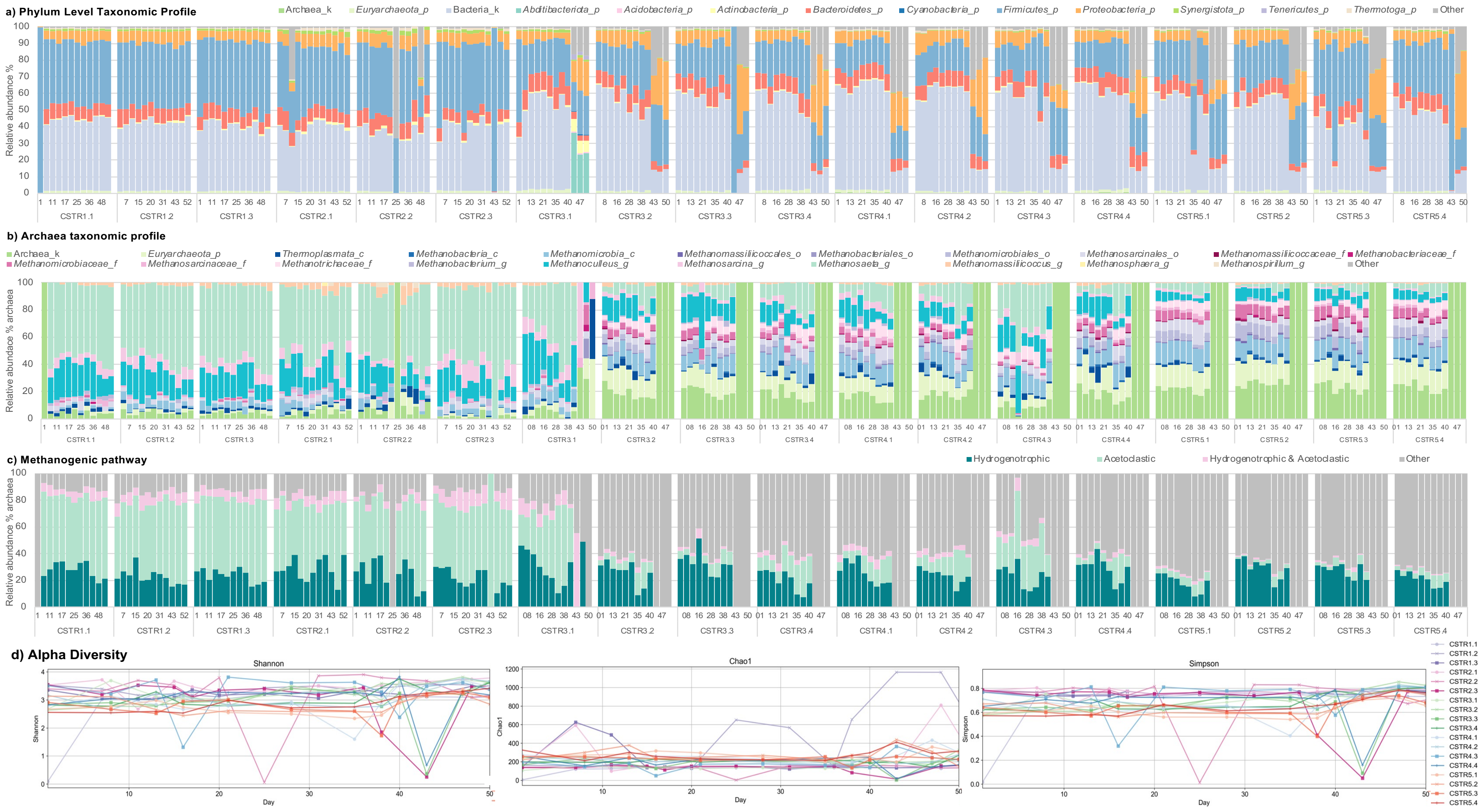

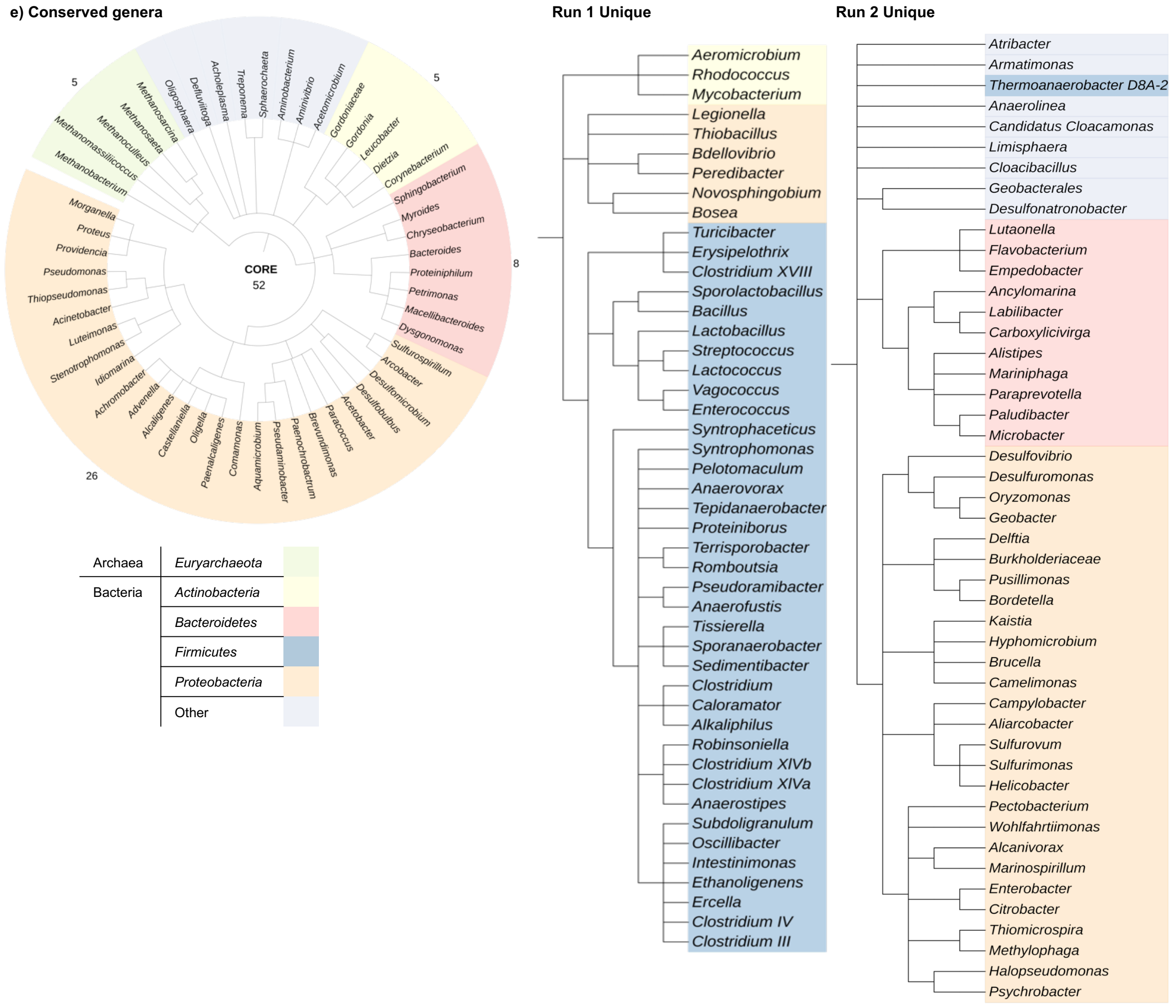

**Figure 5.** CSTR taxonomy **(a)** Phylum level relative abundance (% of total population), phyla with >1% relative abundance for at least one sample are represented and phyla representing <1% of the total population for any sample are aggregated into 'other' **(b)** Relative abundance of archaea (% of total archaea population). Taxa with >1% relative abundance for at least one sample are represented and taxa representing <1% of the total population for any sample are aggregated into 'other'. Taxa are represented at the kingdom (_k), phylum (_p), class (_c), order (_o), family (_f) and genus (_g) level. **(c)** Relative abundance of archaea categorised by methanogenic pathway, where other represents ambiguous taxa. **(d)** Alpha diversity for Shannon, Simpson's and Chao1 diversity indices over time, days. **(e)** Core microbiome of species shared across all reactors and unique species from experimental run 1 and 2. Archaea are highlighted in green, *Actinobacteria* in yellow, *Bacteroidetes* in red, *Firmicutes* in blue, *Proteobacteria* in orange, and other bacteria are light purple.

### 3.3.3. Taxonomic profile appeared to be decoupled from reactor performance.

Despite having the same feedstock, operational conditions and starting inoculum, CSTR3.1 had a distinct taxonomic profile compared to other run 2 reactors, but comparable biogas production and COD removal (Figure 4b, Figure 5). Run 1 reactors had a higher abundance of *Methanosaeta* ($p$=0.000), corresponding to lower acetic acid concentration, suggesting active acetoclastic methanogenesis (Figure 5b). *Methanosaeta* is an obligate acetoclastic methanogen associated with low organic loading rate (OLR), acetic acid, and solid and ammonium contents [27]. Additionally, *Methanosaeta* are reported to have an alkaliphilic optimum pH of 7.0 to 9.0 (Figure 1d). The higher pH and lower acetic acid concentration of run 1 compared to group 3 and 4 could explain the elevated *Methanosaeta* (Figure 4c, Figure 5b).

Notably, CSTR3.1 had similar performance to run 2 reactors but a taxonomic profile that was more comparable to run 1; for example, CSTR3.1 and run 1 reactors had comparable *Methanosaeta* abundance (Figure 5). This result indicates that reactor performance was decoupled from the microbiome profile. Despite low OLR, acetic acid concentration and high pH, *Methanosaeta* content was reduced in group 5 reactors, indicating potentially less reliance on acetoclastic methanogenesis (Figure 5b-c). However, second-stage reactors had a lower percentage of classified genera compared to first- and single-stage reactors making it difficult to conclude the true relative abundance of acetoclastic methanogens (Figure 5).

#### 3.3.4. Despite some fluctuations, reactors had similar and stable species richness and evenness.

Shannon, Simpson's and Chao1 alpha diversity indices were compared as measures of community evenness and richness (Figure 5d). There was a significant drop in Shannon and Simpson's diversity for only CSTR3.3 on day 43, which recovered by day 48 (Figure 5d). By day 28, Simpson's diversity was similar for all reactors (Figure 5d). Chao1 diversity was similar for CSTR1.1 and Group 3; however, CSTR1.2 and CSTR1.3 showed large fluctuations, resulting in periods of elevated Chao1 diversity index values (Figure 5d). Reactor CSTR1.2 had elevated Chao1 diversity past day 20, whereas CSTR1.3 had elevated Chao1 diversity between days 1 and 15, indicating more rare species than other reactors (Figure 5d). Species richness and evenness were reduced in group 5 reactors compared to group 1-4 reactors, indicating that alpha diversity plays a crucial role in reactor resilience to stress conditions. This result agrees with the literature that diversity improves performance (Figure 5d) [28].

## 3.4. Microbiome dynamics and interaction with bioreactor parameters

### 3.4.1. Denser networks were associated with increased biogas production.

Run 1 reactor samples were found to represent less dense networks than those in run 2 (Figure 6a). These results suggest that network density contributes to reactor performance, supporting previous findings from Dalantai *et al.,* [29], who reported that sparser networks were associated with reactor instability. Time series analysis of node centrality showed that groups 3 and 5 had a higher node degree than groups 1, 2 and 4 (Figure 6b). Group 4 showed greater variation in node betweenness and a lower clustering coefficient than other reactor groups (Figure 6a). Interestingly, run 1 had a higher node event centrality than run 2, suggesting that node event centrality could be important in reactor performance (p= 0.000, Figure 6b).

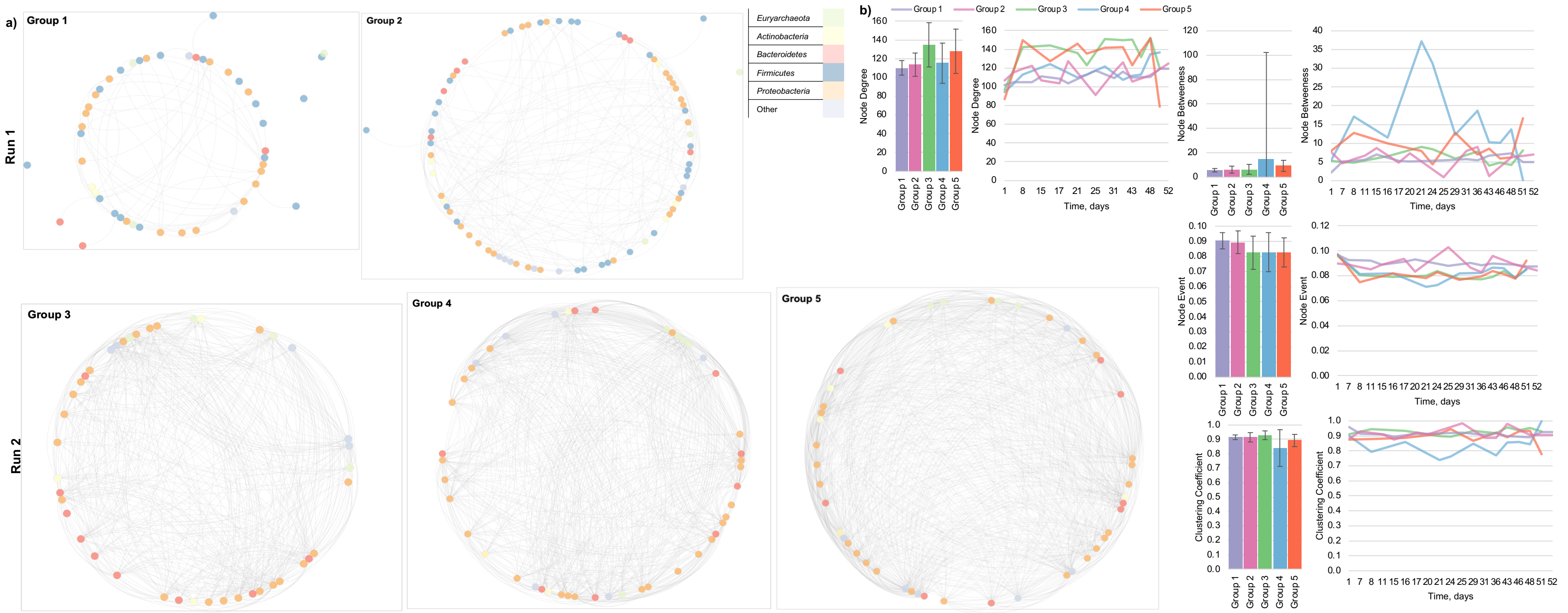

c)

| CSTR | Nodes | Links | Average path distance | R2 power law | Max degree | Max betweenness | Max clustering coefficient |
|---|---|---|---|---|---|---|---|
| 1.1 | 48 | 64 | 3.90 | 0.638 | *Anaerostipes* | *Syntrophomonas* | *Clostridium* XIVa |
| 1.2 | 87 | 2579 | 1.31 | 0.075 | *Methanobacteri8um* | *Methanosaeta* | *Anaerostipes* |
| 1.3 | 85 | 2348 | 1.34 | 0.055 | *Providencia* | *Desulfomicrobium* | *Syntrophomonas* |
| 2.1 | 77 | 1994 | 1.32 | 0.098 | *Methanosaeta* | *Petrimonas* | *Tepidanaerobacter* |
| 2.2 | 75 | 2700 | 1.03 | 0.748 | *Aquamicrobium* | *Aquamicrobium, Bacillus, Clostridium* IV*, Methanosaeta, Providencia* | *Petrimonas* |
| 2.3 | 36 | 44 | 4.63 | 0.849 | *Acetomicrobium, Lactobacillus* | *Acetomicrobium* | *Methanosarcina* |
| 3.1 | 82 | 453 | 1.94 | 0.696 | *Acetobacter* | *Acetobacter* | *Comamonas, Desulfomicrobium, Sphingobacterium* |
| 3.2 | 134 | 6417 | 1.28 | 0.007 | *Sulfurimonas* | *Sulfurimonas* | *Myroides* |
| 3.3 | 108 | 586 | 2.63 | 0.530 | *Sulfurimonas* | *Sulfurimonas* | *Oligosphaera* |
| 3.4 | 80 | 355 | 2.95 | 0.530 | *Sulfurimonas* | *Sulfurimonas* | *Macellibacteroides* |
| 4.1 | 93 | 291 | 3.14 | 0.706 | *Providencia, Sphaerochaeta* | *Sphaerochaeta* | *Pseudomonas, Acetobacter* |
| 4.2 | 139 | 6851 | 1.29 | 0.028 | *Sphaerochaeta* | *Morganella* | *Methanoculleus* |
| 4.3 | 121 | 5706 | 1.21 | 0.137 | *Sphaerochaeta* | *Sphaerochaeta* | *Defluviitoga* |
| 4.4 | 121 | 553 | 3.35 | 0.614 | *Providencia* | *Providencia* | *Methanomassiliicoccus* |
| 5.1 | 133 | 7240 | 1.18 | 0.052 | *Stenotrophomonas, Proteiniphilum* | *Stenotrophomonas, Proteiniphilum* | *Oligosphaera* |
| 5.2 | 138 | 7871 | 1.17 | 0.093 | *Comamonas* | *Proteus* | *Acetomicrobium* |
| 5.3 | 135 | 7423 | 1.18 | 0.006 | *Bacteroides, Sulfurimonas, Providencia* | *Bacteroides, Sulfurimonas, Providencia* | *Acetomicrobium* |
| 5.4 | 109 | 564 | 2.77 | 0.600 | *Providencia* | *Oligella* | *Morganella, Methanoculleus, Petrimonas, Stenotrophomonas, Empedobacter* |

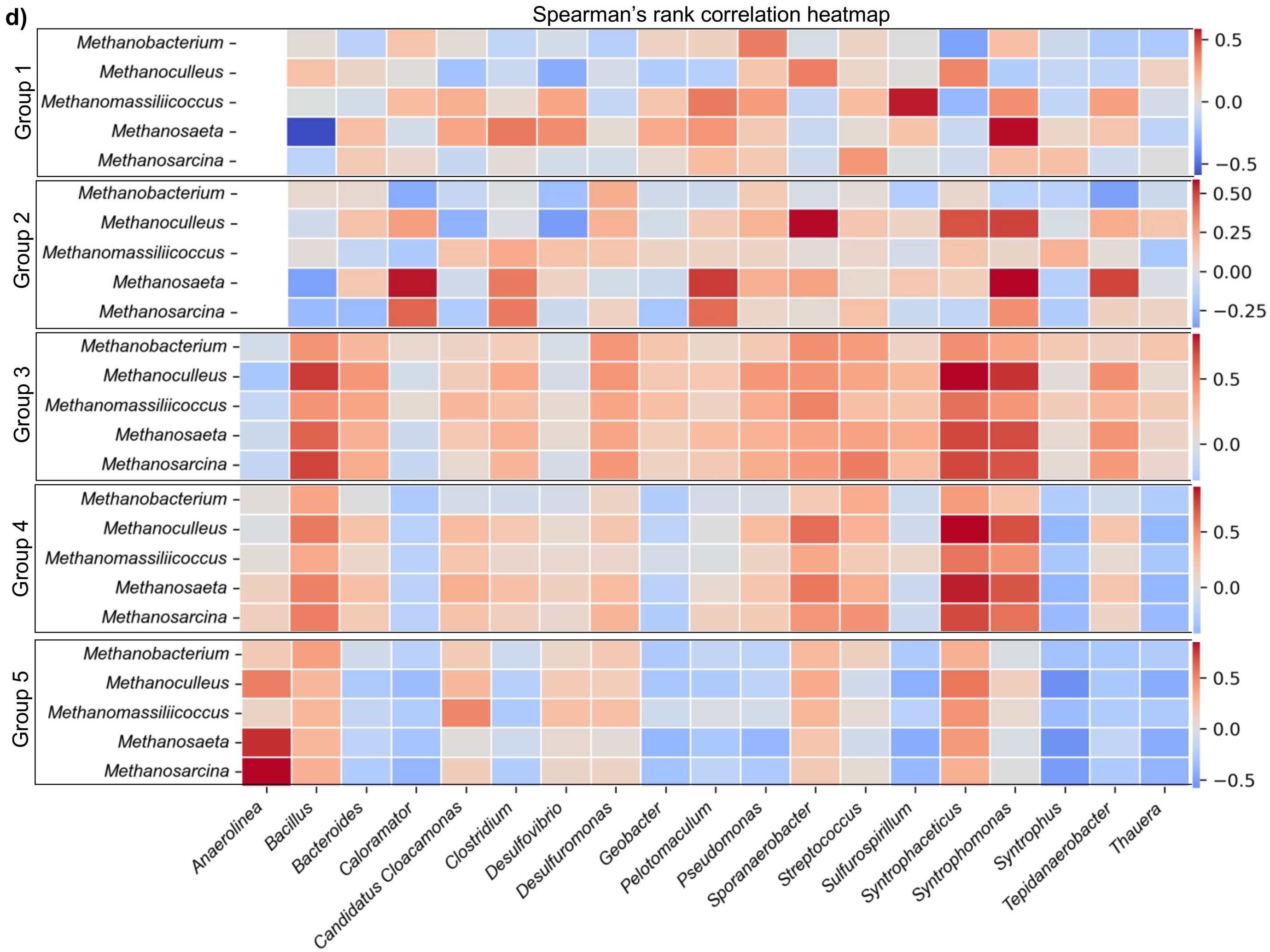

**Figure 6.** Microbiome interaction analysis. **(a)** Time series network analysis of each reactor group at genus level, where nodes represent genera. **(b)** Node centrality is presented where bar charts represent the average, and error bars represent the standard deviation, and line charts represent node centrality over time for node degree, betweenness, clustering coefficient and event centrality. **(c)** Descriptive statistics of network analysis and the identity of nodes with maximum degree, betweenness and centrality are described. **(d)** Spearman rank correlation coefficient of dominant methanogens and syntrophic genera presented as a heatmap for group 3-5 reactors, where *c*orrelations are reported as positive (red) or negative (blue). Blank values indicate insufficient data. Only correlations with sufficient data for at least one genera are presented.

### 3.4.2. Syntrophic relationships were associated with increased biogas production.

Syntrophic bacteria and methanogens, including *Syntrophomonas, Methanosarcina* and *Methanosaeta*, represented group 1 central network nodes (Figure 6c). CSTR2.3 also had high connectivity of syntrophic genera (Figure 6c) and elevated biogas production compared to other group 2 reactors, indicating the potential importance of syntrophic genera connectivity in reactor performance (Figure 4b, Figure 6c). Mesophilic reactors had elevated syntrophic genera compared to thermophilic reactors, and higher connectivity of syntrophic genera and methanogens was associated with improved biogas production. Syntrophic genera belonging to *Firmicutes*, which have been related to stress conditions, such as high OLR, were present across all run 1 reactors, which could have contributed to lower biogas production [30,31]. Spearman's rank correlation analysis revealed significant correlations between thermophilic *Caloramator* and *Methanosaeta* and *Methanosarcina* in thermophilic reactors only (Figure 1, Figure 6d).

Spearman rank correlation analysis revealed a positive correlation between hydrogen-producing syntrophic bacteria and hydrogenotrophic methanogens in single- (group 3) and first-stage (group 4) reactors, which was diminished in the second-stage (group 5) reactors (Figure 6d). Most syntrophic bacteria showed significant correlations to dominant methanogens for single- and first-stage reactors ($p<0.05$), except for *Candidatus Cloacamonas* ($p>0.05$, Figure 6d). However, the positive correlation between *Candidatus Cloacamonas* and *Methanomassiliicoccus* was significant for second-stage reactors ($p=0.000$, Figure 6d). *Anaerolinea* and acetoclastic methanogens were significantly positively correlated; however, the strength of this correlation was lower in single- compared to first- and second-stage reactors ($p<0.05$, Figure 6d).

The relationship between syntrophic genera were less clear in samples from second-stage reactors, which were found to have fewer significant correlations (Figure 6d). However,

*Anaerolinea, Methanosarcina,* and *Methanosaeta* showed significant positive correlations (p=0.000, Figure 6d). *Desulfuromonas* and *Pseudomonas* were also significantly positively correlated to *Methanoculleus, Methanosaeta* and *Methanomassiliicoccus* and *Geobacter* was significantly correlated with *Methanoculleus* (p=0.000, Figure 6d). Importantly, samples in the group 5 reactors exhibited decreased acetic acid concentrations and acetoclastic methanogenesis, which could reduce the strength of syntrophic acetate oxidising bacteria relationships within the reactors (Figure 4c, Figure 6d). This result highlights the importance of syntrophic relationships in reactor performance.

#### 3.4.3. *Methanomassiliicoccus* was associated with increased butyric acid concentration and decreased biogas production.

Group 2 reactors had a greater abundance of *Methanomassiliicoccus* than other reactors (p=0.000, Figure 5b). Interestingly, *Methanomassiliicoccus* were associated with periods of enriched butyric acid, suggesting a potential correlation between *Methanomassiliicoccus* and isobutyric and butyric acid concentrations (Figure 4c, Figure 5b). This result is consistent with data from Nikitina *et al.,* [32], who found that *Methanomassiliicoccus* were enriched under butyric acid dominant VFA accumulation. *Methanomassiliicoccus* has been associated with reduced biogas yields, which agrees with the results that run 1 reactors had lower biogas production than run 2 reactors (Figure 4c, Figure 5) [33].

### 3.5. Effects of chemical and operational parameters on bioreactor performance and biodiversity

#### 3.5.1. Random forest models performed best for predicting anaerobic digestion reactor performance.

Supervised ML models were developed to investigate the potential of operational parameters and reactor chemistry and biology for predicting reactor performance in terms of COD removal and biogas production. Lasso regression models provided a baseline with moderate predictive capability. Random forest regression is a robust ensemble method that creates and merges multiple decision trees for a more accurate and stable prediction. Bagging regression is an ensemble technique that improves the stability and accuracy of ML algorithms by reducing variance and preventing overfitting. Random forest regression showed the best performance with the highest RMSE and lowest $R^2$, reflecting high accuracy and substantial variance explanation (Figure 8b-d). Overall, the evaluation revealed that the random forest outperformed the two other models across all experiments (Figure 8b-d).

### 3.5.2. Influent COD and OLR were determined to be key influences on reactor biodiversity and performance.

The average impact of features on model output was measured using SHAP values. The top five most important features affecting COD removal included influent COD, time, OLR, temperature, and pH (Figure 8b). Influent COD was the most significant feature, with the highest SHAP value (Figure 8b). The most influential factors affecting biogas production were time, temperature, pH, *Oscillibacter* and influent COD (Figure 8c). The horizontal placement of these lines reflects the SHAP values, offering a numerical assessment of their impact on the model's results. Analysing the distribution along the x-axis reveals that higher temperature values are associated with lower cumulative biogas and decreased reactor performance (Figure 8c). Two *Clostridium* species, temperature, time, pH were the top five factors influencing specific daily biogas (Figure 8d). Overall, the predictive models emphasised the importance of operational parameters on reactor performance, and biological genera within *Firmicutes* (*Clostridium* and *Oscillibacter*) on biogas production (Figure 8b-d).

Dissimilarity analysis revealed that influent COD and OLR were the most influential operational parameters on beta diversity (Figure 8a). Influent butyric acid content, and mannitol, arabitol, and maltitol concentrations strongly influenced beta diversity (Figure 8a).

Group 3 reactors were distributed across three main clusters, which appeared to be strongly influenced by individual sugars and sugar alcohols. Group 1 formed one central cluster and was strongly influenced by total sugars and sugar alcohol content (Figure 8a). Influent isobutyric, butyric, and acetic acid content appeared to strongly influence diversity (Figure 8a). Notably, individual VFAs appeared to have a more substantial influence on beta diversity than total VFA content, highlighting the importance of detailed chemical analysis for understanding microbiome diversity (Figure 8a).

Feedstock composition was identified as a critical driver of biological parameters across all reactors. The use of ML models, particularly random forests, has proven highly effective in this work. The agreement between the dissimilarity analysis and ML models emphasised the robustness of the ML models and highlighted that influent COD and OLR were highly influential on microbiome diversity and reactor performance (Figure 8).

**a)** **Combined beta dissimilarity analysis**

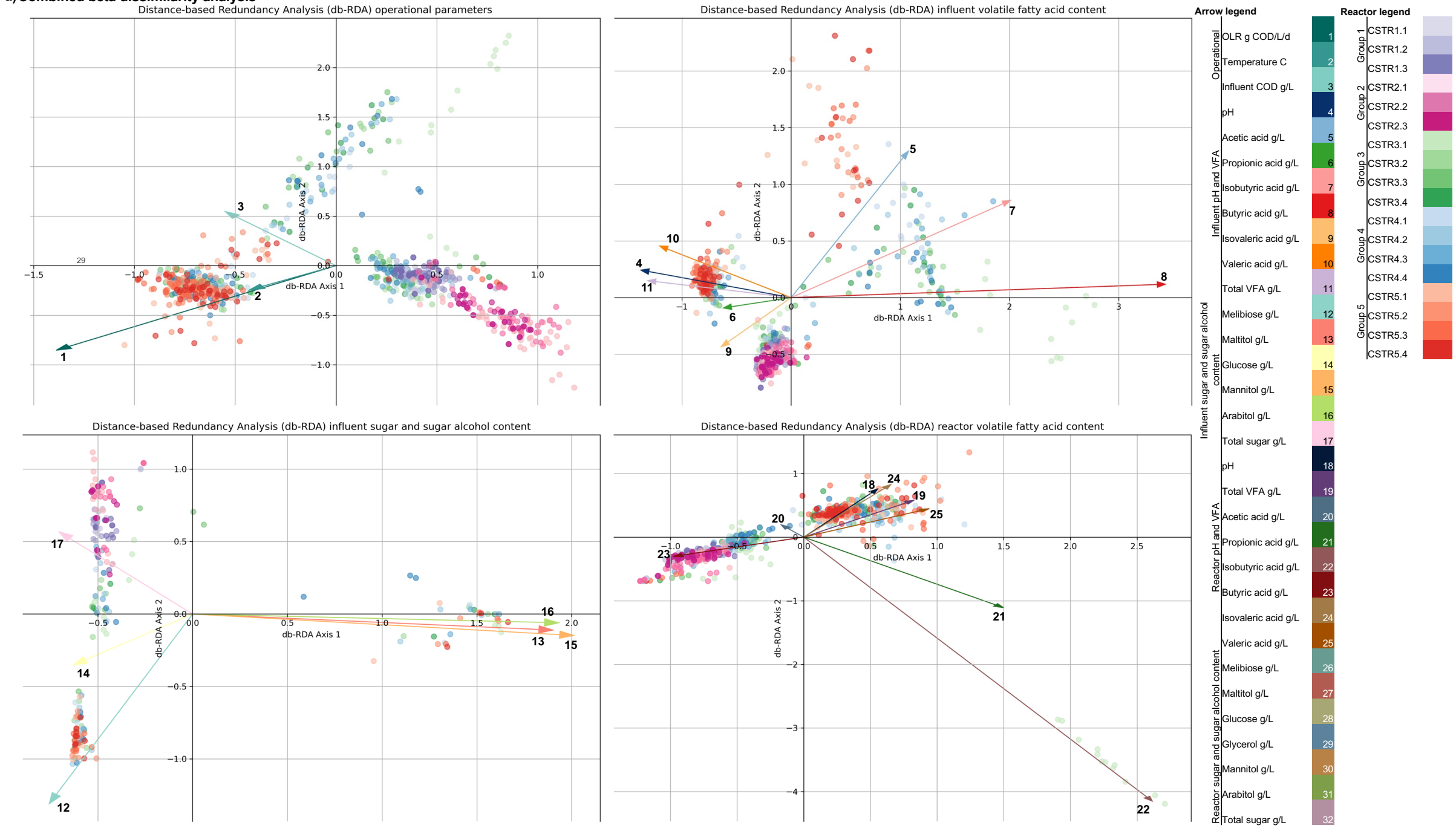

**b)**

| Inputs | Outputs |
|---|---|
| Organic Loading Rate | COD removal |
| Temperature | |
| pH | |
| Influent COD | |
| Sugar alcohol | |
| Amino acid | |
| Relative Abundance of Genera | |

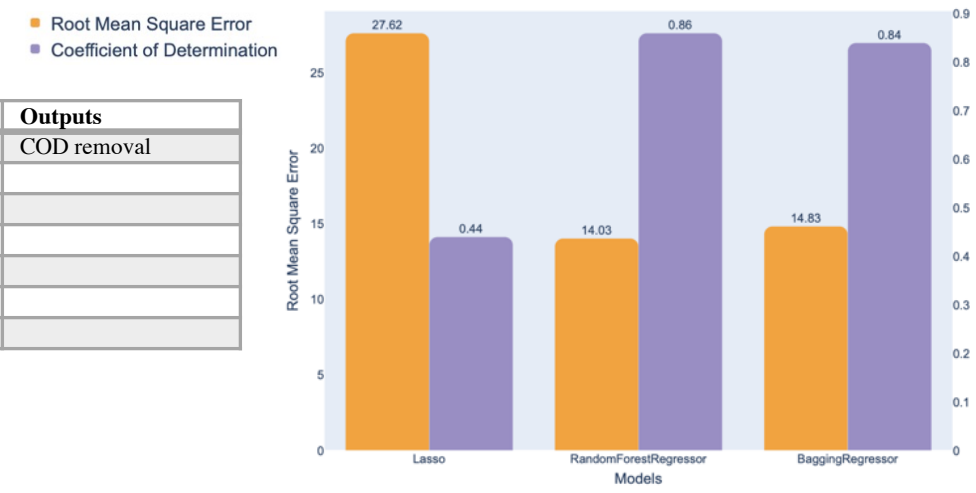

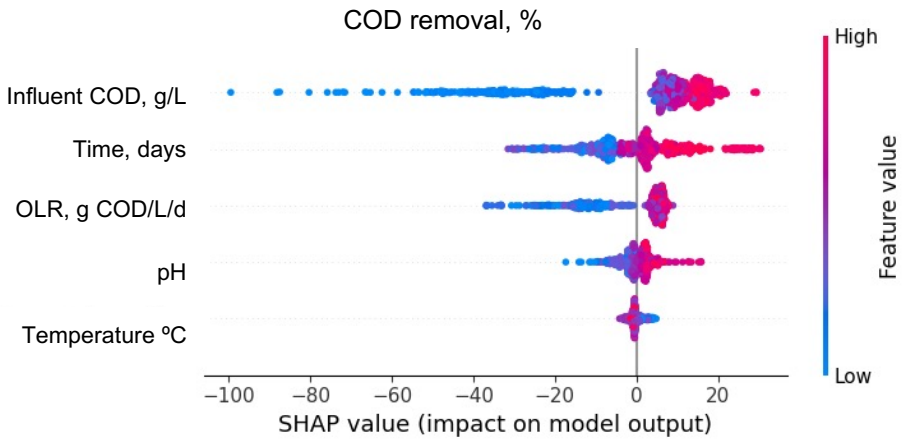

**c)**

| Inputs | Outputs |
|---|---|
| Day | Cumulative Biogas |
| Organic Loading Rate | |
| Temperature | |
| pH | |
| Influent COD | |
| Relative Abundance of Genera | |

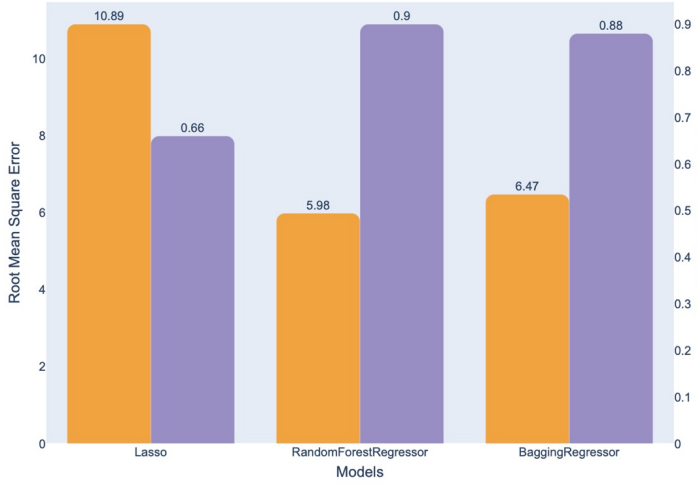

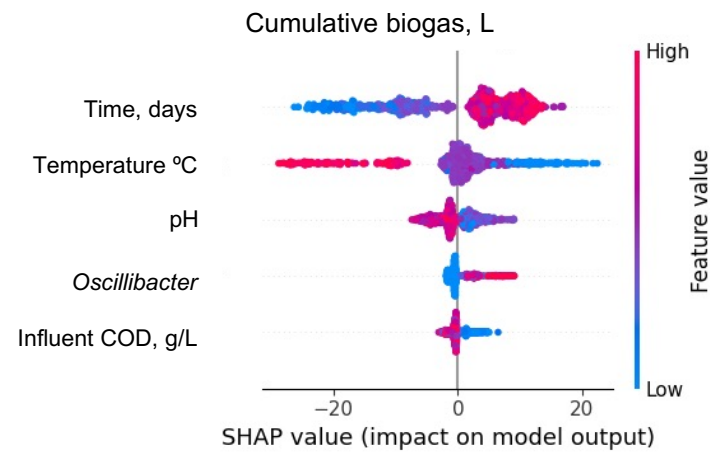

**d)**

| Inputs | Output |
|---|---|
| Day | Biogas L/g COD |
| Organic Loading Rate | |
| Temperature | |
| pH | |
| Influent COD | |
| Relative Abundance of Genera | |

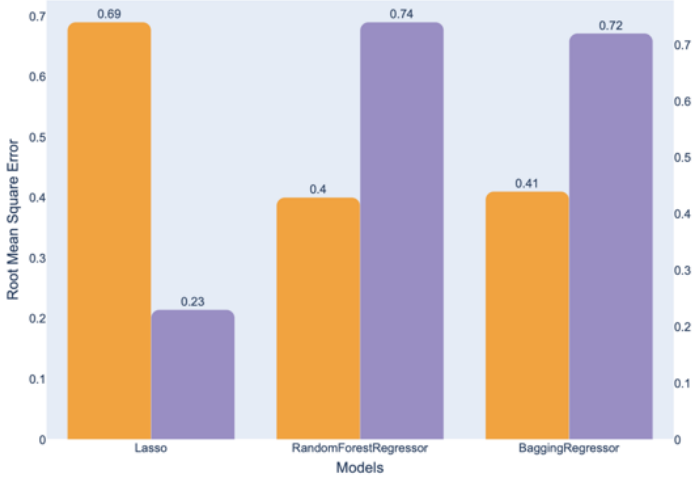

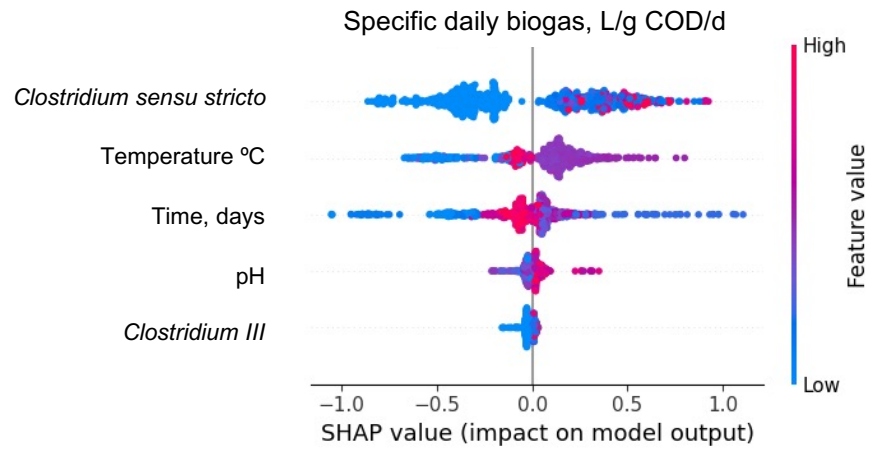

**Figure 8.** Effect of operational and chemical parameters on microbiome biodiversity and performance. **(a)** Dissimilarity analysis of mesophilic (M1 to M3, purple) and thermophilic (T1 to T3, pink) reactors assessing the influence of operational parameters, influent VFA content, influent sugar and sugar alcohol content, and reactor VFA content. Machine learning predictive models includes Lasso regression, random forests and bagging regression**.** The inputs and outputs of each model are described in the tables, the RMSE and $R^2$ values used to evaluate the models are presented as bar charts and factor influence is displayed in terms of SHAP values for each model predicting reactor performance in terms of **(b)** COD removal. **(c)** Cumulative biogas, L. **(d)** Specific daily biogas, L biogas/g COD/d. *Abbreviations: organic loading rate (OLR), chemical oxygen demand (COD), volatile fatty acid (VFA), root mean squared error (RMSE).

## 3.6. Discussion

Biological repeats of various anaerobic digestion reactors fed with MFWW were monitored over a time series to evaluate the effect of feedstock composition on the microbial community underpinning anaerobic digestion across different operational parameters and time scales.

Influent COD and OLR were also identified as key influences on reactor performance by ML and dissimilarity analysis. Dissimilarity analysis showed strong clustering of reactor type by operational parameters, implying a strong impact on microbial diversity (Figure 8a). Additionally, individual VFAs were found to have a stronger influence on beta diversity than pH or total VFA, as did individual sugars and sugar alcohols compared to total sugar and sugar alcohol concentrations (Figure 8a). Isobutyric acid content in particular was found to have a major influence on microbiome biodiversity, as were influent acetic, isobutyric, and butyric acid levels (Figure 8a).

Bioreactor chemistry highlighted distinct VFA profiles between runs 1 and 2 (Figure 4c). VFA profiles in run 1 consisted of a majority of isobutyric acid, which was correlated to the relative abundance of *Methanomassiliicoccus* [32]. The detection of both acetoclastic and hydrogenotrophic methanogens in various reactors underscores the flexibility and resilience of AD microbial communities [34]. It is important to note that *Methanomassiliicoccus* lack the Wood-Ljungdahl pathway for methyl group oxidation to carbon dioxide [35]. *Methanomassiliicoccus* have instead been found to use hydrogen-dependent methylotrophic methanogenesis pathways, utilising methylamines and methanol, which highlights an unusual metabolic pathway compared to classic hydrogenotrophic or acetoclastic methanogens [36,38]. Additionally, *Methanomassiliicoccus* have been associated with reduced biogas production [33], and were significantly correlated with syntrophic bacteria only in group 5, which had the lowest biogas production (Figure 6d). Therefore, this study provides further evidence to support the hypothesis that *Methanomassiliicoccus* is associated with butyric acid and reduced biogas production. Moreover, this study has also highlighted the importance of *Firmicutes* in relation to reactor performance. The relative abundance of *Firmicutes* was elevated and showed little conservation over time in run 1 compared to run 2 samples. This result implies that a relatively low abundance of *Firmicutes* and a dynamic *Firmicutes* population was associated with increased biogas production. Furthermore, ML models identified three genera of *Firmicutes* (*Oscillibacter*, and two *Clostridium*) as primary factors that influence biogas production (Figure 8c-d). These combined results implicate *Firmicutes* as key bacteria for influencing in biogas production (Figure 5, Figure 8).

A shift in taxonomic profile observed between days 43 and 50 for almost all reactors corresponded to increased influent COD and increased OLR (Figure 4c, Figure 5). Dissimilarity analysis and ML outcomes revealed that OLR had the most substantial effect on microbiome diversity, and that influent COD was a key factor in modifying reactor performance (Figure 8). The shift in taxonomy was associated with improved performance and microbiome diversity stability, indicating that flexibility within the microbiome improves performance in response to fluctuating feedstock composition (Figure 4b, Figure 5).

Despite having identical feedstock (MFWW), operational conditions and starting inoculum, CSTR3.1 had a distinct taxonomic profile compared to other single-stage reactors but comparable biogas production and COD removal (Figure 4b, Figure 5). This result indicates that reactor COD removal and biogas production was decoupled from the microbiome profile. Syntrophic relationships showed stronger correlations in single- and first-stage compared to second-stage reactors (Figure 6d). Additionally, network analysis revealed that syntrophic species had higher node connectivity in mesophilic compared to thermophilic reactors (Figure 6). Several bacterial genera are involved in syntrophic relationships essential for AD. *Syntrophobacter* converts butyrate and propionate to acetate, which is then utilised by methanogens such as *Methanosarcina* and *Methanosaeta* [37]. Similarly, *Syntrophomonas wolfei* degrades butyrate in association with hydrogen-utilising methanogens, maintaining low hydrogen partial pressure, which is critical for the energetics of the reaction [38]. Additionally, *Smithella* species grow on propionate only in the presence of methanogenic bacteria that remove hydrogen and formate, facilitating propionate oxidation [36]. *Candidatus Cloacamonas*, part of the WWE1 candidate phylum, and *Syntrophomonas* are involved in the fermentation of amino acids and butyrate, respectively, indicating their significant roles in syntrophic degradation processes (Figure 1) [39]. Furthermore, the presence of syntrophic bacteria like *Thermotoga* and *Thermacetogenium* in thermophilic digesters supports syntrophic acetate oxidation under high-temperature and high-ammonia conditions, highlighting the adaptability and importance of these bacteria in diverse AD environments [27]. Additionally, *Cytophaga*, *Herbaspirillum*, *Symbiobacterium*, *Comamonas*, and *Allochromatium* have been associated with enhanced biogas production, although their exact roles remain unclear [40,41].

A greater understanding of the microbiome underpinning AD could improve reactor stability and reduce start-up times [28,42,43]. The well-described process of acetoclastic methanogenesis has formed the basis for much top-down AD microbiome optimisation. However, our results agree with the findings that the obligate acetoclastic methanogen *Methanosaeta* has been associated with reactor dysfunction, and acetoclastic methanogenesis is not always the dominant pathway for methane production [41,43-45].

# 4. Conclusions and Future Directions

This study investigated the effect of complex feedstock (MFWW) from the microbial protein industry on microbiome diversity and reactor performance in a time series. ML models and dissimilarity analysis highlighted the importance of crude and detailed chemical analysis for influencing reactor biodiversity and performance. The findings also indicated that *Methanomassiliicoccus* is associated with high butyric acid concentrations and reduced biogas production, providing new insights into the newly discovered methanogen. Finally, *Firmicutes* were implicated in reactor performance and were found to be highly dynamic in reactors with high biogas production.

Future research should focus on building upon the findings of this study by elongating the run times, including untargeted chemical analysis and additional -omics tools. HPLC analysis was limited to known compounds, including VFAs and sugars and sugar alcohols, which failed to identify all compounds of interest within the reactors. Untargeted analysis would be recommended to identify all compounds of interest within the reactors. Furthermore, individual compound manipulation of characterised feedstocks (MFWW) could be conducted to validate the findings of the dissimilarity analysis and ML models. Preliminary exploration of bioreactor chemistry and operational parameters as predictors of microbiome structure was used to investigate whether microbial community assembly is deterministic or stochastic (Supplementary Materials 3.3). However, more data will be required to further explore this topic. Moreover, advanced time series analysis was challenging to implement. An extended run time of at least 6 months would be recommended to further explore the temporal trends and whether community can be predicted based on feedstock composition and operational conditions. Finally, we recommend additional -omics to investigate the role of structure and function in relation to reactor chemistry and performance parameters. Meta-transcriptomics is a common approach used in microbiome studies to elucidate the functional profile of the microbiome and should be applied to further explore some interesting relationships observed during the experimental run, such as the metabolic basis for the elevated production of valeric acid in CSTR3.1. A meta-pan genomic approach could also be applied to compare reactor metagenomes and their relationship to bioreactor chemistry.

This work developed robust predictive ML models based on AD chemical and biological fingerprinting. The combined ML approach derived actionable insights for the potential of *Firmicutes*, *Methanomassiliicoccus* and butyric acid as biological and chemical markers to monitor and predict system performance. By integrating biological, chemical, and computational methods, researchers can develop robust monitoring methods that consider whole systems approaches to enhance AD performance. Detailed feedstock analysis in tandem with the application of metagenomic sequencing and reactor chemical analysis are key methods for capturing the complexity of these communities. By investigating substrates ranging from single molecules to complex macroparticles, we may be able uncover microbial structure and function trends. By effectively managing operational parameters and bioreactor chemistry, it is possible to harness the full potential of AD, ensuring efficient wastewater treatment and resource recovery.

## Data Availability

All sequencing data have been submitted to the European Nucleotide Archive (ENA) under the project ID PRJEB80086. Sequences reported in this study are deposited in the EMBL-EBI database under accession numbers ERS22193437-ERS22335023. The correspondence between accession numbers, sequences and metadata is provided in Supplementary Database 1. Machine learning models are open source and available at https://github.com/MGuo-Lab/AnaerobicDigestionML.

## Declaration of competing interests

The authors declare no known conflicts of interest.

## Author contributions

EP and MG conceived the study. EP acquired the data. EP and XS conducted the experiments, analysed and interpreted the data. EP, XS, PE, MT and MG drafted or revised the manuscript.

# Acknowledgements

We are grateful to Marlow Ingredients and the research and development team, especially Dr Mark Taylor, for supplying the feedstock for the experiments. EP and MG would like to acknowledge the UK Engineering and Physical Sciences Research Council (EPSRC) and Monde Nissin Corporate for providing fundings under the EPSRC iCASE programme.

# Temporal Dynamics of Microbial Communities in Anaerobic Digestion: Influence of Temperature and Feedstock Composition on Reactor Performance and Stability: Supplementary Materials and Methods

Ellen Piercy [a], Xinyang Sun [a], Miao Guo [a]*.

[a] Department of Engineering, Faculty of Natural, Mathematical & Engineering Sciences, King's College London, Strand Campus, London, WC2R 2LS, UK. E-mail: miao.guo@kcl.ac.uk.

# Table of Contents

# 1. Supplementary Introduction and Literature Review

## 1.1. Reported substrate utilisation and categorisation

**Supplementary Table 1.1.** Functional categorisation of substrates reported to support the growth of species involved in AD. The functional categorisation was based on the molecular formula, condensed formula, ionic charge, and elemental composition in terms of carbon (C), hydrogen (H), nitrogen (N), oxygen (O), sulphur (S), chlorine (Cl) and iron (Fe), molecular weight, and the number of functional groups. Ionic charge is presented on a colour scale of red (strongly negative) to blue (strongly positive). The number of functional groups is colour-coded from light blue (low) to dark blue (high), where black represents a given functional group's absence and NA indicates information unavailable.

| Functional Category | Name | Molecular formula | Condensed Formula | Ionic Charge | Elemental composition | | | | | | | Molecular weight | R-COOH, carboxylic acid | R-COO$^-$, carboxylate | R-OH, hydroxyl | R-NH$_2$, amine | R-NH | R-N-R | R-NH$_3^+$ | R-NO$_3^-$ | Nitrile, CN | R-N-(C=O)-R, amide | R-C=O, ketone | R-HC=O, aldehyde | R-O-C=O, Ester | R-CH$_3$, methyl | R-O-CH$_3$, oxymethyl | R-Cl | R-Fe$^{3+}$ | R-S | C=C | Benzene | Phenyl | Pyridine | Imidazole | Indole | Oleate | pyranose | furanose | pyrrolizidine | Glucosamine | Purine | Pteridine |
|---|---|---|---|---|---|---|---|---|---|---|---|---|---|---|---|---|---|---|---|---|---|---|---|---|---|---|---|---|---|---|---|---|---|---|---|---|---|---|---|---|---|---|
| | | | | | C | H | N | O | S | Cl | Fe | | | | | | | | | | | | | | | | | | | | | | | | | | | | | | | | |
| Reducing agent | Hydrogen | $H_2$ | $H_2$ | 0 | 0 | 2 | 0 | 0 | 0 | 0 | 0 | 2.02 | | | | | | | | | | | | | | | | | | | | | | | | | | | | | | | |
| Oxidising agent | Carbon dioxide | $CO_2$ | $CO_2$ | 0 | 1 | 0 | 0 | 2 | 0 | 0 | 0 | 44.01 | | | | | | | | | | | | | | | | | | | | | | | | | | | | | | | |
| | Fe(III) | $Fe^{3+}$ | $Fe^{3+}$ | 3 | 0 | 0 | 0 | 0 | 0 | 0 | 1 | 55.85 | | | | | | | | | | | | | | | | | 1 | | | | | | | | | | | | | | |
| | Carbon monoxide | $CO$ | $CO$ | 0 | 1 | 0 | 0 | 1 | 0 | 0 | 0 | 28.01 | | | | | | | | | | | | | | | | | | | | | | | | | | | | | | | |
| Redox agent | Nitrate | $NO_3^-$ | $NO_3^-$ | -1 | 0 | 0 | 1 | 3 | 0 | 0 | 0 | 62.00 | | | | | | | | 1 | | | | | | | | | | | | | | | | | | | | | | | |
| Methylating agent | Dichloromethane | $C_2H_2Cl_2$ | $CH_2Cl_2$ | 0 | 2 | 2 | 0 | 0 | 0 | 2 | 0 | 96.94 | | | | | | | | | | | | | | | | 1 | | | | | | | | | | | | | | | |
| Ethylating agent | Chloroethane | $C_2H_5Cl$ | $CH_3CH_2Cl$ | 0 | 2 | 5 | 0 | 0 | 0 | 1 | 0 | 64.51 | | | | | | | | | | | | | | | | 2 | | | | | | | | | | | | | | | |
| | Oxalate | $C_2O_4^{(2-)}$ | $(COO^-)_2$ | -2 | 2 | 0 | 0 | 4 | 0 | 0 | 0 | 88.02 | | 2 | | | | | | | | | | | | | | | | | | | | | | | | | | | | | |
| | Formate | $CHO_2^-$ | $HCOO^-$ | -1 | 1 | 1 | 0 | 2 | 0 | 0 | 0 | 45.02 | | 1 | | | | | | | | | | | | | | | | | | | | | | | | | | | | | |
| | Acetate | $C_2H_3O_2^-$ | $CH_3COO^-$ | -1 | 2 | 3 | 0 | 2 | 0 | 0 | 0 | 59.04 | | 1 | | | | | | | | | | | | | | | | | | | | | | | | | | | | | |
| | Propionate | $C_3H_5O_2^-$ | $CH_3CH_2COO^-$ | -1 | 3 | 5 | 0 | 2 | 0 | 0 | 0 | 73.07 | | 1 | | | | | | | | | | | | | | | | | | | | | | | | | | | | | |
| | Butyrate | $C_4H_7O_2^-$ | $CH_3CH_2CH_2COO^-$ | -1 | 4 | 7 | 0 | 2 | 0 | 0 | 0 | 87.10 | | 1 | | | | | | | | | | | | | | | | | | | | | | | | | | | | | |
| | Isobutyrate | $C_4H_7O_2^-$ | $(CH_3)_2CHCOO^-$ | -1 | 4 | 7 | 0 | 2 | 0 | 0 | 0 | 87.10 | | 1 | | | | | | | | | | | | 1 | | | | | | | | | | | | | | | | | |
| | Valerate | $C_5H_9O_2^-$ | $CH_3(CH_2)_3COO^-$ | -1 | 5 | 9 | 0 | 2 | 0 | 0 | 0 | 101.12 | | 1 | | | | | | | | | | | | | | | | | | | | | | | | | | | | | |
| | Caproate | $C_6H_{11}O_2^-$ | $CH_3(CH_2)_4COO$- | -1 | 6 | 11 | 0 | 2 | 0 | 0 | 0 | 115.15 | | 1 | | | | | | | | | | | | | | | | | | | | | | | | | | | | | |
| | Haptonoate/Enanthic acid | $C_7H_{13}O_2^-$ | $CH_3(CH_2)_5COO^-$ | -1 | 7 | 13 | 0 | 2 | 0 | 0 | 0 | 129.18 | | 1 | | | | | | | | | | | | | | | | | | | | | | | | | | | | | |
| | Isoheptanoate | $C_7H_{13}O_2^-$ | $CH_3(CH_2)_3CH(COO^-)CH_3$ | -1 | 7 | 13 | 0 | 2 | 0 | 0 | 0 | 129.18 | | 1 | | | | | | | | | | | | | | | | | | | | | | | | | | | | | |
| | Octonoate/Caprylate | $C_8H_{15}O_2^-$ | $CH_3(CH_2)_6COO^-$ | -1 | 8 | 15 | 0 | 2 | 0 | 0 | 0 | 143.20 | | 1 | | | | | | | | | | | | | | | | | | | | | | | | | | | | | |
| | Nonanoate/Pelargonate | $C_9H_{17}O_2^-$ | $CH_3(CH_2)_7COO^-$ | -1 | 9 | 17 | 0 | 2 | 0 | 0 | 0 | 157.23 | | 1 | | | | | | | | | | | | | | | | | | | | | | | | | | | | | |
| | Decanoate/Caprate | $C_{10}H_{19}O_2^-$ | $CH_3(CH_2)_8COO^-$ | -1 | 10 | 19 | 0 | 2 | 0 | 0 | 0 | 171.26 | | 1 | | | | | | | | | | | | | | | | | | | | | | | | | | | | | |
| | Dodecanoate/Laurate | $C_{12}H_{23}O_2^-$ | $CH_3(CH_2)_{10}COO^-$ | -1 | 12 | 23 | 0 | 2 | 0 | 0 | 0 | 199.31 | | 1 | | | | | | | | | | | | | | | | | | | | | | | | | | | | | |
| | Tetradecanoate/Myristate | $C_{14}H_{27}O_2^-$ | $CH_3(CH_2)_{12}COO$ | -1 | 14 | 27 | 0 | 2 | 0 | 0 | 0 | 227.36 | | 1 | | | | | | | | | | | | | | | | | | | | | | | | | | | | | |
| | Palmitate | $C_{16}H_{31}O_2^-$ | $CH_3(CH_2)_{14}COO^-$ | -1 | 16 | 31 | 0 | 2 | 0 | 0 | 0 | 255.42 | | 1 | | | | | | | | | | | | | | | | | | | | | | | | | | | | | |
| | Octadecanoate/Stearate | $C_{18}H_{35}O_2^-$ | $CH_3(CH_2)_{16}COO^-$ | -1 | 18 | 35 | 0 | 2 | 0 | 0 | 0 | 283.47 | | 1 | | | | | | | | | | | | | | | | | | | | | | | | | | | | | |
| | Linoleate | $C_{18}H_{31}O_2^-$ | $COO^-(CH_2)CH=CHCH_2CH=CH(CH_2)_4CH_3$ | -1 | 18 | 31 | 0 | 2 | 0 | 0 | 0 | 279.44 | | 1 | | | | | | | | | | | | | | | | | 2 | | | | | | | | | | | | |
| | Crotonate/2-butenoate/b-methylacrylate | $C_4H_5O_2^-$ | $CH_3HC=CHCOO^-$ | -1 | 4 | 5 | 0 | 2 | 0 | 0 | 0 | 85.08 | | 1 | | | | | | | | | | | | | | | | | 1 | | | | | | | | | | | | |
| | Cinnamate | $C_9H_7O_2^-$ | $C_6H_5CH=CHCOO^-$ | -1 | 9 | 7 | 0 | 2 | 0 | 0 | 0 | 147.15 | | 1 | | | | | | | | | | | | | | | | | 1 | | 1 | | | | | | | | | | |
| | Pyruvate | $C_3H_3O_3^-$ | $CH_3COCOO^-$ | -1 | 3 | 3 | 0 | 3 | 0 | 0 | 0 | 87.05 | | 1 | | | | | | | | | 1 | | | | | | | | | | | | | | | | | | | | |
| | Lactate | $C_3H_5O_3^-$ | $CH_3CH(OH)COO^-$ | -1 | 3 | 5 | 0 | 3 | 0 | 0 | 0 | 89.07 | | 1 | 1 | | | | | | | | | | | | | | | | | | | | | | | | | | | | |
| | Fumarate | $C_4H_2O_4^-$ | $HOOC(CH)=(CH)COO^-$ | -1 | 4 | 2 | 0 | 4 | 0 | 0 | 0 | 114.06 | 1 | 1 | | | | | | | | | | | | | | | | | 1 | | | | | | | | | | | | |
| | Glutarate | $C_5H_7O_4^-$ | $COOH(CH_2)_3COO^-$ | -1 | 5 | 7 | 0 | 4 | 0 | 0 | 0 | 131.11 | 1 | 1 | | | | | | | | | | | | | | | | | | | | | | | | | | | | | |
| | Gluconate | $C_6H_{11}O_7^-$ | $(OH)CH_2(CHOH)_4COO^-$ | -1 | 6 | 11 | 0 | 7 | 0 | 0 | 0 | 195.15 | | 1 | 5 | | | | | | | | | | | | | | | | | | | | | | | | | | | | |
| | Succinate/Butanedioate | $C_4H_4O_4^{2-}$ | $(COO^-)_2(CH_2)_2$ | -2 | 4 | 4 | 0 | 4 | 0 | 0 | 0 | 116.07 | | 2 | | | | | | | | | | | | | | | | | | | | | | | | | | | | | |
| | Malate | $C_4H_4O_5^{2-}$ | $(COO^-)CH_2CH(OH)COO^-$ | -2 | 4 | 4 | 0 | 5 | 0 | 0 | 0 | 132.07 | | 2 | 1 | | | | | | | | | | | | | | | | | | | | | | | | | | | | |
| | Tartrate | $C_4H_4O_6^{2-}$ | $^-OOC(CH)_2COO^-$ | -2 | 4 | 4 | 0 | 6 | 0 | 0 | 0 | 148.07 | | 2 | | | | | | | | | | | | | | | | | | | | | | | | | | | | | |
| | Adipate | $C_6H_{10}O_4^{2-}$ | $(CH_2)_4(COO^-)_2$ | -2 | 6 | 10 | 0 | 4 | 0 | 0 | 0 | 146.14 | | 2 | | | | | | | | | | | | | | | | | | | | | | | | | | | | | |
| Carboxylate | Citrate | $C_6H_5O_7^{2-}$ | $COO^-CH_2C(OH)(COO^-)CH_2COO^-$ | -2 | 6 | 5 | 0 | 7 | 0 | 0 | 0 | 189.10 | | 3 | 1 | | | | | | | | | | | | | | | | | | | | | | | | | | | | |
| | Glyoxylate | $C_2HO_3^-$ | $COH(CH)OO^-$ | -1 | 2 | 1 | 0 | 3 | 0 | 0 | 0 | 73.03 | | 1 | 1 | | | | | | | | | | | | | | | | | | | | | | | | | | | | |
| | Glycolate | $C_2H_3O_3^-$ | $HO(CH2)COO^-$ | -1 | 2 | 3 | 0 | 3 | 0 | 0 | 0 | 75.04 | | 1 | 1 | | | | | | | | | | | | | | | | | | | | | | | | | | | | |
| | Glycerate | $C_3H_5O_4^-$ | $CH_2(OH)CH(OH)COO^-$ | -1 | 3 | 5 | 0 | 4 | 0 | 0 | 0 | 105.07 | 1 | 1 | 2 | | | | | | | | | | | | | | | | | | | | | | | | | | | | |
| | Hydroxybutyrate | $C_4H_7O_3^-$ | $CH_3CH_2CH(OH)COO^-$ | -1 | 4 | 7 | 0 | 3 | 0 | 0 | 0 | 103.10 | | 1 | 1 | | | | | | | | | | | | | | | | | | | | | | | | | | | | |
| | 2-Hydroxybutyrate | $C_4H_7O_3^-$ | $CH_3CH_2CH(OH)COO^-$ | -1 | 4 | 7 | 0 | 3 | 0 | 0 | 0 | 103.10 | | 1 | 1 | | | | | | | | | | | | | | | | | | | | | | | | | | | | |
| | 3-Hydroxybutyrate | $C_4H_7O_3^-$ | $CH_3CH(OH)CH_2COO^-$ | -1 | 4 | 7 | 0 | 3 | 0 | 0 | 0 | 103.10 | | 1 | 1 | | | | | | | | | | | | | | | | | | | | | | | | | | | | |
| | 4-Hydroxybutyrate | $C_4H_7O_3^-$ | $HO(CH_2)_3COO^-$ | -1 | 4 | 7 | 0 | 3 | 0 | 0 | 0 | 103.10 | | 1 | 1 | | | | | | | | | | | | | | | | | | | | | | | | | | | | |
| Hydroxy carboxylate | Ferulate/Ethyl 4-hydroxy-3-methoxycinnamate | $C_{10}H_{10}O_4$ | $HOC_6H_3CH=CHCOO^-$ | 0 | 10 | 10 | 0 | 4 | 0 | 0 | 0 | 194.18 | | 1 | 1 | | | | | | | | | | | | | | | | 1 | | 1 | | | | | | | | | | |
| | Methylpropionate | $C_4H_7O_2^-$ | $CH_3CH_2(COO^-)CH_3$ | -1 | 4 | 7 | 0 | 2 | 0 | 0 | 0 | 87.10 | | 1 | | | | | | | | | | | | 1 | | | | | | | | | | | | | | | | | |
| | 3-methyl-butyrate | $C_5H_9O_2^-$ | $CH_3(CH_2)_2(COO^-)CH_3$ | -1 | 5 | 9 | 0 | 2 | 0 | 0 | 0 | 101.12 | | 1 | | | | | | | | | | | | 1 | | | | | | | | | | | | | | | | | |
| Methyl carboxylate | 4-Methyl-valerate/4-methylpentanoate | $C_6H_{11}O_2^-$ | $CH_3(CH_2)_3(COO^-)CH_3$ | -1 | 6 | 11 | 0 | 2 | 0 | 0 | 0 | 115.15 | | 1 | | | | | | | | | | | | 1 | | | | | | | | | | | | | | | | | |
| | Syringate/4-hydroxy-3,5-dimethoxybenzoate/3,5-dimethoxy-4-hydroxybenzoate | $C_9H_9O_5$ | $COO^-C=CHC(OCH_3)=C(OH)C(OCH_3)=CH$ | 0 | 9 | 9 | 0 | 5 | 0 | 0 | 0 | 197.16 | | 1 | 1 | | | | | | | | | | | | 2 | | | | 3 | | 1 | | | | | | | | | | |
| | 3-methoxycinnamate | $C_{10}H_9O_3^-$ | $(CH_3)C_6H_4CH=CHCOO^-$ | -1 | 10 | 9 | 0 | 3 | 0 | 0 | 0 | 177.18 | | 1 | | | | | | | | | | | | | 1 | | | | 1 | | 1 | | | | | | | | | | |
| | 4-methyoxycinnamate/p-methoxycinnamate | $C_{10}H_9O_3^-$ | NA | -1 | 10 | 9 | 0 | 3 | 0 | 0 | 0 | 177.18 | | 1 | | | | | | | | | | | | | 1 | | | | 1 | | 1 | | | | | | | | | | |
| | 3,4-dimethoxycinnamate | $C_{11}H_{11}O_4^-$ | NA | -1 | 11 | 11 | 0 | 4 | 0 | 0 | 0 | 207.20 | | 1 | | | | | | | | | | | | | 1 | | | | 1 | | 1 | | | | | | | | | | |
| Methyloxy carboxylate | 3,4,5-trimethoxycinnamate | $C_{12}H_{13}O_5^-$ | NA | -1 | 12 | 13 | 0 | 5 | 0 | 0 | 0 | 237.23 | | 1 | | | | | | | | | | | | | 1 | | | | 1 | | 1 | | | | | | | | | | |
| | Vanillate/4-hydroxy-3-methoxybenzoate | $C_8H_8O_4$ | $COO^-C=CHC(OCH_3)=C(OH)HC=CH$ | 0 | 8 | 8 | 0 | 4 | 0 | 0 | 0 | 168.15 | | 1 | 1 | | | | | | | | | | | | 1 | | | | 3 | | 1 | | | | | | | | | | |
| Hydoxyl methyloxy carboxylate | Sinapate/3,5-dimethoxy-4-hydroxycinnamate | $C_{11}H_{12}O_5^-$ | $COO^-HC=CH(C=CHC(OCH_3)=COHC(OCH_3)=CH)$ | -1 | 11 | 12 | 0 | 5 | 0 | 0 | 0 | 224.21 | | 1 | 1 | | | | | | | | | | | | 2 | | | | 4 | | 1 | | | | | | | | | | |
| | Phenyl acetate | $C_8H_8O_2$ | $CH3COO(C_6H_5)$ | 0 | 8 | 8 | 0 | 2 | 0 | 0 | 0 | 136.15 | | 1 | | | | | | | | | | | 1 | 1 | | | | | | | 1 | | | | | | | | | | |
| Phenyl carboxylate | 3-phenylpropionate/Hydrocinnamic acid | $C_9H_{10}O_2$ | $COOH(CH_2)_2(C_6H_5)$ | 0 | 9 | 10 | 0 | 2 | 0 | 0 | 0 | 150.18 | 1 | | | | | | | | | | | | | | | | | | | | 1 | | | | | | | | | | |
| | 4-amino-butyrate/γ-aminobutyrate | $C_4H_8NO_2^-$ | $NH_2(CH_2)_3COO^-$ | -1 | 4 | 8 | 1 | 2 | 0 | 0 | 0 | 102.11 | | 1 | | 1 | | | | | | | | | | | | | | | | | | | | | | | | | | | |
| | Aspartate | $C_4H_6NO_4^-$ | $COOHCH2CH(NH2)COO^-$ | -1 | 4 | 6 | 1 | 4 | 0 | 0 | 0 | 132.09 | 1 | 1 | | 1 | | | | | | | | | | | | | | | | | | | | | | | | | | | |
| | Aminovalerate/Aminopentanoate | $C_5H_{11}NO_2$ | $NH_2(CH_2)_4COO^-$ | 0 | 5 | 11 | 1 | 2 | 0 | 0 | 0 | 117.15 | | 1 | | 1 | | | | | | | | | | | | | | | | | | | | | | | | | | | |
| | Glutamate | $C_5H_8NO_4^-$ | $COO^-(CH_2)_2CH(NH_3)COO^-$ | -1 | 5 | 8 | 1 | 4 | 0 | 0 | 0 | 146.12 | | 2 | | | | | 1 | | | | | | | | | | | | | | | | | | | | | | | | |
| Aminocarboxylate | Benzoylamino acetate/Hippurate | $C_9H_{10}NO_2^-$ | $(C_6H_5)CH_2NHCH_2COO^-$ | -1 | 9 | 10 | 1 | 2 | 0 | 0 | 0 | 164.18 | | 1 | | 1 | | | | | | | | | | | | | | | | 1 | | | | | | | | | | | |
| | Methanol | $CH_4O$ | $CH_3OH$ | 0 | 1 | 4 | 0 | 1 | 0 | 0 | 0 | 32.04 | | | 1 | | | | | | | | | | | | | | | | | | | | | | | | | | | | |
| | Ethanol | $C_2H_6O$ | $CH_3CH_2OH$ | 0 | 2 | 6 | 0 | 1 | 0 | 0 | 0 | 46.07 | | | 1 | | | | | | | | | | | | | | | | | | | | | | | | | | | | |
| | Ethylene glycol | $C_2H_6O_2$ | $(CH_2OH)_2$ | 0 | 2 | 6 | 0 | 2 | 0 | 0 | 0 | 62.07 | | | 2 | | | | | | | | | | | | | | | | | | | | | | | | | | | | |
| | 1-propanol | $C_3H_8O$ | $CH_3(CH_2)_2OH$ | 0 | 3 | 8 | 0 | 1 | 0 | 0 | 0 | 60.09 | | | 1 | | | | | | | | | | | | | | | | | | | | | | | | | | | | |
| | 2-propanol/Isopropanol | $C_3H_8O$ | $(CH_3)_2CHOH$ | 0 | 3 | 8 | 0 | 1 | 0 | 0 | 0 | 60.09 | | | 1 | | | | | | | | | | | 1 | | | | | | | | | | | | | | | | | |
| | 1,2-propanediol/Propylene glycol | $C_3H_8O_2$ | $CH_3(CH_2OH)_2$ | 0 | 3 | 8 | 0 | 2 | 0 | 0 | 0 | 76.09 | | | 2 | | | | | | | | | | | | | | | | | | | | | | | | | | | | |
| | 1,3-propanediol | $C_3H_8O_2$ | $CH_2(CH_2OH)_2$ | 0 | 3 | 8 | 0 | 2 | 0 | 0 | 0 | 76.09 | | | 2 | | | | | | | | | | | | | | | | | | | | | | | | | | | | |
| | 1-butanol | $C_4H_{10}O$ | $CH_3(CH_2)_3OH$ | 0 | 4 | 10 | 0 | 1 | 0 | 0 | 0 | 74.12 | | | 1 | | | | | | | | | | | | | | | | | | | | | | | | | | | | |
| | 2-butanol/isobutanol | $C_4H_{10}O$ | $CH_3CH_2CH(OH)CH_3$ | 0 | 4 | 10 | 0 | 1 | 0 | 0 | 0 | 74.12 | | | 1 | | | | | | | | | | | | | | | | | | | | | | | | | | | | |
| | 2-pentanol | $C_5H_{12}O$ | $CH_3CH(OH)(CH_2)_2CH_3$ | 0 | 5 | 12 | 0 | 1 | 0 | 0 | 0 | 88.15 | | | 1 | | | | | | | | | | | | | | | | | | | | | | | | | | | | |
| | 3-pentanol | $C_5H_{12}O$ | $CH_3CH_2CH(OH)CH_2CH_3$ | 0 | 5 | 12 | 0 | 1 | 0 | 0 | 0 | 88.15 | | | 1 | | | | | | | | | | | | | | | | | | | | | | | | | | | | |
| | Cyclopentanol | $C_5H_{10}O$ | $(CH_2)_4CH(OH)$ | 0 | 5 | 10 | 0 | 1 | 0 | 0 | 0 | 86.13 | | | 1 | | | | | | | | | | | | | | | | | | | | | | | | | | | | |
| Alcohol | Adonitol | $C_5H_{12}O_5$ | $CH_2(OH)CH(OH)_3CH_2(OH)$ | 0 | 5 | 12 | 0 | 5 | 0 | 0 | 0 | 152.14 | | | 5 | | | | | | | | | | | | | | | | | | | | | | | | | | | | |
| Amino alcohol | Ethanolamine | $C_2H_7NO$ | $HOCH_2CH_2NH_2$ | 0 | 2 | 7 | 1 | 1 | 0 | 0 | 0 | 61.08 | | | 1 | 1 | | | | | | | | | | | | | | | | | | | | | | | | | | | |
| | Glycerol | $C_3H_8O_3$ | $(CH_2OH)_2CHOH$ | 0 | 3 | 8 | 0 | 3 | 0 | 0 | 0 | 92.09 | | | 3 | | | | | | | | | | | | | | | | | | | | | | | | | | | | |
| | Erythritol | $C_4H_{10}O_4$ | $(CH_2OH)_2(HCOH)_2$ | 0 | 4 | 10 | 0 | 4 | 0 | 0 | 0 | 122.12 | | | 4 | | | | | | | | | | | | | | | | | | | | | | | | | | | | |
| | Arabitol | $C_5H_{12}O_5$ | $(HCOH)_3(CH_2OH)_2$ | 0 | 5 | 12 | 0 | 5 | 0 | 0 | 0 | 152.14 | | | 5 | | | | | | | | | | | | | | | | | | | | | | | | | | | | |
| | Inositol/Cyclohexanehexol | $C_6H_{12}O_6$ | $(CHOH)_6$ | 0 | 6 | 12 | 0 | 6 | 0 | 0 | 0 | 180.15 | | | 6 | | | | | | | | | | | | | | | | | | | | | | | | | | | | |
| | Galactitol/Dulcitol | $C_6H_{14}O_6$ | $(HO)CH_2(HCOH)_4CH_2(OH)$ | 0 | 6 | 14 | 0 | 6 | 0 | 0 | 0 | 182.17 | | | 6 | | | | | | | | | | | | | | | | | | | | | | | | | | | | |
| | Mannitol | $C_6H_{14}O_6$ | $(CH_2OH)_2(CHOH)_4$ | 0 | 6 | 14 | 0 | 6 | 0 | 0 | 0 | 182.17 | | | 6 | | | | | | | | | | | | | | | | | | | | | | | | | | | | |
| Sugar alcohol | Sorbitol/Glucitol | $C_6H_{14}O_6$ | $(CH_2OH)_2(CHOH)_4$ | 0 | 6 | 14 | 0 | 6 | 0 | 0 | 0 | 182.17 | | | 6 | | | | | | | | | | | | | | | | | | | | | | | | | | | | |
| | Ribose | $C_5H_{10}O_5$ | $(CH_2OH)CH_2(CHOH)_3O$ | 0 | 5 | 10 | 0 | 5 | 0 | 0 | 0 | 150.13 | | | 4 | | | | | | | | | 1 | | | | | | | | | | | | | | | | | | | |
| | Xylose | $C_5H_{10}O_5$ | $(CHOH)_5O$ | 0 | 5 | 10 | 0 | 5 | 0 | 0 | 0 | 150.13 | | | 5 | | | | | | | | | | | | | | | | | | | | | | | | | | | | |
| Pentose | Arabinose | $C_5H_{10}O_5$ | $(HO)CH_2(CHOH)_3COOH$ | 0 | 5 | 10 | 0 | 5 | 0 | 0 | 0 | 150.13 | 1 | | 4 | | | | | | | | | 1 | | | | | | | | | | | | | | | | | | | |
| | Galactose | $C_6H_{12}O_6$ | $CH_2(OH)(HCOH)_4COOH$ | 0 | 6 | 12 | 0 | 6 | 0 | 0 | 0 | 180.15 | 1 | | 5 | | | | | | | | | 1 | | | | | | | | | | | | | | | | | | | |
| | Glucose | $C_6H_{12}O_6$ | $H(C=O)(CHOH)_5H$ | 0 | 6 | 12 | 0 | 6 | 0 | 0 | 0 | 180.15 | | | 5 | | | | | | | | | 1 | | | | | | | | | | | | | | | | | | | |
| | Mannose | $C_6H_{12}O_6$ | $CH_2OH(CHOH)_5O$ | 0 | 6 | 12 | 0 | 6 | 0 | 0 | 0 | 180.15 | | | 6 | | | | | | | | | 1 | | | | | | | | | | | | | | | | | | | |
| | Tagatose | $C_6H_{12}O_6$ | $CH_2(OH)(CHOH)_3C=OCH_2OH$ | 0 | 6 | 12 | 0 | 6 | 0 | 0 | 0 | 180.15 | | | 5 | | | | | | | | 1 | | | | | | | | | | | | | | | | | | | | |
| | Sorbose | $C_6H_{12}O_6$ | $CH_2(OH)(CHOH)_4CH_2O$ | 0 | 6 | 12 | 0 | 6 | 0 | 0 | 0 | 180.15 | | | 5 | | | | | | | | 1 | | | | | | | | | | | | | | | | | | | | |
| Hexose | Fructose | $C_6H_{12}O_6$ | NA | 0 | 6 | 12 | 0 | 6 | 0 | 0 | 0 | 180.15 | | | 1 | | | | | | | | 1 | | | | | | | | | | | | | | | | | | | | |
| | Methylmannoside/methyl mannopyranoside | $C_7H_{14}O_6$ | $(HCOH)_2CH_2(OCH_3)COH(O)CHO(CH_3)$ | 0 | 7 | 14 | 0 | 6 | 0 | 0 | 0 | 194.18 | | | 5 | | | | | | | | | | | 1 | 1 | | | | | | | | | | | 1 | | | | | |
| | Methylglucoside | $C_7H_{14}O_6$ | $(HCOH)_3(COHCH_3)O(COHCH_3)$ | 0 | 7 | 14 | 0 | 6 | 0 | 0 | 0 | 194.18 | | | 5 | | | | | | | | | | | 1 | | | | | | | | | | | | 1 | | | | | |
| Methyl hexose | Methylglucopyranoside | $C_7H_{14}O_6$ | $(HCOH)_3CH_2(CH_2OH)O(HCOCH_3)$ | 0 | 7 | 14 | 0 | 6 | 0 | 0 | 0 | 194.18 | | | 4 | | | | | | | | | | | 1 | | | | | | | | | | | | 1 | | | | | |
| | Glucopyranoside | $C_6H_{11}O_6$ | $(HCOH)_3CH_2(H_2COH)OCO$ | 0 | 6 | 11 | 0 | 6 | 0 | 0 | 0 | 179.15 | | | 5 | | | | | | | | | | | | | | | | | | | | | | | 1 | | | | | |
| Methyloxy hexose | Mannopyranoside | $C_6H_{11}O_6$ | $(HCOH)_4COCH_3$ | 0 | 6 | 11 | 0 | 6 | 0 | 0 | 0 | 179.15 | | | 4 | | | | | | | | | | | | | | | | | | | | | | | 1 | | | | | |
| Amino hexose | Glucosamine | $C_6H_{13}NO_5$ | $CH_2(OH)CHOCH(OH)CH(NH_2)(HCOH)_2$ | 0 | 6 | 13 | 1 | 5 | 0 | 0 | 0 | 179.17 | | | 5 | 1 | | | | | | | | | | | | | | | | | | | | | | 1 | | | | | |
| | Acetylglucosamine | $C_8H_{15}NO_6$ | $(HCOH)_2CH(CH_2OH)OCH_2(OH)CH_2(NHCOCH_3)$ | 0 | 8 | 15 | 1 | 6 | 0 | 0 | 0 | 221.21 | | | 4 | | 1 | | | | | 1 | | | | | | | | | | | | | | | | 1 | | | | | |
| Aminoacetylhexose | Acetylgalactosamine | $C_8H_{15}NO_6$ | $(HCOH)_2CH(CH_2OH)OCH_2(OH)CH(NHCOCH_3)$ | 0 | 8 | 15 | 1 | 6 | 0 | 0 | 0 | 221.21 | | | 4 | | 1 | | | | | 1 | | | | | | | | | | | | | | | | 1 | | | | | |
| | Rhamnose/6-deoxy-L-mannopyranose | $C_6H_{12}O_5$ | $(CH_3)CH_2(CHOH)_4O$ | 0 | 6 | 12 | 0 | 5 | 0 | 0 | 0 | 164.16 | | | 4 | | | | | | | | | | | | | | | | | | | | | | | | | | | | |
| Deoxy hexose | Fucose | $C_6H_{12}O_5$ | $CH_3(HCOH)_4COOH$ | 0 | 6 | 12 | 0 | 5 | 0 | 0 | 0 | 164.16 | 1 | | 4 | | | | | | | | | | | | | | | | | | | | | | | | 1 | | | | |
| | Lactose | $C_{12}H_{22}O_{11}$ | NA | 0 | 12 | 22 | 0 | 11 | 0 | 0 | 0 | 342.29 | | | 1 | | | | | | | | | | | | | | | | | | | | | | | | | | | | |
| | Melibiose | $C_{12}H_{22}O_{11}$ | NA | 0 | 12 | 22 | 0 | 11 | 0 | 0 | 0 | 342.29 | | | 1 | | | | | | | | | | | | | | | | | | | | | | | | | | | | |
| | Maltose | $C_{12}H_{22}O_{11}$ | NA | 0 | 12 | 22 | 0 | 11 | 0 | 0 | 0 | 342.29 | | | 1 | | | | | | | | | | | | | | | | | | | | | | | | | | | | |
| | Sucrose/Saccharose | $C_{12}H_{22}O_{11}$ | Glc(α-1,2-β)Fruf | 0 | 12 | 22 | 0 | 11 | 0 | 0 | 0 | 342.29 | | | 1 | | | | | | | | | | | | | | | | | | | | | | | | | | | | |
| | Trehalose | $C_{12}H_{22}O_{11}$ | Glc(α-1,1-α)Glc | 0 | 12 | 22 | 0 | 11 | 0 | 0 | 0 | 342.29 | | | 1 | | | | | | | | | | | | | | | | | | | | | | | | | | | | |
| | Turanose | $C_{12}H_{22}O_{11}$ | Glc(α-1,3)Fruf | 0 | 12 | 22 | 0 | 11 | 0 | 0 | 0 | 342.29 | 1 | | 1 | | | | | | | | | | | | | | | | | | | | | | | | | | | | |
| | Cellobiose | $C_{12}H_{22}O_{11}$ | $(C_6H_7(OH)_4O)_2O$ | 0 | 12 | 22 | 0 | 11 | 0 | 0 | 0 | 342.29 | | | 8 | | | | | | | | | | | | | | | | | | | | | | | | | | | | |
| Disaccharide | Gentiobiose | $C_{12}H_{22}O_{11}$ | NA | 0 | 12 | 22 | 0 | 11 | 0 | 0 | 0 | 342.29 | | | 1 | | | | | | | | | | | | | | | | | | | | | | | | | | | | |
| | Maltotriose | $C_{18}H_{32}O_{16}$ | NA | 0 | 18 | 32 | 0 | 16 | 0 | 0 | 0 | 504.43 | | | 1 | | | | | | | | | | | | | | | | | | | | | | | | | | | | |
| | Melezitose | $C_{18}H_{32}O_{16}$ | NA | 0 | 18 | 32 | 0 | 16 | 0 | 0 | 0 | 504.43 | | | 1 | | | | | | | | | | | | | | | | | | | | | | | | | | | | |
| Trisaccharide | Raffinose | $C_{18}H_{32}O_{16}$ | Gal(α-1,6)Glc(α-1,2β)Fruf | 0 | 18 | 32 | 0 | 16 | 0 | 0 | 0 | 504.43 | | | 1 | | | | | | | | | | | | | | | | | | | | | | | | | | | | |
| | Hemicellulose/polyose | $(C_5H_8O_4)_n$ | NA | 0 | 5 | 8 | 0 | 4 | 0 | 0 | 0 | 132.11 | | | | | | | | | | | | | | | | | | | | | | | | | | | | | | | |
| | Xylan | $C_5H_{10}O_5$ | Xyl(β-2,2-β)Xyl | 0 | 5 | 10 | 0 | 5 | 0 | 0 | 0 | 150.13 | | | 1 | | | | | | | | | | 1 | | 1 | | | | | | | | | | | | | | | | |
| | Amidon/starch | $C_6H_7N_3O$ | $C_5H_4N\text{-}CONH(NH_2)$ | 0 | 6 | 7 | 3 | 1 | 0 | 0 | 0 | 137.14 | | | 1 | | | | | | | | | | | | | | | | | | | | | | | | | | | | |
| | Cellulose | $(C_6H_{10}O_5)_n$ | NA | 0 | 6 | 10 | 0 | 5 | 0 | 0 | 0 | 162.14 | | | | | | | | | | | | | | | | | | | | | | | | | | | | | | | |
| | Lichenin | $(C_6H_{10}O_5)_n$ | NA | 0 | 6 | 10 | 0 | 5 | 0 | 0 | 0 | 162.14 | | | 1 | | | | | | | | | | | | | | | | | | | | | | | | | | | | |
| | Starch | $(C_6H_{10}O_5)_n$ | Glc(α-1,4)Glc | 0 | 6 | 10 | 0 | 5 | 0 | 0 | 0 | 162.14 | | | 1 | | | | | | | | | | | | 1 | | | | | | | | | | | | | | | | |
| | Pectin | $C_6H_{10}O_7$ | $(HCOH)_4OCH(COOH)$ | 0 | 6 | 10 | 0 | 7 | 0 | 0 | 0 | 194.14 | 1 | | 4 | | | | | | | | | | | | | | | | | | | | | | | | | | | | |
| | Arabinoxylan | $(C_{10}H_{16}O_8)_n$ | NA | 0 | 10 | 16 | 0 | 8 | 0 | 0 | 0 | 264.23 | | | | | | | | | | | | | | | | | | | | | | | | | | | | | | | |
| | Glucan | $C_{18}H_{32}O_{16}$ | NA | 0 | 18 | 32 | 0 | 16 | 0 | 0 | 0 | 504.43 | | | 1 | | | | | | | | | | | | | | | | | | | | | | | | | | | | |
| | Laminarin | $C_{18}H_{32}O_{16}$ | NA | 0 | 18 | 32 | 0 | 16 | 0 | 0 | 0 | 504.43 | | | 1 | | | | | | | | | | | | | | | | | | | | | | | | | | | | |
| | Pullulan | $C_{18}H_{32}O_{16}$ | NA | 0 | 18 | 32 | 0 | 16 | 0 | 0 | 0 | 504.43 | | | 1 | | | | | | | | | | | | | | | | | | | | | | | | | | | | |
| | Arabinogalactan/Galactoarabinan | $C_{20}H_{36}O_{14}$ | NA | 0 | 20 | 10 | 0 | 14 | 0 | 0 | 0 | 474.28 | | | 1 | | | | | | | | | | | | 1 | | | | | | | | | | | 1 | | | | | |
| | Glycogen | $C_{24}H_{42}O_{21}$ | NA | 0 | 24 | 42 | 0 | 21 | 0 | 0 | 0 | 666.57 | | | 1 | | | | | | | | | | | | | | | | | | | | | | | | | | | | |
| | Amylopectin | $C_{30}H_{52}O_{26}$ | NA | 0 | 30 | 52 | 0 | 26 | 0 | 0 | 0 | 828.71 | | | 1 | | | | | | | | | | | | | | | | | | | | | | | | | | | | |
| Polysaccharide | Maltodextrin | $C_{6n}H_{10n+2}O_{5n+1}$ | NA | 0 | NA | NA | 0 | NA | 0 | 0 | 0 | NA | | | 1 | | | | | | | | | | | | | | | | | | | | | | | | | | | | |
| Aminopolysaccharide | Chitin | $(C_8H_{13}O_5N)_n$ | NA | 0 | 8 | 13 | 1 | 5 | 0 | 0 | 0 | 203.19 | | | | | 1 | | | | | 1 | | | | 1 | | | | | | | | | | | | | | | 1 | | |
| | Cyclodextrin/Cycloheptaamylose/Cyclomaltoheptaose/Cycloheptaglucan | $(C_6H_{10}O_5)_7$ | $(C_6H_{10}O_5)_7$ | 0 | 6 | 10 | 0 | 5 | 0 | 0 | 0 | 162.14 | | | 5 | | | | | | | | | | | | | | | | | | | | | | | | | | | | |
| | Dextrin | $C_{18}H_{32}O_{16}$ | NA | 0 | 18 | 32 | 0 | 16 | 0 | 0 | 0 | 504.43 | | | 1 | | | | | | | | | | | | | | | | | | | | | | | | | | | | |
| | Galactooligosaccharides | $C_{18}H_{32}O_{16}$ | NA | 0 | 18 | 32 | 0 | 16 | 0 | 0 | 0 | 504.43 | | | 1 | | | | | | | | | | | | | | | | | | | | | | | | | | | | |
| | Fructooligosaccharides | $C_{72}H_{122}O_{61}$ | $Fruf(2\text{-}1)_{11}\text{-}Glc$ | 0 | 72 | 122 | 0 | 61 | 0 | 0 | 0 | 1963.69 | | | 1 | | | | | | | | | | | | | | | | | | | | | | | | | | | | |
| Oligosaccharide | Inulin | $C_{6n}H_{10n+2}O_{5n+1}$ | NA | 0 | NA | NA | 0 | NA | 0 | 0 | 0 | NA | | | 1 | | | | | | | | | | | | | | | | | | | | | | | | 1 | | | | |

[Supplementary Table 1.1 continued]

| Functional Category | Name | Molecular formula | Condensed Formula | Ionic Charge | Elemental composition C | H | N | O | S | Cl | Fe | Molecular weight |
|---|---|---|---|---|---|---|---|---|---|---|---|---|
| | Arbutin | $C_{12}H_{16}O_7$ | NA | 0 | 12 | 16 | 0 | 7 | 0 | 0 | 0 | 272.25 |
| Glycoside | Salicin | $C_{13}H_{18}O_7$ | NA | 0 | 13 | 18 | 0 | 7 | 0 | 0 | 0 | 286.28 |
| Methylglycoside | Methyl a-glycoside | $C_7H_{14}O_6$ | $(HCOH)_3COO(CH_3)CH_2(CH_2OH)$ | 0 | 7 | 14 | 0 | 6 | 0 | 0 | 0 | 194.18 |
| Ester | Polyoxyethylene sorbitan mono-oleate | $C_{64}H_{124}O_{26}$ | NA | 0 | 64 | 124 | 0 | 26 | 0 | 0 | 0 | 1309.65 |
| | Glycine | $C_2H_5NO_2$ | $NH_2CH_2COOH$ | 0 | 2 | 5 | 1 | 2 | 0 | 0 | 0 | 75.07 |
| | Alanine | $C_3H_7NO_2$ | $CH_3CH(NH_2)COOH$ | 0 | 3 | 7 | 1 | 2 | 0 | 0 | 0 | 89.09 |
| | Serine | $C_3H_7NO_3$ | $HOCH_2CH(NH_2)COOH$ | 0 | 3 | 7 | | 3 | 0 | 0 | 0 | 91.08 |
| | Cysteine | $C_3H_7NO_2S$ | $HSCH_2CH(NH_2)COOH$ | 0 | 3 | 7 | 1 | 2 | 1 | 0 | 0 | 121.16 |
| | Threonine | $C_4H_9NO_3$ | $CH_3CH(OH)CH(NH_2)COOH$ | 0 | 4 | 9 | 1 | 3 | 0 | 0 | 0 | 119.12 |
| | Asparagine | $C_4H_8N_2O_3$ | $H_2NCOCH_2CH(NH_2)COOH$ | 0 | 4 | 8 | 2 | 3 | 0 | 0 | 0 | 132.12 |
| | Aspartic acid | $C_4H_7NO_4$ | $COOHCH_2CH(NH_2)COOH$ | 0 | 4 | 7 | 1 | 4 | 0 | 0 | 0 | 133.10 |
| | Proline | $C_5H_9NO_2$ | $NH(CH_2)_3CHCOOH$ | 0 | 5 | 9 | 1 | 2 | 0 | 0 | 0 | 115.13 |
| | Valine | $C_5H_{11}NO_2$ | $(CH_3)_2CHCH(NH_2)COOH$ | 0 | 5 | 11 | 1 | 2 | 0 | 0 | 0 | 117.15 |
| | Glutamine | $C_5H_{10}N_2O_3$ | $H_2NCO(CH_2)_2CH(NH_2)COOH$ | 0 | 5 | 10 | 2 | 3 | 0 | 0 | 0 | 146.14 |
| | Glutamic acid | $C_5H_9NO_4$ | $COOH(CH_2)_2CH(NH_2)COOH$ | 0 | 5 | 9 | 1 | 4 | 0 | 0 | 0 | 147.13 |
| | Leucine | $C_6H_{13}NO_2$ | $(CH_3)_2CHCH_2CH(NH_2)COOH$ | 0 | 6 | 13 | 1 | 2 | 0 | 0 | 0 | 131.17 |
| | Isoleucine | $C_6H_{13}NO_2$ | $CH_3CH_2CH(CH_3)CH(NH_2)COOH$ | 0 | 6 | 13 | 1 | 2 | 0 | 0 | 0 | 131.17 |
| | Lysine | $C_6H_{14}N_2O_2$ | $H_2N(CH_2)_4CH(NH_2)COOH$ | 0 | 6 | 14 | 2 | 2 | 0 | 0 | 0 | 146.19 |
| | Methionine | $C_6H_{14}N_2O_2$ | $CH_3S(CH_2)_2CH(NH_2)COOH$ | 0 | 6 | 14 | 2 | 2 | 0 | 0 | 0 | 146.19 |
| | Histidine | $C_6H_9N_3O_2$ | $NHCH{=}NCH{=}CCH_2CH(NH_2)COOH$ | 0 | 6 | 9 | 3 | 2 | 0 | 0 | 0 | 155.15 |
| | Arginine | $C_6H_{14}N_4O_2$ | $HN{=}C(NH_2)NH(CH_2)_3CH(NH_2)COOH$ | 0 | 6 | 14 | 4 | 2 | 0 | 0 | 0 | 174.20 |
| | Phenylalanine | $C_9H_{11}NO_2$ | $PhCH_2CH(NH_2)COOH$ | 0 | 9 | 11 | 1 | 2 | 0 | 0 | 0 | 165.19 |
| | Tyrosine | $C_9H_{11}NO_3$ | $HOPhCH_2CH(NH_2)COOH$ | 0 | 9 | 11 | 1 | 3 | 0 | 0 | 0 | 181.19 |
| Amino acid | Tryptophan | $C_{11}H_{12}N_2O_2$ | $PhNHCH{=}CH_2CH(NH_2)COOH$ | 0 | 11 | 12 | 2 | 2 | 0 | 0 | 0 | 204.23 |
| Carboxylic acid | Ketoglutaric acid | $C_5H_6O_5$ | $(COOH)C{=}O(CH_2)_2COOH$ | 0 | 5 | 6 | 0 | 5 | 0 | 0 | 0 | 146.10 |
| Pyridinecarboxylic acid | Nicotinic acid/Vitamin B3 | $C_6H_5NO$ | pyridine-COOH | 0 | 6 | 5 | 1 | 1 | 0 | 0 | 0 | 107.11 |
| | Glucosaminic acid | $C_6H_{12}NO_6$ | $CH_2(OH)(HCOH)_3CH(NH_2)COOH$ | 0 | 6 | 12 | 1 | 6 | 0 | 0 | 0 | 194.16 |
| Aminocarboxylic acid | Pantothenic acid/Vitamin B5 | $C_9H_{17}NO_5$ | $CH_2(OH)C(CH_3)_2CH(OH)C({=}O)NH(CH_2)_2COOH$ | 0 | 9 | 17 | 1 | 5 | 0 | 0 | 0 | 219.23 |
| | Aminobenzoyl-triglutamic acid/4-aminobenzoyl-pentanedioic acid | $C_{22}H_{28}N_4O_{13}$ | NA | 0 | 22 | 28 | | 13 | 0 | 0 | 0 | 500.45 |
| | Pteroyl-triglutamic acid/Folic acid/vitamin B9 | $C_{29}H_{33}N_9O_{12}$ | pteridine ring - para-amino-benzoic acid - glutamic acid | 0 | 29 | 33 | 9 | 12 | 0 | 0 | 0 | 699.63 |
| Aminobenzoylcarboxylic acid | Aminobenzoylpolyglutamic acid | NA | NA | NA | NA | NA | NA | NA | NA | NA | NA | NA |
| | Monimethylamine | $CH_5N$ | $CH_3NH_2$ | 0 | 1 | 5 | 1 | 0 | 0 | 0 | 0 | 31.06 |
| | Dimethylamine | $C_2H_7N$ | $NH(CH_3)_2$ | 0 | 2 | 7 | 1 | 0 | 0 | 0 | 0 | 45.08 |
| Methyl amine | Trimethylamine | $C_3H_9N$ | $N(CH_3)_3$ | 0 | 3 | 9 | 1 | 0 | 0 | 0 | 0 | 59.11 |
| Polyamine | Putrescine | $C_4H_{12}N_2$ | $(CH_2)_4(NH_2)_2$ | 0 | 4 | 12 | 2 | 0 | 0 | 0 | 0 | 88.15 |
| | Tryptone | $C_3H_5NO$ | $H_2C{=}CO(NH_2)$ | 0 | 3 | 5 | 1 | 1 | 0 | 0 | 0 | 71.08 |
| Peptide | Peptone | NA | NA | 0 | 13 | 24 | 0 | 4 | 0 | 0 | 0 | 244.33 |
| Hydroxypyridine | Pyridoxamine/Vitamin B6 | $C_8H_{12}N_2O_2$ | $NH_2(CH_2)C{=}C(OH)C(CH_3){=}NC{=}C(H_2COH)$ | 0 | 8 | 12 | 2 | 2 | 0 | 0 | 0 | 168.19 |
| Benzoate | Benzoate | $C_7H_5O_2^-$ | NA | -1 | 7 | 5 | 0 | 2 | 0 | 0 | 0 | 121.11 |
| | 3-hydroxybenzoate | $C_7H_5O_3^-$ | NA | -1 | 7 | 5 | 0 | 3 | 0 | 0 | 0 | 137.11 |
| Hydroxybenzoate | 4-hydroxy-benzoate | $C_7H_5O_3^-$ | NA | -1 | 7 | 5 | 0 | 3 | 0 | 0 | 0 | 137.11 |
| Methyoxybenzoate | 3,4,5-trimethyoxybenzoate | $C_{10}H_{12}O_5^-$ | $(CH_3O)_3$-benzoate | -1 | 10 | 12 | 0 | 5 | 0 | 0 | 0 | 212.20 |
| Phenol | Phenol | $C_6H_6O$ | $(C_6H_5)OH$ | 0 | 6 | 6 | 0 | 1 | 0 | 0 | 0 | 94.11 |
| Pyrrolizine | Heliotrine | $C_{16}H_{27}NO_5$ | NA | 0 | 16 | 27 | 1 | 5 | 0 | 0 | 0 | 313.39 |
| Nucleoside | Inosine | $C_{10}H_{12}N_4O_5$ | NA | 0 | 10 | 12 | 4 | 5 | 0 | 0 | 0 | 268.23 |
| Methylglycine | Betaine/Trimethylglycine | $C_5H_{11}NO_2$ | $(CH_3)_3N^+CH_2COO^-$ | 0 | 5 | 11 | 1 | 2 | 0 | 0 | 0 | 117.15 |
| | 4-OHB/4-hydroxybutanoic acid | $C_4H_8O_3$ | $COOH(CH_2)_3OH$ | 0 | 4 | 8 | 0 | 3 | 0 | 0 | 0 | 104.10 |
| Fatty acid | Oleic acid | $C_{18}H_{34}O_2$ | $CH_3(CH_2)_7CH{=}CH(CH_2)_7COOH$ | 0 | 18 | 34 | 0 | 2 | 0 | 0 | 0 | 282.46 |

| Name | R-COOH, carboxylic acid | R-COO$^-$, carboxylate | R-OH, hydroxyl | R-NH$_2$, amine | R-NH | R-N-R | R-NH$_3^+$ | R-NO$_3^-$ | Nitrile, CN | R-N-(C=O)-R, amide | R-C=O, ketone | R-HC=O, aldehyde | R-O-C=O, Ester | R-CH$_3$, methyl | R-O-CH$_3$, oxymethyl | R-Cl | R-Fe$^{3+}$ | R-S | C=C | Benzene | Phenyl | Pyridine | Imidazole | Indole | Oleate | pyranose | furanose | pyrrolizidine | Glucosamine | Purine | Pteridine |
|---|---|---|---|---|---|---|---|---|---|---|---|---|---|---|---|---|---|---|---|---|---|---|---|---|---|---|---|---|---|---|---|
| Arbutin | | | 1 | | | | | | | | | | | | | | | | | 1 | 1 | | | | | | | | | | |
| Salicin | | | 1 | | | | | | | | | | | | | | | | | 1 | | | | | | | | | | | |
| Methyl a-glycoside | | | 4 | | | | | | | | | | 1 | 1 | 1 | | | | | | | | | | | 1 | | | | | |
| Polyoxyethylene sorbitan mono-oleate | | | | | | | | | | | | | 1 | | | | | | | | | | | | 1 | | | | | | |
| Glycine | 1 | | | 1 | | | | | | | | | | | | | | | | | | | | | | | | | | | |
| Alanine | 1 | | | 1 | | | | | | | | | | 1 | | | | | | | | | | | | | | | | | |
| Serine | 1 | | 1 | 1 | | | | | | | | | | | | | | | | | | | | | | | | | | | |
| Cysteine | 1 | | | 1 | | | | | | | | | | | | | | 1 | | | | | | | | | | | | | |
| Threonine | 1 | | 1 | 1 | | | | | | | | | | 1 | | | | | | | | | | | | | | | | | |
| Asparagine | 1 | | | 2 | | | | | | 1 | | | | | | | | | | | | | | | | | | | | | |
| Aspartic acid | 1 | | | 1 | | | | | | | | | | | | | | | | | | | | | | | | | | | |
| Proline | 1 | | | 1 | | | | | | | | | | | | | | | | | | | | | | | | | | | |
| Valine | 1 | | | 1 | | | | | | | | | | 1 | | | | | | | | | | | | | | | | | |
| Glutamine | 1 | | | 2 | | | | | | 1 | | | | | | | | | | | | | | | | | | | | | |
| Glutamic acid | 2 | | | 1 | | | | | | | | | | | | | | | | | | | | | | | | | | | |
| Leucine | 1 | | | 1 | | | | | | | | | | 1 | | | | | | | | | | | | | | | | | |
| Isoleucine | 1 | | | 1 | | | | | | | | | | 1 | | | | | | | | | | | | | | | | | |
| Lysine | 1 | | | 2 | | | | | | | | | | | | | | | | | | | | | | | | | | | |
| Methionine | 1 | | | 1 | | | | | | | | | | 1 | | | | 1 | | | | | | | | | | | | | |
| Histidine | 1 | | | 1 | 1 | 1 | | | | | | | | | | | | | | | | | 1 | | | | | | | | |
| Arginine | 1 | | | 1 | 2 | | | | 1 | | | | | | | | | | | | | | | | | | | | | | |
| Phenylalanine | 1 | | | 1 | | | | | | | | | | | | | | | | | 1 | | | | | | | | | | |
| Tyrosine | 1 | | 1 | 1 | | | | | | | | | | | | | | | | 1 | | | | | | | | | | | |
| Tryptophan | 1 | | | 1 | 1 | | | | | | | | | | | | | | | | | | | 1 | | | | | | | |
| Ketoglutaric acid | 2 | | | | | | | | | | 1 | | | | | | | | | | | | | | | | | | | | |
| Nicotinic acid/Vitamin B3 | 1 | | | | | | | | | | | | | | | | | | | | | 1 | | | | | | | | | |
| Glucosaminic acid | 1 | | 4 | 1 | | | | | | | | | | | | | | | | | | | | | | | | | | | |
| Pantothenic acid/Vitamin B5 | 1 | | 2 | | | | | | | 1 | | | | 2 | | | | | | | | | | | | | | | | | |
| Aminobenzoyl-triglutamic acid/4-aminobenzoyl-pentanedioic acid | 1 | | | 1 | 1 | | | | | 1 | | | | | | | | | | 1 | | | | | | | | | | | |
| Pteroyl-triglutamic acid/Folic acid/vitamin B9 | 1 | | | 1 | 1 | | | | | 1 | | | | | | | | | | | | | | | | | | | | | 1 |
| Aminobenzoylpolyglutamic acid | 1 | | | 1 | 1 | | | | | 1 | | | | | | | | | | 1 | | | | | | | | | | | |
| Monimethylamine | | | | 1 | | | | | | | | | | 1 | | | | | | | | | | | | | | | | | |
| Dimethylamine | | | | | 1 | | | | | | | | | 1 | | | | | | | | | | | | | | | | | |
| Trimethylamine | | | | | | 1 | | | | | | | | 2 | | | | | | | | | | | | | | | | | |
| Putrescine | | | | 2 | | | | | | | | | | | | | | | | | | | | | | | | | | | |
| Tryptone | | | | 1 | | | | | | | | | 1 | | | | | | 1 | | | | | | | | | | | | |
| Peptone | NA | NA | NA | | | | | | | | | | | | | | | | | | | | | | | | | | | | |
| Pyridoxamine/Vitamin B6 | | | 2 | 1 | | 1 | | | | | | | | 1 | | | | | 2 | | | 1 | | | | | | | | | |
| Benzoate | 1 | | | | | | | | | | | | | | | | | | | 1 | | | | | | | | | | | |
| 3-hydroxybenzoate | 1 | | 1 | | | | | | | | | | | | | | | | | 1 | | | | | | | | | | | |
| 4-hydroxy-benzoate | 1 | | 1 | | | | | | | | | | | | | | | | | 1 | | | | | | | | | | | |
| 3,4,5-trimethyoxybenzoate | 1 | | | | | | | | | | | | | | 3 | | | | | 1 | | | | | | | | | | | |
| Phenol | | | 1 | | | | | | | | | | | | | | | | | 1 | | | | | | | | | | | |
| Heliotrine | | | 1 | | | | | | | | | | 1 | | | | | | | | | | | | | | | 1 | | | |
| Inosine | | | | | | | | | | | | | | | | | | | | | | | | | | | 1 | | | 1 | |
| Betaine/Trimethylglycine | | 1 | | | | 1 | | | | | | | | 2 | | | | | | | | | | | | | | | | | |
| 4-OHB/4-hydroxybutanoic acid | 1 | | 1 | | | | | | | | | | | | | | | | | | | | | | | | | | | | |
| Oleic acid | 1 | | | | | | | | | | | | | | | | | | 1 | | | | | | | | | | | | |

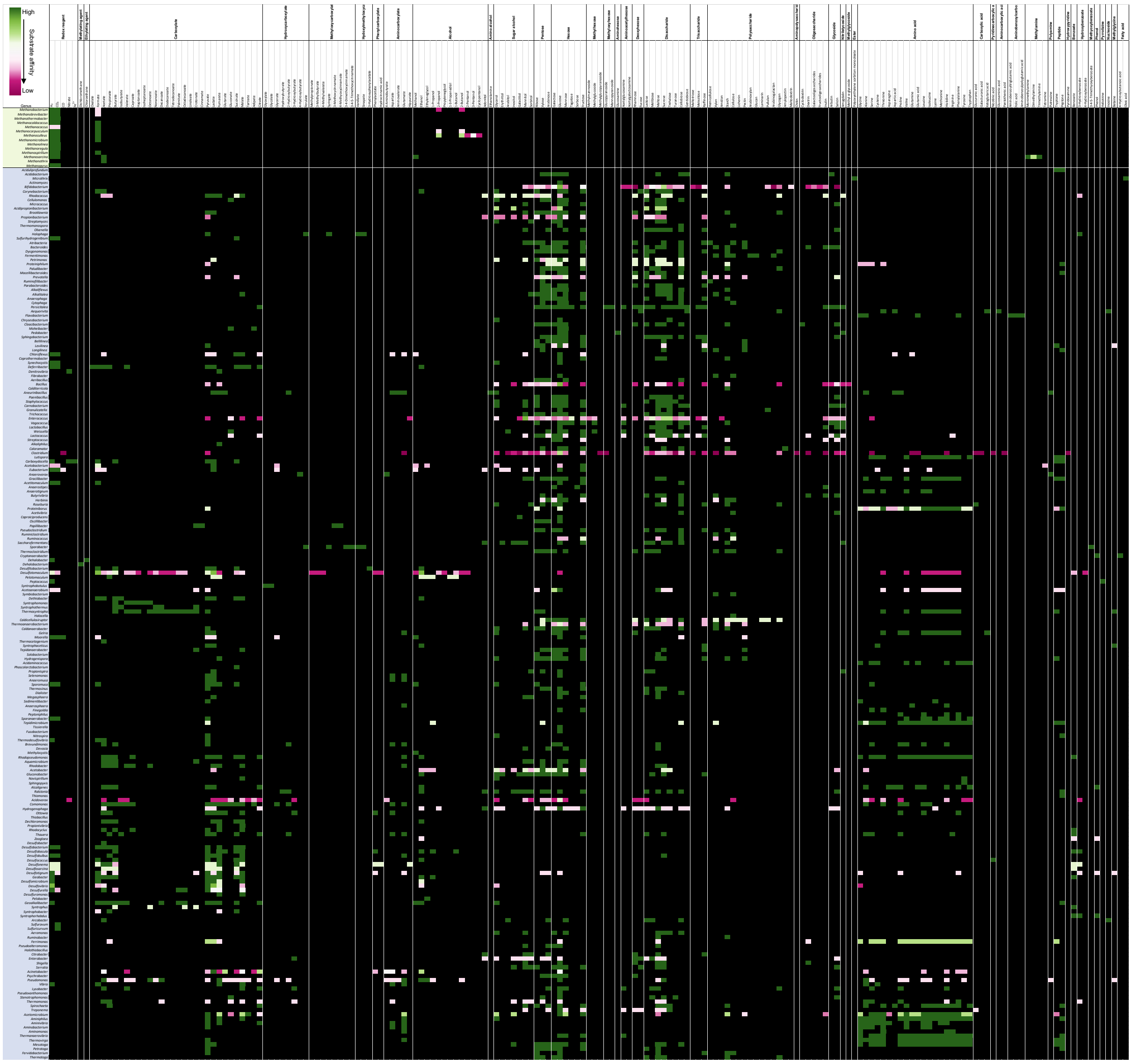

**Supplementary Figure 1.1.** AD substrates, temperatures and pH ranges reported to support microbial growth of different genera from a literature review of 197 papers **a)** Substrates reported to support microbial growth. 187 substrates were investigated. Affinity was based on substrate utilisation where high affinity (green) means that most species within that genera are capable of growth on that substrate, and pink indicates only a proportion of species are capable of growth on that substrate. Black indicates no reported growth on a given substrate. Substrates are categorised by key functional groups [1-195].

## 1.2. Literature review of genera reported to be involved in syntrophic interactions in anaerobic digestion

**Table 1.6.** Genera reported to be involved in syntrophic relationships in anaerobic digestion. Reported growth conditions are noted including pH range, optimum pH, temperature (temp) range (°C), optimum temp (°C). Whether the genus has been reported to be capable of DIET is indicated by Y for yes, or NA for information not available. Potential function reported in literature is noted and references are provided. *Abbreviations Direct Interspecies Electron Transfer (DIET); Hydrogen Formate Interspecies Transfer (HFIT); Syntrophic Acetate Oxidising Bacteria (SAOB), data not available (NA). [11,14,23,56,67,79,92,99,121,123,127,132,141,142,176,177,196,199-215]

| GENUS | DIET | REPORTED FUNCTION | REF |
|---|---|---|---|
| *Methanomassiliicoccus* | NA | Hydrogenotrophic methanogen | 199,200 |
| *Methanobacterium* | NA | Hydrogenotrophic methanogen | 56,123,141,177,199,201-203 |
| *Methanobrevibacter* | NA | Hydrogenotrophic methanogen | 23 |
| *Methanothermobacter* | NA | Hydrogenotrophic methanogen | 199,204 |
| *Methanoculleus* | NA | Hydrogenotrophic methanogen, acetate turnover | 92,99,132,204 |
| *Methanolinea* | NA | Hydrogenotrophic methanogen | 203,205 |
| *Methanospirillum* | NA | Hydrogenotrophic methanogen | 23,56,99,123,203 |
| *Methanosarcina* | Y | Acetate turnover | 56,92,177,203,205,206 |
| *Methanothrix* | Y | Obligate acetoclastic methanogen | 56,99,123,141,203,206,207 |
| *Bifidobacterium* | NA | Acid formation and hydrogen release | 14 |
| *Bacillus* | NA | Acid formation and hydrogen release | 14 |
| *Exiguobacterium* | NA | SAOB | 199,208 |
| *Streptococcus* | Y | Acid formation and hydrogen release | 14 |
| *Butyrivibrio* | NA | Acid formation and hydrogen release | 14 |
| *Caloramator* | Y | | 56 |
| *Clostridium* | NA | SAOB | 14,92,207,209 |
| *Desulfotomaculum* | NA | Propionate oxidiser, sulfate-reducer, acetate consumer | 14,23 |
| *Pelotomaculum* | NA | Acetogenic propionate oxidiser | 23,121,123,204 |
| *Syntrophomonas* | Y | 4-8 carbon short-chain fatty acid degrader, propionate/acetate producer, SAOB | 14,56,67,79,121,123,127,142,176,202,204,206,207,209 |
| *Syntrophothermus* | NA | Acetogen | 204 |
| *Caldanaerobacter* | NA | SAOB | 199 |
| *Thermacetogenium* | NA | SAOB | 176,204 |
| *Syntrophaceticus* | NA | | 92,99 |
| *Tepidanaerobacter* | NA | SAOB | 56,199 |
| *Propionispira* | NA | Acetogenesis | 206 |
| *Anaeroarcus* | Y | Acetate and propionate production, amino acid degradation | 56,216 |
| *Sporanaerobacter* | Y | | 56,206 |
| *Bacteroides* | Y | Acid formation and hydrogen release | 14,56,205 |
| *Prosthecochloris* | Y | Sulfide/sulfur reducer | 142,217 |
| *Sulfurospirillum* | Y | | 56,218 |
| *Candidatus Cloacamonas* | NA | Hydrogen producer, amino acid fermenter, butyrate/propionate oxidation | 123,150,219 |
| *Anaerolinea* | NA | Short chain fatty acid degrader, acetate producer | 129,220 |
| *Coprothermobacter* | NA | SAOB | 199 |
| *Deferribacter* | Y | | 56 |
| *Thermodesulfovibrio* | NA | Syntrophic mediator of lactate degradation | 204 |
| *Thauera* | Y | | 56 |
| *Shewanella* | Y | | 56,67,221 |
| *Pseudomonas* | Y | Acid formation and hydrogen release, protein fermentation | 14,196 |
| *Vibrio* | NA | | 67 |
| *Tepidiphilus* | NA | SAOB | 199 |
| *Cloacibacillus* | NA | SAOB, $H_2/CO_2$ producer, acetic acid production, amino acid fermentation | 210,211 |
| *Synergistes* | NA | Amino acid fermenter, VFA producer | 205,212 |
| *Candidatus Desulfofervidus* | Y | Sulfate-reducer, methane oxidation | [56,222] |
| *Desulfobacterium* | Y | | [56] |
| *Desulfobacula* | Y | | [56] |
| *Desulfovibrio* | Y | Hydrogen-producing acetogen, lactate and ethanol degradation | [14,23,56,207] |
| *Desulfuromonas* | Y | | [56] |
| *Geoalkalibacter* | Y | | [56,142] |
| *Geobacter* | Y | HFIT, ethanol degradation | [56,142,207,213] |
| *Smithella* | Y | Propionate oxidiser, produces acetate and butyrate | [23,79,132,214] |
| *Syntrophus* | Y | HFIT | [121,123,142,205] |
| *Syntrophobacter* | NA | Butyrate/ propionate to acetate conversion, propionate oxidiser | [11,14,23,121,205,207] |
| *Mesotoga* | NA | Acetate oxidiser | [204] |
| *Pseudothermotoga* | NA | Acetate oxidiser | [204,215] |
| *Thermotoga* | NA | Acetate oxidiser | [176] |

# 2. Supplementary Materials and Methods

## 2.1. Digester description

Two experimental runs were designed to address the latter two research objectives using continuously stirred tank reactor (CSTR) experiments and Quorn™ MFWW as feedstock. In run 1 mesophilic and thermophilic reactors treating MFWW were run, and in run 2 single- and two-stage CSTRs treating MFWW were run (Figure 2b).

CSTRs were selected due to their widespread application (Figure 1f). Anaero technology patented single- and multi-stage CSTRs with temperature and gas sensors were used in this experiment. The patented reactor technology enabled advanced research of AD as the bioreactors mimic commercial-scale AD processes and the inclusion of gas composition sensors, and real-time biogas and temperature detection offer advanced system monitoring. Furthermore, Anaero technology flexible modular bioreactor design enabled the development of a two-stage system for comparison of single- and multi-stage systems.

CSTRs were run with a consistent 14.7 day HRT, 30rpm mixing, and 0.32L/d influent flow rate (Supplementary Table 2.1). OLR was based on influent COD (g COD/L/d). All CSTRs were inoculated with inoculum sourced from a UK-based mesophilic AD plant treating mixed food waste. CSTR's were inoculated with 5.5L of inoculum with continuous mixing of the initial inoculum between inoculations to ensure homogeneity between reactors. Before starting the experimental run, reactors were operated for approximately two months at low OLR (0.18L/d influent flow rate) to ensure all food waste was eliminated and the inoculum was acclimated to the new feedstock. We quantified volatile soluble solids (VSS) as a measure of active biomass on the first and last day of the experimental run. Bioreactor chemistry and influent composition were monitored through the experiment, sampling in triplicate according to Figure 2d-e, including pH, VFA, COD, and sugar and sugar alcohol concentration (Supplementary Table 2.1). Metagenomic sequencing of 16s rRNA was conducted to investigate the microbiome underpinning AD, and OTU assignment and taxonomic classification were carried out over the course of the experiment (Figure 2). Performance was measured in terms of COD removal efficiency, cumulative biogas production, and specific daily biogas production (Figure 2).

**Table 2.1.** CSTR experimental design, including operational parameters, bioreactor chemistry, feedstock composition, and performance parameters. *Abbreviations: continuously stirred tank reactor (CSTR); organic loading rate (OLR), hydraulic retention time (HRT), volatile fatty acid (VFA), chemical oxygen demand (COD).

| Parameter type | Parameter | Value | Unit |
|---|---|---|---|
| **Operational** | Flow rate | 0.32 | L/d |
| | OLR | 0.19 to 1.34 | g COD/L/day |
| | HRT | 14.7 | days |
| | Temperature | 37 or 50 | ℃ |
| | Mixing speed | 30.0 | rpm |
| **Chemical** | pH | Variable | n/a |
| **(Influent & reactor)** | VFA | Variable | g/L |
| | Sugar and sugar alcohol | Variable | g/L |
| **Biological** | 16s rRNA sequencing | Variable | n/a |
| **Performance** | Biogas | Variable | L/g COD/d |
| | COD removal | Variable | % |

### 2.1.1. Run 1

All reactors within run 1 had the same influent through individual peristaltic pumps. Reactors passively outputted effluent through a swan neck outlet. Reactors were sampled in a time-series over a 52-day run time to investigate temporal scales (Figure 2d). Each reactor configuration was run in triplicate, where CSTR1.1-3 represent mesophilic and CSTR2.1-3 represent thermophilic reactors (Figure 2b).

#### 2.1.2. Run 2

All reactors within run 2 had the same influent through individual peristaltic pumps. Single- and second-stage reactors passively outputted effluent through a swan neck outlet. For two-stage reactors, the first-stage effluent transferred to the second-stage through a hose connection (Figure 2c). Reactors were sampled in a time series over a 50-day run time to investigate temporal scales (Figure 2e). Each reactor configuration was run in quadruplicate, where CSTR3.1-4 represent single-stage reactors, CSTR4.1-4 represent first-stage reactors, and CSTR5.1-4 represent second-stage reactors (Figure 2b). DNA quality for day 28 was not sufficient to allow sequencing (Figure 2e).

### 2.2. Volatile Suspended Solid Content

Total suspended solids (TSS) and volatile suspended solids (VSS) were measured according to APHA standard methods using a 0.22µm glass fibre filter and a vacuum filtration system (Eq. S1, Eq. S2) [223].

Where TSS denotes the total soluble solids (mg/L), $W_1$ represents the dish weight (g), $W_2$ represents the heated sample and dish weight (g) and SV represents the sample volume (mL). VSS denotes volatile soluble solids of total soluble solids (mg/L), $W_2$ represents heated sample weight (g), $W_3$ represents the variable ash and dish weight (g), and SV represents the sample volume (mL).

$$TSS = \frac{W_2 - W_1}{SV} \times 10^6 \quad \text{(S1)}$$

$$VSS = \frac{W_3 - W_2}{SV} \times 10^6 \quad \text{(S2)}$$

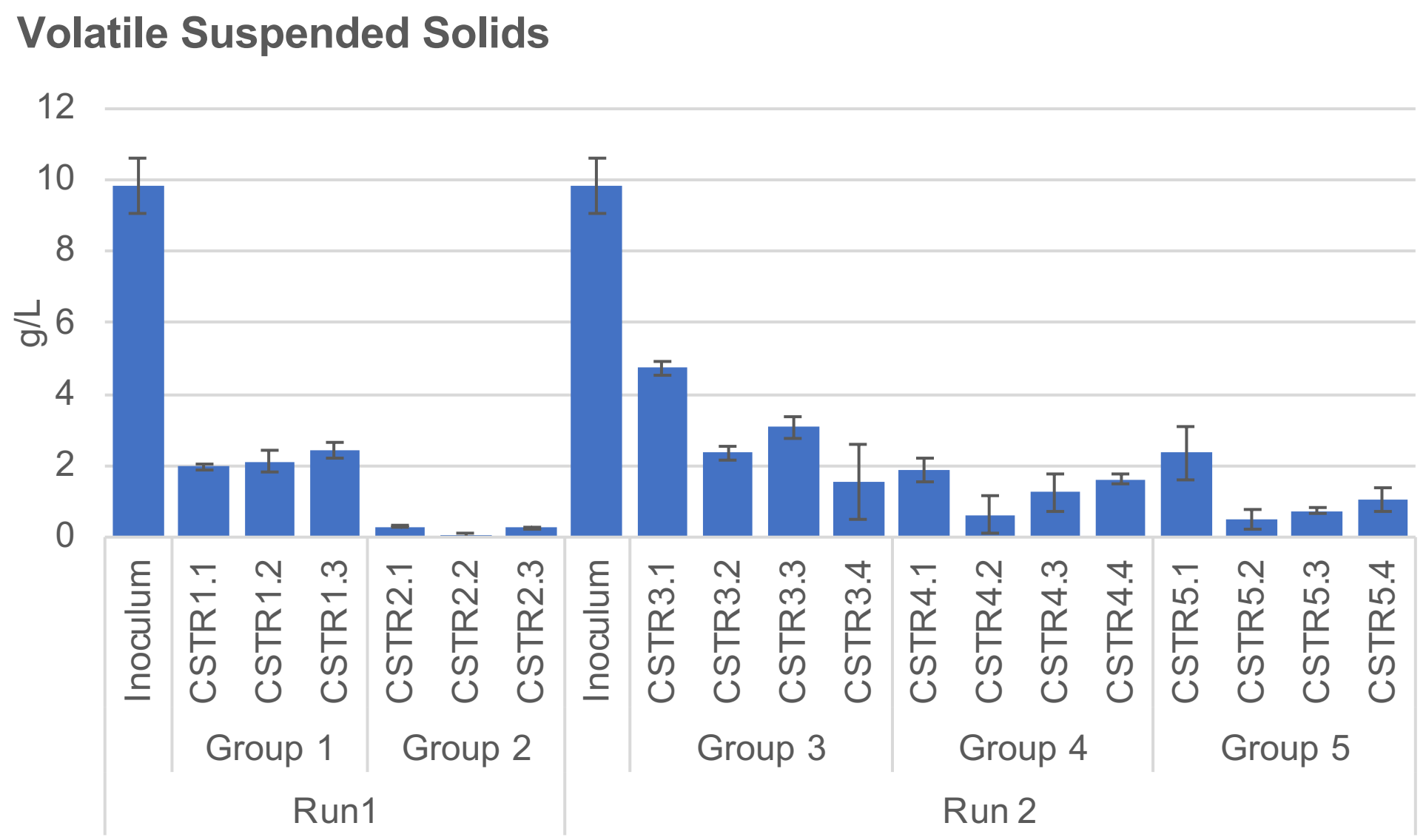

**Supplementary Figure 2.1.** Volatile suspended solid concentration, g/L of inoculum and experimental reactors. *Abbreviations: continuously stirring tank reactor (CSTR).

## 2.3. Chemical oxygen demand

For high range COD (200mg/L to 15,000mg/L), a 0.2mL aliquot of the test sample was added to separate vials containing 61% sulfuric acid, silver sulfate, and potassium dichromate (Vario HR/COD 200mg/L to 15,000mg/L test reagent, Lovibond). For medium range (20mg/L to 1,500mg/L) and low range (3mg/L to 150mg/L) a 2mL aliquot of test sample was added to separate vials containing 61% sulfuric acid, silver sulfate, and potassium dichromate (Vario MR/COD 200mg/L to 1,500mg/L test reagent, and Vario LR/COD 3mg/L to 150mg/L test reagent, Lovibond).

HPLC-grade water was used as a negative control. 5000mg/L (high range) or 100mg/L (medium and low range) COD standards were used as a positive control (Lovibond). Closed vials were mixed by inverting 2-3 times and incubated at 150°C for 2hrs in the thermoreactor (Thermoreactor RD 125, Lovibond) according to standard methods (ISO, 1989; Moore et al., 1949). Samples were allowed to cool to room temperature and mixed by inverting 2-3 times. Solids were allowed to settle, and samples were measured at 600nm wavelength using the MD-600 spectrophotometer (Lovibond).

COD removal (%) was calculated from measured effluent and influent COD, g/L (Eq. S3)

$Inf_{COD}$ denotes influent chemical oxygen demand (g/L), and $Eff_{COD}$ represents effluent chemical oxygen demand (g/L).

$$COD\ removal\ efficiency = \frac{Inf_{COD}}{Eff_{COD}} \qquad \text{(S3)}$$

## 2.4. High performance liquid chromatography

HPLC analysis was conducted using an LC-40D XR solvent delivery pump (Shimadzu) and CTO 40C column oven (Shimadzu). A sample volume of 10µL was injected using a SIL-40C XR Autosampler UHPLC autoinjector (Shimadzu). The mobile phase was degassed by a DGU-405 degassing unit (5 channel; Shimadzu). The column and autosampler were purged with mobile phase before and after every analysis. Calibration curves were constructed by plotting peak area against compound concentration. Linearity of the line of best fit was calculated by the least square regression method within the LabSolutions software. The Limits of Detection (LOD) and Limits of Quantification (LOQ) were calculated using LabSolutions software. These values were used to verify the quality of the calibration curve. Samples (0.6 mL) were filtered through a 0.22µm pore filter (Claristep® Filtration system, Satorius) into a clean 1.5mL HPLC vial. Sample analysis was carried out in triplicate and a mean value with standard deviation was calculated.

### 2.4.1. Sugar and sugar alcohol

HPLC analysis was conducted using an SPD-M40 photodiode array (PDA) detector (Shimadzu), a RID-20A refractive index detector operating at 40°C (Shimadzu), and a Rezek RCM-Monosaccharide $Ca^{2+}$ column (8µm, 100mm x 7.8mm internal diameter; Phenomenex) maintained at 80°C. HPLC grade water was used as a mobile phase with a flow rate of 0.2mL/min and an isocratic elution run time of 1hr. Instrument parameters for sugar and sugar alcohol HPLC analysis are detailed in Supplementary Table 2.4.1 [224].

Standard stock solutions (10g/L) were prepared using ultrapure HPLC grade water (Arium® Pro Ultrapure Water Systems, Satorius). Five dilution levels (100, 200, 500, 1000 and 2000mg/L) were run in triplicate for the 7 reference compound stock solutions: alpha-D(+)-Melibiose (neat, Ehrenstorfer GMBH), Glucose (1000mg/L in pure water, TraceCERT® Supelco® Merck), D-(+)-Xylose (neat, Ehrenstorfer GMBH), Maltitol (neat, Supelco® Merck), Glycerol (neat, Supelco® Merck), Mannitol (neat, Supelco® Merck), and D-(+)-Arabitol (neat, Ehrenstorfer GMBH). Sample analysis was carried out using the same method as the standard solutions for sugar and sugar alcohol content (Supplementary Table 2.4.1). Peak identification was carried out using the calibration curve for sugar and sugar alcohol (Supplementary Figure 2.4.1).

**Supplementary Table 2.4.1.** HPLC sugar and sugar alcohol content instrument parameters.

| | |
|---|---|
| Detectors | Photodiode Array (210nm) and Refractive Index Detector (40ºC) |
| Column | Phenomenex Rezek-RCM Monosaccharide $Ca^{2+}$ |
| Column Oven Temperature, ºC | 80 |
| Mobile Phase | Ultrapure HPLC-grade Water |
| Flow rate, mL/min | 0.2 |
| Time, min | 60 |
| Injection volume, µL | 10 |

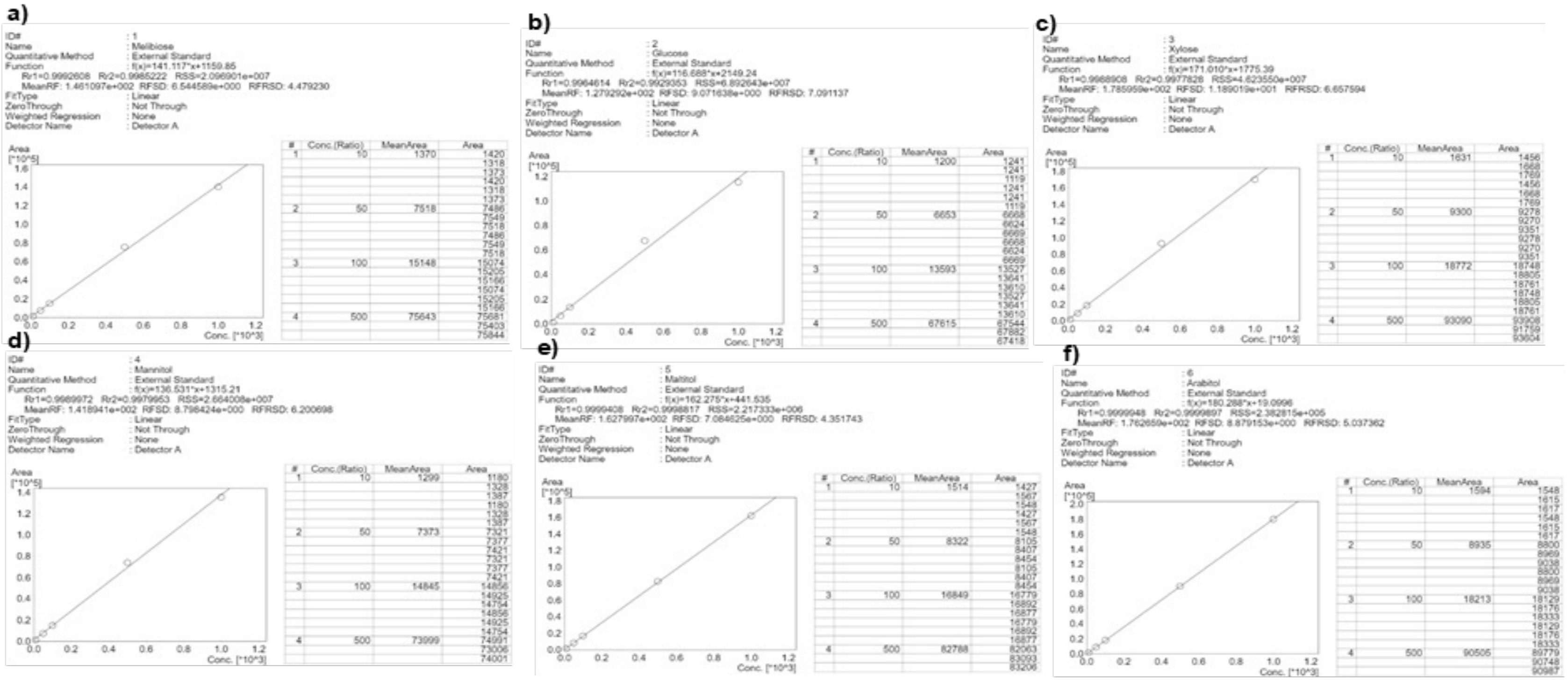

| Compound | $R^2$ |
|---|---|
| Melibiose | 0.999 |
| Glucose | 0.993 |
| Xylose | 0.998 |
| Mannitol | 0.998 |
| Maltitol | >0.999 |
| Arabitol | >0.999 |

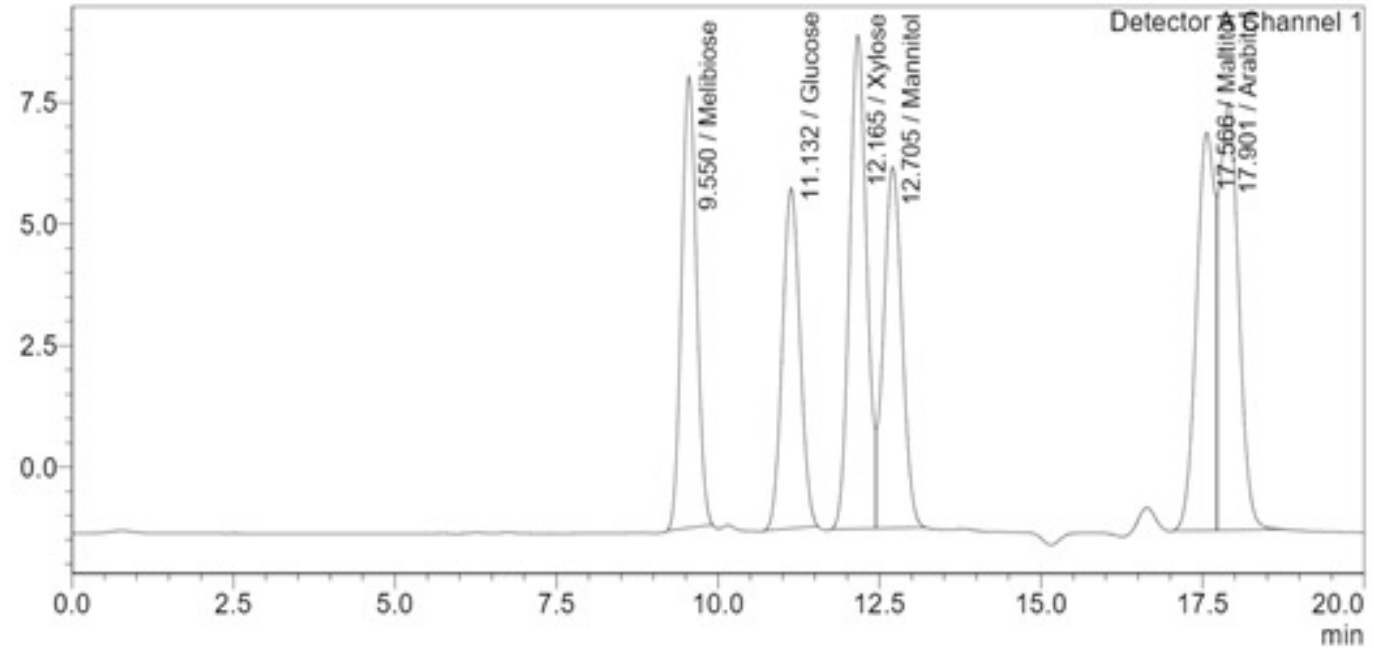

**Supplementary Figure 2.4.1..** Sugar and sugar alcohol calibration curves. **(a)** Melibiose. **(b)** Glucose. **(c)** Xylose. **(d)** Mannitol. **(e)** Maltitol. **(f)** Arabitol. Five levels of concentration were used: 10, 50, 100, 500 and 1000mg/L were used for level 1, 2, 3, 4 and 5 respectively. Chromatogram of standard solutions. Analysis was conducted in triplicate and repeated. Both runs shows high reproducibility and were both included in the final calibration curve. R2 values are provided for each compound calibration.

### 2.4.2. Volatile fatty acids

A sample volume of 20µL was injected using the autoinjector (Shimadzu) and run through an Aminex® HPX-87H column (8µm, 300mm x 7.8mm internal diameter; BioRad) maintained at 55°C. $H_2SO_4$ (0.008M HPLC grade) was used as a mobile phase with a flow rate of 0.6mL/min and an isocratic elution run time of 40 minutes. Instrument parameters are detailed in Supplementary Table 2.4.2 [225].

**Supplementary Table 2.4.2**. Instrument parameters used for HPLC analysis of VFAs.

| **Detectors** | **Photodiode Array (205nm)** |
|---|---|
| **Column** | Biorad Aminex HPX-87H |
| **Column Oven Temperature, ºC** | 55 |
| **Mobile Phase** | 0.008N $H_2SO_4$ |
| **Flow rate, mL/min** | 0.6 |
| **Time, min** | 40 |
| **Injection volume, $\mu$L** | 20 |

Standard stock solutions (10g/L) were prepared using ultrapure HPLC grade water (Arium® Pro Ultrapure Water Systems, Satorius). Five dilution levels (0.01, 0.05, 0.10, 0.50, 1.00 and 2.00g/L) were run in triplicate for the 7 reference compound stock solutions: formic acid (≥99%, HiPerSolv CHROMANORM®), acetic acid (≥99.8%, HiPerSolv CHROMANORM®), propionic acid (neat, Supelco® Merck), iso-butyric acid (neat, EHRENSTORFER GMBH), butyric acid (neat, Supelco® Merck), iso-valeric acid (neat, Supelco® Merck) and valeric acid (neat, Supelco® Merck). Calibration curves were constructed by plotting peak area against standard concentration (Supplementary Figure 2.4.2). Sample analysis was carried out using the same method as the standard solutions for volatile fatty acid content (Supplementary Table 2.4.2). Peak identification was carried out using the calibration curve for sugar and sugar alcohol (Supplementary Figure 2.4.2). Acetic acid was used as a positive control and water was used as a negative control for sample analysis.

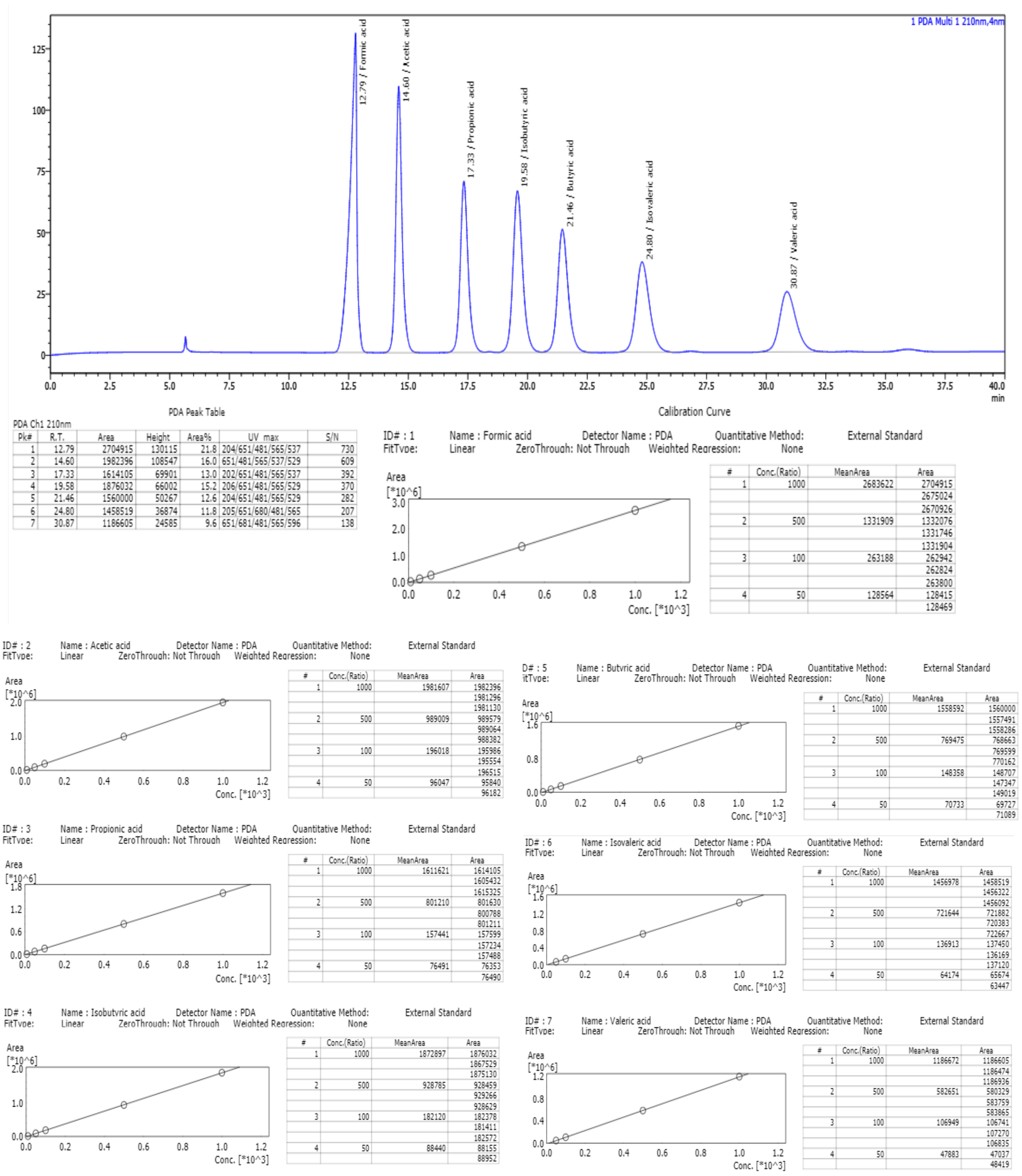

| Compound | $R^2$ |
|---|---|
| Formic acid | >0.999 |
| Acetic acid | >0.999 |
| Propionic acid | >0.999 |
| Iso-butyric acid | >0.999 |
| Butyric acid | >0.999 |
| Iso-valeric acid | >0.999 |
| Valeric acid | >0.999 |

**Supplementary Figure 2.4.2.** Combined VFA chromatogram and calibration curves. Calibration curves were constructed using 7 standards, acetic acid, propionic acid, iso-butyric acid, butyric acid, iso-valeric acid and valeric acid. $R^2$ values are given for each compound calibration.

### 2.4.3. Biomethane potential reactors

Biomethane potential (BMP) CSTR reactors were developed by Anaero technology based on operational research in the UK food waste industry. Each machine consists of a hot water bath incubator for temperature control, 15 1L CSTRs with a centralised mixer, a gas flow meter, a gas composition sensor, built-in bespoke monitoring software (algorithms and controller coded in Python) and cloud data storage, enabling real-time biogas production and composition data.

## 2.5. DNA extraction, 16S rRNA amplification, barcoding and sequencing

### 2.5.1. DNA extraction DNeasy® PowerSoil® Pro Kit

2mL of each CSTR sample (in triplicate) was centrifuged at 10,000rpm for 5 minutes, the supernatant was decanted, and the pellet was used for the DNA extraction, which was carried out using the DNeasy® PowerSoil® Pro Kit according to standard protocol (QIAGEN GmbH, Germany). Samples were suspended in solution C6 (QIAGEN GmbH, Germany) and quantified using a NanoDrop™ One (Thermo Scientific, Waltham, MA). Quantified samples were immediately stored at -20°C prior to sequencing.

### 2.5.2. Illumina 16S rRNA amplicon sequencing

Illumina 16S rRNA V3-V4 amplicon sequencing was conducted externally at the WHRI-Genome-Centre at Queen Mary's University, United Kingdom.

#### 2.5.2.1. DNA sample quantification and quality control

Sample DNA concentration was quantified using a Nanodrop, with Milli-Q water as a negative control and blank. DNA quality was checked using a Qubit Flex Fluorometer. A working solution of 1:200 dilution of Qubit Fluorescent reagent (reagent lot no. 2600187) and Qubit

Dilution Buffer (reagent lot no. 2600189) was prepared. 2 standards were used (Standard lot number 2600187), as well as calf thymus DNA as a positive control. 190μL of the working solution with 10μL of the standards, and 199μL of the working solution with 1μL of the calf thymus DNA, or sample DNA were added to Quibit Flex 8-tube strips. Tubed were capped and vortexed briefly before measuring using the Qubit Flex Fluorometer.

### 2.5.2.2. Sample PCR using Platinum SuperFi II DNA Polymerase

PCR reagents, forward and reverse primers (Supplementary Figure 2.5.1c) and the sample DNA were combined according to Supplementary Figure 2.5.1a, and vortexed briefly to ensure good mixing, and briefly centrifuged. The PCR cycle was run using a tetrad thermocycle (LG190) according to the program detailed in Supplementary Figure 2.5.1b, with the tetrad lid set to 105ºC.

a)

| Reagent | 1X volume, μL | Component Lot No. |
|---|---|---|
| 5X SuperFi II DNA Buffer | 2.00 | 013011305 |
| Forward and reverse primer, 10μM | 0.20 | 10/01/2024 |
| PCR nucleotide mix, 10μM | 0.20 | 10/01/2021 |
| Platinum SuperFi II DNA Polymerase, 2U/μL | 0.25 | 01299609<br>01337237 |
| RNase-free water | 5.35 | 175027101 |

b)

| Step | Temperature ºC | Time, secs | Cycles |
|---|---|---|---|
| Initial denaturation | 95 | 120 | 1 |
| Denaturation | 95 | 30 | 30 |
| Primer annealing | 55 | 30 | |
| Elongation | 72 | 30 | |
| Final Elongation | 72 | 300 | 1 |
| Hold | 4 | Hold | ∞ |

c)

| | | | | | | |
|---|---|---|---|---|---|---|
| PCR Primer Box 7 | D1 | 16S V3-V4 Forward | ACACTGACGACATGGTTCTACACCTACGGGNGGCWGCAG | Bacteria | 16s v3-v4 | CS1 |
| PCR Primer Box 7 | D2 | 16S V3-V4 Reverse | TACGGTAGCAGAGACTTGGTCTGACTACHVGGGTATCTAATCC | Bacteria | 16s v3-v4 | CS2 |

d)

| Reagent | 1X volume, μL | Component Lot No. |
|---|---|---|
| 10X FastStart high fidelity reaction buffer (without $MgCl_2$) | 1.0 | 74602900 |
| 25mM $MgCl_2$ | 1.8 | 72402300 |
| DMSO | 0.5 | 61443900 |
| 10mM PCR-grade nucleotide mix | 0.2 | 67206126 |
| 5U/μL FastStart high fidelity enzyme blend | 0.1 | 67206126 |
| RNase-free water | 0.4 | 1540233300 |

e)

| Step | Temperature ºC | Time, secs | Cycles |
|---|---|---|---|
| Initial denaturation | 95 | 600 | 1 |
| Denaturation | 95 | 15 | 15 |
| Primer annealing | 60 | 30 | |
| Elongation | 72 | 60 | |
| Final Elongation | 72 | 180 | 1 |
| Hold | 4 | Hold | ∞ |

**Supplementary Figure 2.5.1.** PCR amplification. **(a)** PCR mixture components. **(b)** PCR cycle program. **(c)** Forward and reverse PCR primers for 16S V3-V4 amplicon sequencing. **(d)** Barcoding PCR mixture. **(e)** Amplicon barcoding PCR.

#### 2.5.2.3. Agarose gel electrophoresis

Once the PCR was completed samples were run on a 2% agarose gel (lot no. ESS520-B110700) mixed with 1X TBE Buffer, and GelRed Nucleic acid stain. The gel was added to a gel electrophoresis tank and filled with 1X TBE buffer. Either HyperLadder™ 100bp (Bioline) with a size range of 100 to 1013 bp, or HyperLadder™ 1 Kb (Bioline) with a size range of 200 to 10,037 bp were used as a DNA ladder as appropriate depending on sample DNA size. 2µL of the DNA ladder was loaded into the first and last well of each gel. 1µL of 5X DNA loading buffer and 4µL of the DNA sample were loaded into individual gel channels, and the gel electrophoresis was run at 400V for 30 minutes. The gel was visualised using UVP-BioDoc.

#### 2.5.2.4. Automated Amplicon barcoding PCR

Amplicon barcoding PCR was conducted using BiomekFX robot and automated software (Supplementary Figure 2.5.1d). A master mix of reagents were prepared according to Figure 3.8 and mixed using vortexing and centrifuging. 1X volume of the master mix was added to each sample from the previous PCR. The thermocycling was conducted (Supplementary Figure 2.5.1e). Four wells were run on a D1000 screen tape to check the barcoding had been successful (buffer lot number 0006745353, ScreenTape lot number 0203008-293, Agilent).

#### 2.5.2.5. Library pooling and cleaning

Libraries were cleaned and pooled using AMPure XP Beads. 0.9X the pool volume of AMPure XP beads were added to the library DNA and incubated at room temperature for 10 minutes. The mixture was then spun briefly, and a magnet was used to separate the beads from the liquid to allow supernatant removal. 200µL of freshly prepared 80% ethanol was added to the pellet and incubated for 30 seconds at room temperature. The ethanol was discarded, and this cleaning step was repeated for a total of three times. The tube was then briefly spun, and a magnet was used to remove any residual ethanol. Beads were air dried for 5 to 10 minutes, resuspended in 150uL of EB buffer (lot number 160051261) and incubated at room temperature for 10 minutes. A magnet was used to separate the beads and liquid, and the supernatant was transferred to a clean 1.5mL Lo-bind tube. Quality control was performed on the cleaned pool DNA using the Qubit Fluorometer and D1000 ScreenTape Assay (Supplementary Figure 2.5.1).

#### 2.5.2.6. Illumina MiSeq library denaturation and loading

An equal volume of fresh 0.2N NaOH and DNA library (4nM) were combined in a 1.5mL LoBind microcentrifuge tube and mixed by vortexing and centrifuging. The mixture was incubated at room temperature for 10 minutes. The same volume as used for the library of 200mM Tris-HCL was added to neutralise the NaOH and mixed by briefly vortexing and centrifuging. HT1 Buffer (Lot No. 20789296) was added to a total volume of 1mL, and the sample was mixed by briefly vortexing and centrifuging and stored on ice. The denatured library was diluted using HT1 buffer to the molarity required for loading a total volume of 600uL. 20pM PhiX (Lot No. 20777170) was used as a control. The loading cartridge was mixed by inverting 10 times, and gently tapping, and 600uL of the denatured library was added to the cartridge. Sequencing was run using Illumina MiSeq v3 kit and the run quality was assessed based on quality score (% bases> Q30), raw cluster density (k/mm$^2$), data output (Gbp), read number (Lane 1) and % PhiX aligned (Supplementary Table 2.5.1). Quality control was also assessed using FastQC in the Apocrita environment.

**Supplementary Table 2.5.1.** Quality control of illumina MiSeq v3 kit sequencing

| Metric | Result | Pass/Fail | Comments |
|---|---|---|---|
| **Quality score, % bases > Q30** | 87.17 | Pass | |
| **Raw Cluster Density, k/mm$^2$** | 883±27 | Fail | Good number of reads achieved with high Q30 score |
| **Data Output, Gbp** | 12.69 | Pass | |

| | | |
|---|---|---|
| **Read Number (Lane 1)** | 20,529,0<br>84 | Pass |
| **PhiX Aligned, %** | 18.04 | Pass |

### 2.5.3. Taxonomic classification (Mothur, Epi2me)

Illumina sequencing data taxonomic classification was conducted using the Mothur pipeline available in Galaxy referenced against the Silva_v4 database [226,227]. Additional pre-processing steps were conducted which are presented in Supplementary Figure 2.5.2. Species relative abundance (%) was calculated from taxonomic classification counts divided by total reads. The taxonomic profile was assessed on the phylum level for all reactors. *Firmicutes, Proteobacteria*, and *Bacteroidetes* are commonly reported to significantly affect AD [228,229]. To further investigate the taxonomic profile of these key phyla, the relative abundance of *Firmicutes, Bacteroidetes*, and *Proteobacteria* is presented at the class, order, and family level. The taxonomic profile was investigated in terms of the relative abundance of total archaea for the lowest common ancestor down to the genus level. The relative abundance of methanogens as a percentage of the total archaea population was compared by methanogenic pathway, including obligate acetoclastic (*Methanosaeta*), acetoclastic and hydrogenotrophic (*Methanosarcina*), and hydrogenotrophic. The relative abundance of syntrophic species reported in the literature was compared across reactors. Correlations between syntrophic bacteria and methanogens were investigated by calculating Spearman's rank correlation coefficients.

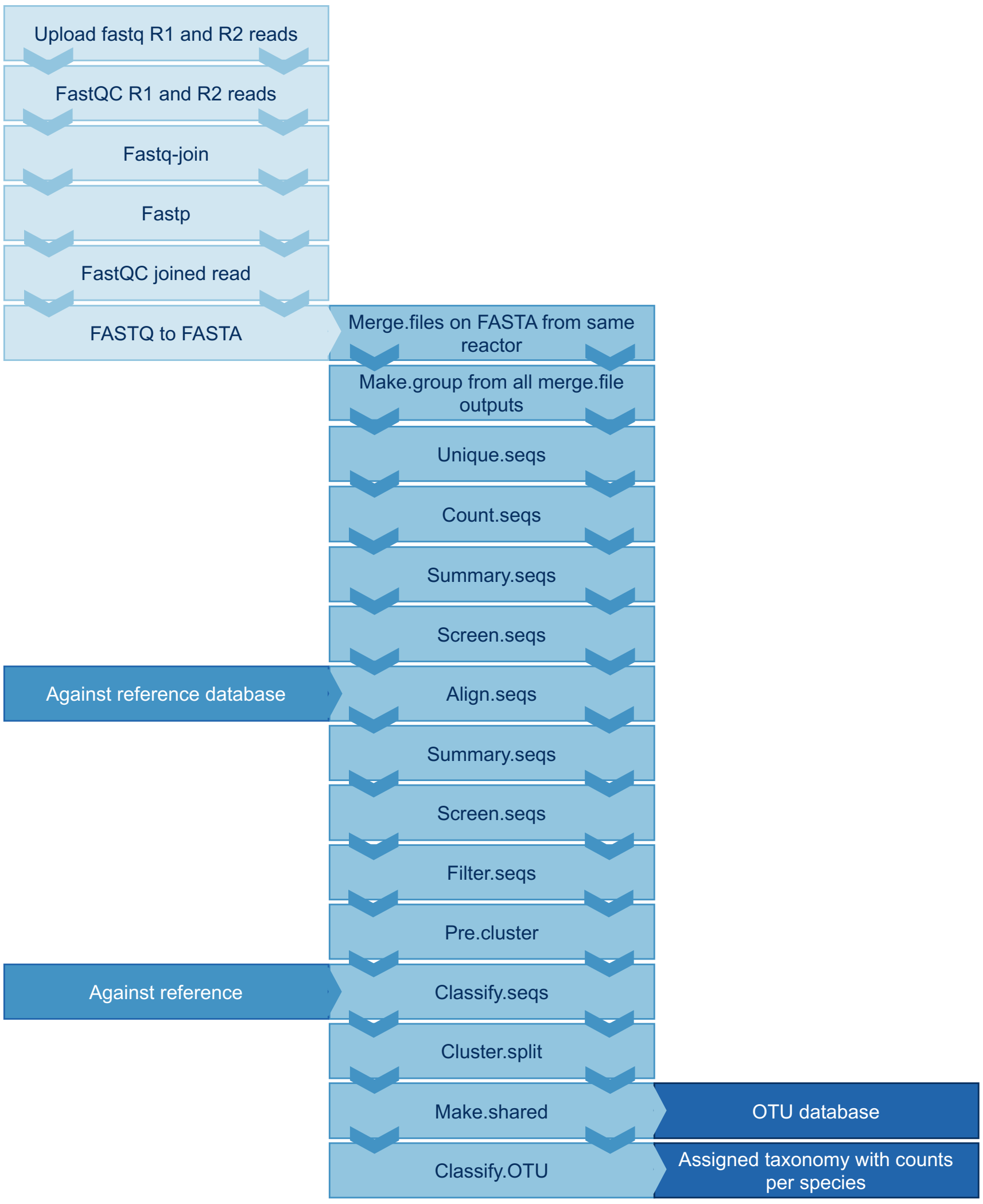

**Supplementary Figure 2.5.2.** Bioinformatic pipeline for metagenomic analysis using a pipeline within galaxy based on Mothur pipeline with added quality control step prior to merging FASTA files [230]. Sequences were aligned against SilvaV4 [227]. Outputs include an OTU database, and taxonomic assignment with read counts.

## 2.6. Network analysis

The Molecular Ecology Network Analysis (MENA) pipeline was used to conduct network analysis [231,232]. Relative abundance of each reactor was analysed across 17 time points. Genes present in nine out of 17 samples per reactor were retained, and missing data was filled with zero values. A logarithm transformation for non-compositional data was applied. The similarity matrix was constructed based on Pearson correlation in time series, allowing one-time point lag for samples following a time series without replicates for individual reactors, and with biological replicates for reactor groups. The calculation order was based on a decrease in the cut-off from the top, and a regression Poisson distribution was applied. A Chi-square test on the Poisson distribution was applied to calculate the cut-off of $p<0.001$. Global network properties, individual nodes' centrality, module separation, and modularity calculation were performed in MENA [232]. A greedy modularity, leading eigenvector and short-random walks optimisation separations were used, and Z and P values were calculated for all nodes. Networks were analysed using regular power, exponential, and truncated power law. The network was randomised, and the network properties were then recalculated to validate the structure. The network file was imported into Cytoscape for visualisation, filtering by genus-level taxonomic classification to simplify the visualisation [231].

## 2.7. Alpha diversity

Alpha diversity was measured using Shannon, Simpson's and Chao1 diversity indices. The Shannon index estimates species richness and species evenness based on the number of unique species, and relative abundances (Eq. S4). The Shannon diversity Index, alpha diversity was also characterised using Simpson's Diversity Index to assess species evenness and richness (Eq. S5). The Simpson index is more sensitive to the most abundant species within the community and tends to emphasise the dominance of common or abundant species. Therefore, communities with fewer dominant species will have a lower Simpson diversity. Simpson's Diversity Index values range from 0 to 1, where 1 indicates complete diversity and 0 indicates no diversity. The Chao1 Diversity Index, an estimator of species richness in terms of the number of unique species present in a given community, without taking into account relative abundances [233]. Chao1 provides an estimate of the minimum number of species present in a community, considering both the observed species and an estimate of the unobserved species based on the number of rare species, specifically those observed once or twice (Eq. S6). Significance was calculated using a one-way ANOVA with post hoc-Tukey test for reactor averages.

Where $H'$ denotes the Shannon diversity index and $p_i$ denotes the proportion of individual species $i$ relative to the total number of individuals within the community, $S_{ob}$ denotes the number of observed species, $F_1$ denotes the number of species only observed once, and $F_2$ denotes the number of species only observed twice.

$$Shannon = -\sum(p_i log p_i) \quad (S4)$$

$$Simpson's = \sum {p_i}^2 \quad (S5)$$

$$Chao1 = S_{ob} + \frac{{F_1}^2}{2F_2} \quad (S6)$$

## 2.8. Beta diversity

The Bray-Curtis similarity index, a non-metric measure of the compositional dissimilarity between two different samples [12]. This method was used to quantify the degree of similarity in species composition between pairs of different reactors at different time points. The Bray-Curtis similarity index was chosen due to its effectiveness in handling ecological relative abundance data and its ability to account for differences in both species presence and abundance.

The datasets used for Bray-Curtis similarity calculation consisted of microbiome taxonomic relative abundance data obtained from reactor samples (Eq. S7). Each sample comprised a list of species along with their corresponding relative abundance values. Prior to analysis, all data were standardised to ensure comparability across different samples. This standardisation involved normalising the species abundance values to relative abundance to account for differences in sampling effort and total counts.

Where $BC_{ij}$ denotes the Bray-Curtis similarity between two samples $i$ and $j$, $x_{ik}$ denotes the abundance of species $k$ in sample $i$, $x_{jk}$ denotes the abundance of species $k$ in sample $j$, and $S$ denotes the number of species across both samples.

$$BC_{ij} = 1 - \frac{\sum_{k=1}^{S} |x_{ik} - x_{jk}|}{\sum_{k=1}^{S} (x_{ik} + x_{jk})} \quad (S7)$$

The standardised abundance data for each sample were organised into a matrix, where rows corresponded to samples and columns corresponded to species. The Bray-Curtis similarity index was then calculated for all pairs of samples using the pdist function from the scipy.spatial library [234]. The resulting pairwise Bray-Curtis dissimilarity values were converted to similarity values (Eq S7), yielding a similarity matrix with values ranging from 0 (identical communities) to 1 (completely dissimilar communities).

The Bray-Curtis dissimilarity index was implemented to evaluate the dissimilarity in microbial community composition across reactors (Eq. S7). Taxonomic dissimilarity was investigated using principal coordinate analysis (PCoA) and distance-based redundancy analysis (db-RDA), a multivariate statical method used to explore the influence of different variables on the microbial community. The db-RDA was conducted using the Bray-Curtis dissimilarity matrix to investigate the influence of influent and bioreactor chemical variables on microbial community composition. The db-RDA is a constrained ordination method that examines relationships between community dissimilarity and potentially explanatory variables. The db-RDA was performed using the skbio and statsmodels libraries in Python. The dissimilarity matrix served as the response variable, while the influent and bioreactor chemical variables were used as predictors. The results of the db-RDA were visualised in a biplot, where the principal coordinates (axes) represent the variation in microbial community composition explained by the influent and bioreactor chemical variables. The length and direction of the vectors indicate the strength and direction of the relationships between the influent and bioreactor chemical variables and the community composition.

Firstly, PCoA was applied to the Bray-Curtis dissimilarity matrix to reduce the data dimensionality, and db-RDA performs a form of linear regression on the results of the PCoA, treating the variables as predictors to find linear combinations of variables that best explain the variation in the diversity data. The variables investigated in the dissimilarity analysis included:

OLR, temperature, influent COD, pH, acetic acid, propionic acid, isobutyric acid, butyric acid, isovaleric acid, valeric acid, total VFA, melibiose, maltitol, glucose, mannitol, arabitol, and total sugar and sugar alcohol content.

A permutation test was conducted to assess the significance of the influence of measured variables on microbial community composition. The db-RDA was performed repeatedly with randomised data to generate a distribution of test statistics under the null hypothesis of no relationship between community composition and influent and bioreactor chemical variables. The observed test statistics were compared to the distribution obtained from the permutations to determine the significance of the relationships. A non-parametric permutational ANOVA (PERMANOVA) was implemented to test the null hypothesis that centroids and group dispersion were not defined by various influent and bioreactor chemical parameters tested. Statistical significance was tested at the 95% confidence interval ($p<0.05$).

## 2.9. Machine learning models

This section outlines the methodology used to predict reactor performance using three machine learning algorithms: Lasso Regression, Random Forest Regression, and Bagging Regression. The selection of the best performing model was based on the coefficient of determination ($R^2$) score, and the feature importance was evaluated using SHAP values. The data was split into training and testing sets using 5-fold cross-validation to ensure robustness in model evaluation.

### 2.9.1. Data Preprocessing and Cross-Validation

The features included in the dataset represent various parameters of the reactor that are expected to influence its performance. For model evaluation, 5-fold cross-validation was employed. In k-fold cross-validation, the dataset is randomly split into k subsets, and random state was set to 42 for reproducibility. In each iteration, the model is trained on k-1 folds and evaluated on the remaining fold. This process is repeated k times, each time with a different fold serving as the validation set. The performance metric used to evaluate the models was the $R^2$ score, which indicates how well the model explains the variance in the target variable, where higher values (closer to 1) indicate better predictive performance.

### 2.9.2. Lasso Regression

Lasso Regression is a type of linear regression that incorporates L1 regularisation, which penalises the absolute magnitude of the model coefficients. This penalty encourages sparsity in the model, driving some of the coefficients to exactly zero. Lasso can automatically exclude irrelevant or redundant features by shrinking their coefficients to zero. Lasso regression was selected because of its ability to handle high-dimensional data where many features may be correlated with each other, making it prone to overfitting. Additionally, it offers built-in feature selection. The objective of Lasso is to minimise the sum of the residual sum of squares (RSS) with an added penalty proportional to the absolute value of the coefficients. Mathematically, the objective function is presented in Supplementary Equation S8.

Where $N$ denotes the number of observations, $p$ denotes the number of features, $y_i$ denotes the actual target value for the $i$ -th sample, $x_{ij}$ denotes the value of the $j$-th feature for the $i$-th sample, $\beta_0$ denotes the intercept term, $\beta = (\beta_1, \beta_2, \beta_3, \ldots, \beta_p)$ are the regression coefficients, $\lambda$ denotes the regularisation parameter controlling the penalty strength.

$$\min_{\beta} \frac{1}{2N} \sum_{i=1}^{N} (y_i - \beta_0 - \sum_{j=1}^{p} x_{ij}\beta_j)^2 + \lambda \sum_{j=1}^{p} |\beta_j| \quad \text{(S8)}$$

The L1 norm penalty, $\sum_{j=1}^{p} |\beta_j|$, forces some coefficients $\beta_j$ to shrink to zero, effectively excluding those features from the model. For Lasso Regression, the regularisation parameter alpha was set to 0.1. This is a commonly used value, as it provides a good balance between bias and variance. Larger values of alpha increase regularisation, which can lead to underfitting, while smaller values may lead to overfitting.

### 2.9.3. Random Forest Regression

Random Forest Regression is an ensemble learning method based on decision trees, where multiple decision trees are trained on random subsets of the data and the final prediction is averaged across all trees. This approach helps reduce the variance compared to individual decision trees and increases the robustness of the model. Random Forest was selected due to its ability to model non-linear relationships between the features and target variable, and its robustness against overfitting. Since reactor performance is likely influenced by complex, non-linear interactions between multiple operational parameters, Random Forest is well-suited to capture these relationships. The random forest algorithm is based on an ensemble of decision trees, which work together to make predictions. Each decision tree in a random forest is trained on a subset of the training data. For each tree $T$ in the random forest, it takes a random sample $X_T$ from the dataset $X$. After that, it selects a random subset of features $F_T$ from the feature space to find the best split based on certain criteria. For regression, the criterion is typically Mean Squared Error (MSE) (Supplementary Equation 9).

Where $N$ denotes the number of samples in the node, $y_i$ denotes the actual value, and $y$ denotes the predicted mean value of the node.

$$MSE = \frac{1}{N}\sum_{i=1}^{N}(y_i - y)^2 \qquad \text{(S9)}$$

Once each tree in the random forest is trained, they are combined to make a prediction. The final prediction is the average of all tree predictions (Supplementary Equation 10).

Where $M$ denotes the total number of trees and $y_{T_m}$ denotes the prediction from each individual tree.

$$y = \frac{1}{M}\sum_{m=1}^{M} y_{T_m} \quad \text{(S10)}$$

The random forest approach leverages this ensemble of trees to improve model accuracy and reduce overfitting, which individual decision trees are prone to. The model was trained using the default settings. These include Number of trees = 100, Maximum depth of trees = None (trees are grown until all leaves are pure), Minimum number of samples required to split a node = 2, Minimum number of samples required at a leaf node = 1, Criterion used to evaluate splits = Mean Squared Error (MSE), Random state was set to 42 for reproducibility.

### 2.9.4. Bagging Regression

Bagging is an ensemble method that combines multiple instances of the same model, each trained on a random bootstrap sample from the training data. The final prediction is obtained by averaging the predictions of the individual models. Bagging helps reduce variance and improves model stability. Bagging Regression was selected as it is an effective method for improving the performance of base regression models that may be prone to overfitting, such as decision trees. By averaging over many different models trained on different subsets of the data, bagging helps increase prediction accuracy and model robustness. Bagging generates multiple datasets $D^{(b)}$ by bootstrapping the original dataset $D$. Each dataset $D^{(b)}$ is formed by sampling $N$ observations with replacement from $D$, where $N$ is the number of original samples. Each base learner $f^{(b)}(X)$ is trained on its corresponding bootstrapped dataset $D^{(b)}$. The final prediction for bagging regression is obtained by averaging the predictions of all the base learners. For a given input $X$, the prediction $y$ is defined in Supplementary Equation 11.

Where $B$ denotes the total number of bootstrapped datasets (and corresponding base learners) and $f^{(b)}(X)$ denotes the prediction from the $b$-th base learner.

$$y = \frac{1}{B}\sum_{b=1}^{B} f^{(b)}(X) \tag{S11}$$

### 2.9.5. Variance reduction

The averaging of predictions reduces the variance of the model without increasing bias significantly, leading to more robust predictions. The model was trained using the default settings. These include Number of base estimators = 10, Maximum number of samples to train each base estimator = 1.0, Maximum number of features used by each base estimator = 1.0, Bootstrap sampling (`bootstrap`) = True, Random state was set to 42 for reproducibility.

### 2.9.6. Model Evaluation: $R^2$ Score

To evaluate the performance of the models, the $R^2$ score was calculated for each fold in the cross-validation. The $R^2$ score is defined by Supplementary Equation S12. Where: $y_{true}$ are the true target values, $y_{pred}$ are the predicted values, $\overline{y_{pred}}$ is the mean of the true target values. An $R^2$ value closer to 1 indicates that the model explains a large portion of the variance in the target variable, suggesting better performance. The algorithm with the highest mean $R^2$ scores across the 5-fold cross-validation was selected as the final model.

$$R^2 = 1 - \frac{\sum(y_{true} - y_{pred})^2}{\sum(y_{true} - \overline{y_{pred}})^2} \tag{S12}$$

### 2.9.7. Feature Importance Evaluation using SHAP Values

To understand the influence of individual features on the model's predictions, SHAP values were used. SHAP values are based on cooperative game theory and provide a consistent method for explaining the contribution of each feature to a given prediction. They decompose the prediction of a model into contributions from each feature, which allows for a clear understanding of how each input affects the output. SHAP values were used to identify which features had the most significant impact on reactor performance predictions. Since the models used in this study, particularly Random Forest and Bagging, can be seen as "black box" models, SHAP values provide transparency by assigning a numerical importance to each feature, facilitating better interpretability. The SHAP values were computed for the best-performing model (as determined by the $R^2$ score) to visualise and rank feature importances. The SHAP summary plots were used to provide an overall understanding of the relationships between features and reactor performance.

#### 2.9.8. Algorithm Selection

The final model selection was based on the mean $R^2$ score obtained from the 5-fold cross-validation. The algorithm that produced the highest mean $R^2$ score was selected as the best model for predicting reactor performance. Once the best model was identified, SHAP values were used to interpret the model's decisions and assess which features contributed most to the model's predictions. Machine learning models are open source and available at https://github.com/MGuo-Lab/AnaerobicDigestionML.

## 2.10. Nucleotide sequence accession numbers.

All sequencing data have been submitted to the European Nucleotide Archive (ENA) under the project ID PRJEB80086 and under accession numbers ERS22193437-ERS22335023. The correspondence between accession numbers, sequences and metadata is provided in Supplementary Database 1.

# 3. Supplementary Results

## 3.1. Operational parameter monitoring

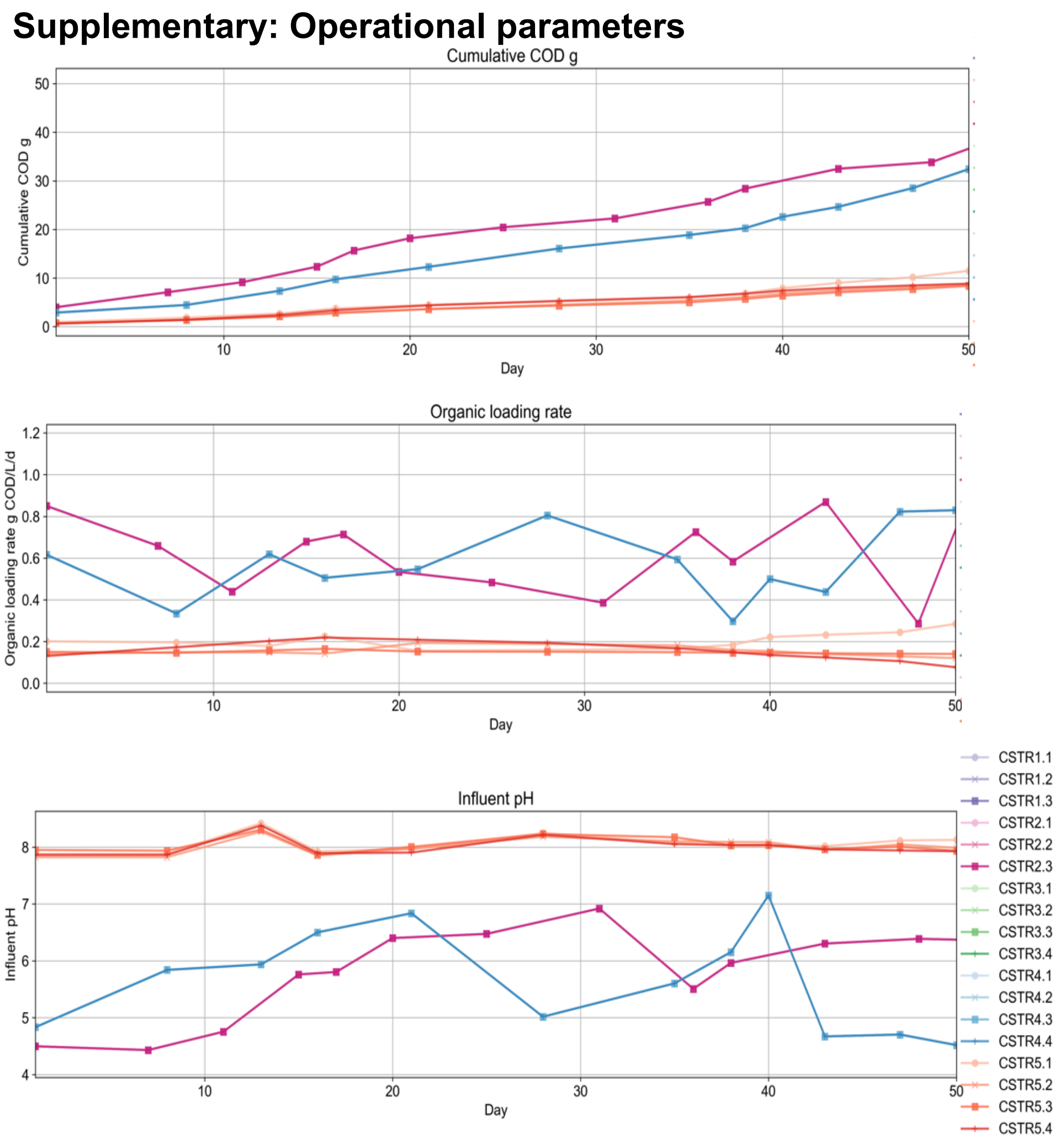

**Supplementary Figure 3.1.** Operational parameter monitoring of experimental runs including cumulative COD loading (g), organic loading rate (g COD/d/L) and influent pH.

## 3.2. Core microbiome construction per group

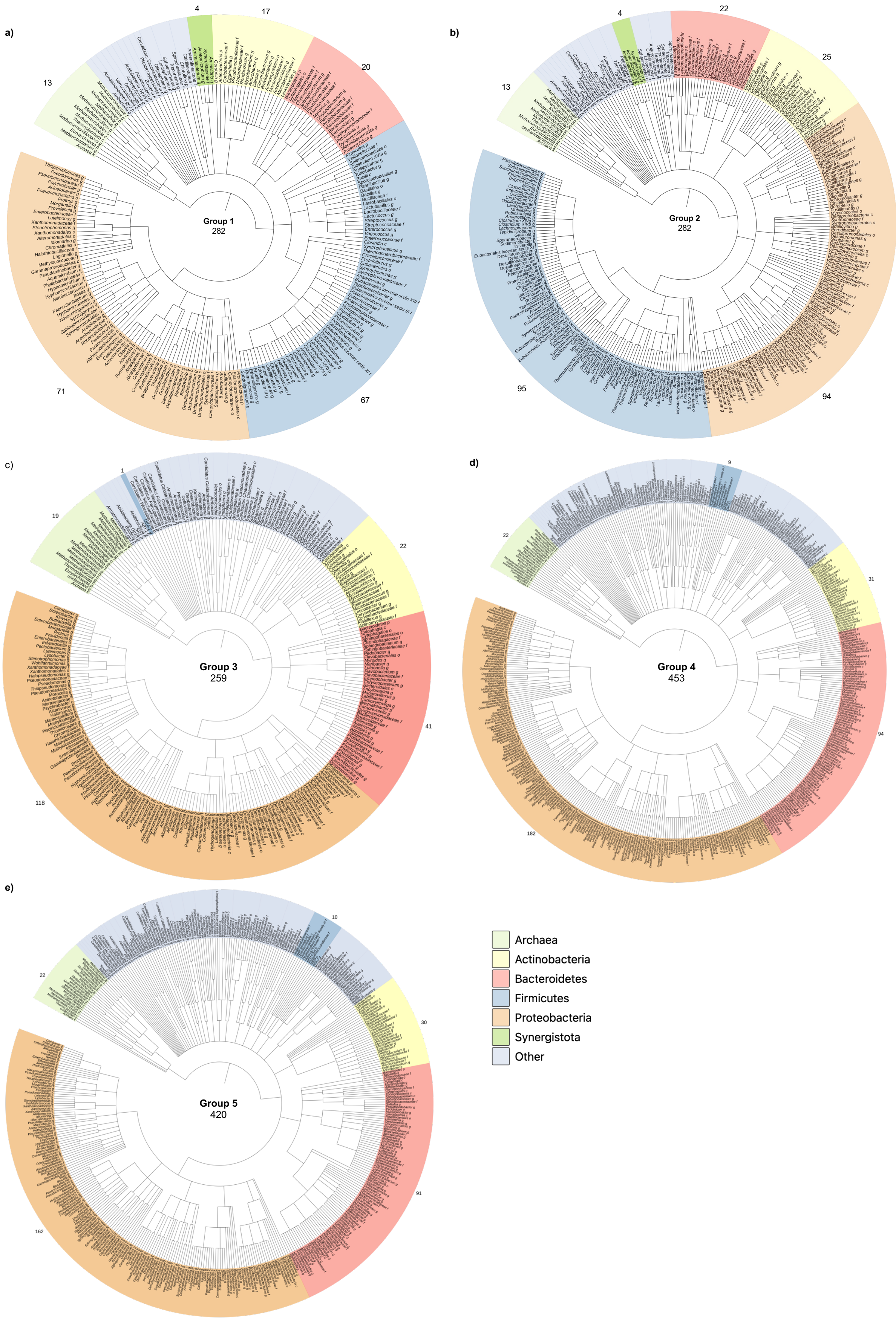

**Figure 3.2.** Core microbiome of genera present across all reactors within **(a)** mesophilic single-stage reactors (run 1), **(b)** thermophilic single-stage reactors (run 1), **(c)** single-stage reactors (run 2), **(d)** first-stage reactors (run 2), and **(e)** second-stage reactors (run 2). Archaea are highlighted as green, *Firmicutes* are highlighted as blue, *Actinobacteria* are highlighted as yellow, *Bacteroidetes* are highlighted as red, *Proteobacteria* are highlighted as orange and other bacterial phyla are highlighted as purple. Phylogenetic trees are constructed using iTOL [235].

## 3.3. Machine learning microbial community prediction

Predictive machine learning models were developed to estimate the relative abundance of microbial phyla and reactor performance. Reactor performance was measured in terms of COD removal, cumulative biogas and specific daily biogas. The input factors included influent COD, temperature, pH, organic loading rate, sugars, sugar alcohols, amino acids, and volatile fatty acids. Three machine learning models were evaluated: Linear Regression, Random Forest, and Bagging Regression.

Exploratory results are displayed for microbiome prediction using predictive machine learning models (Supplementary Figure 3.3-3.5). The findings of this work indicate that temperature, pH, and organic loading rate are the most influential factors in predicting the relative abundance of phyla and reactor performance. The use of machine learning models, particularly Random Forest, has proven to be highly effective in this work for reactor performance prediction. However, further prediction of microbiome composition will require additional data. We recommend an extended run period of at least 6 months, sampling twice per week to generate enough data for future exploration of microbiome prediction based on operational parameters and feedstock composition.

## Output: Phylum or genus level relative abundance

### a) Model inputs and outputs

| INPUTS | OUTPUTS |
|---|---|
| Organic Loading Rate | Relative abundance of phyla |
| Temperature | |
| pH | |
| Day | |
| Influent COD | |

### b) Model performance metrics.

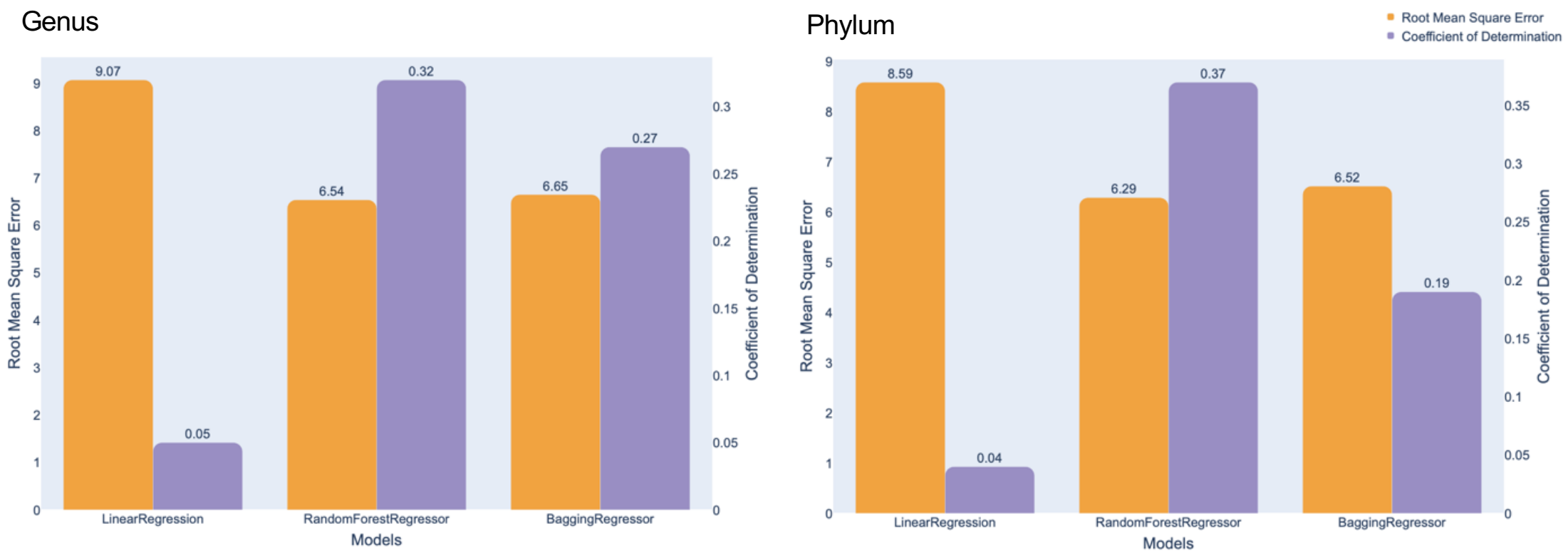

### c) Top factors for genus and phylum relative abundance prediction

Top factors (genus)
- Temperature
- Day
- pH
- Influent COD
- OLR
- *Clostridiaceae*
- *Sulfurimonas*
- *Alarcobacter*

Top factors (phylum)
- Influent COD
- Day
- OLR
- Temperature
- pH
- *Firmicutes*
- *Proteobacteria*
- *Bacteroidetes*

### d) Influence of each parameter on the top 5 most influential genera.

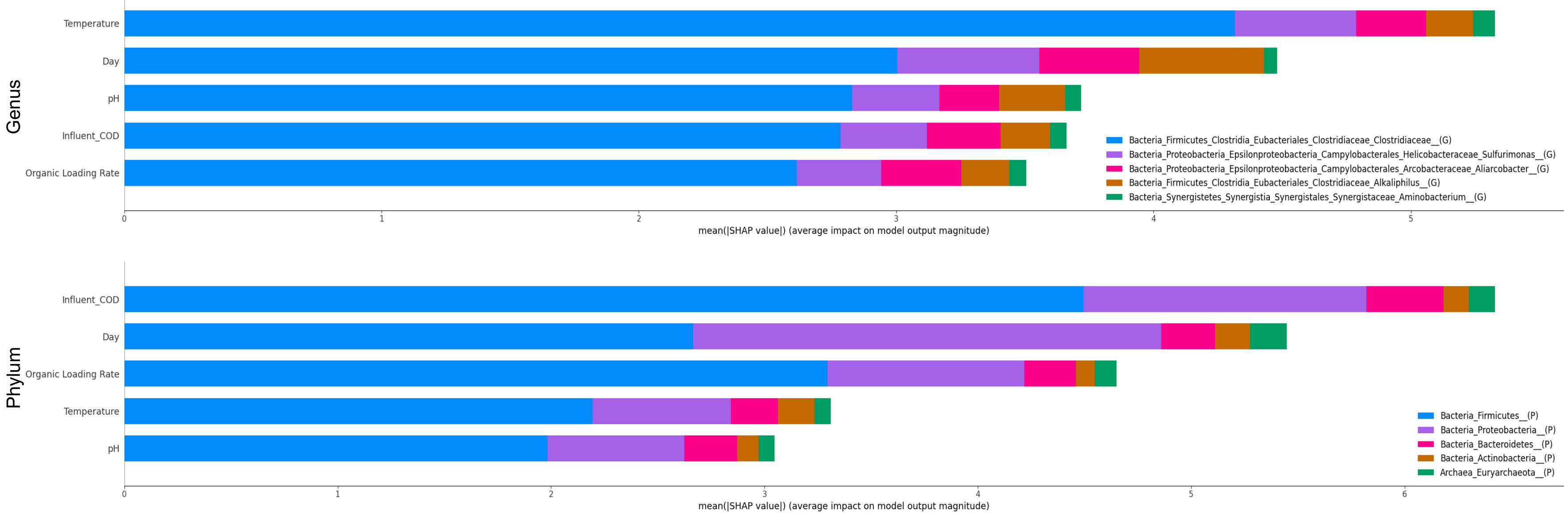

**Supplementary Figure 3.3.** Predictive machine learning models for estimation of relative abundance of the AD microbiome at the genus and phylum taxonomic levels. **(a)** Model inputs and outputs. **(b)** Model performance metrics. **(c)** Top factors influencing the model. **(d)** Influence of each parameter on the top five genera/phyla measured by SHAP values.

## Output: Phylum relative abundance including detailed feedstock composition

### a) Model inputs and outputs

| INPUTS | OUTPUTS |
|---|---|
| Organic Loading Rate | Relative abundance of phyla |
| Temperature | |
| pH | |
| Sugar | |
| Sugar Alcohol | |
| Amino Acid | |
| Volatile Fatty Acid | |
| Influent COD | |
| Day | |

### b) Model performance metrics

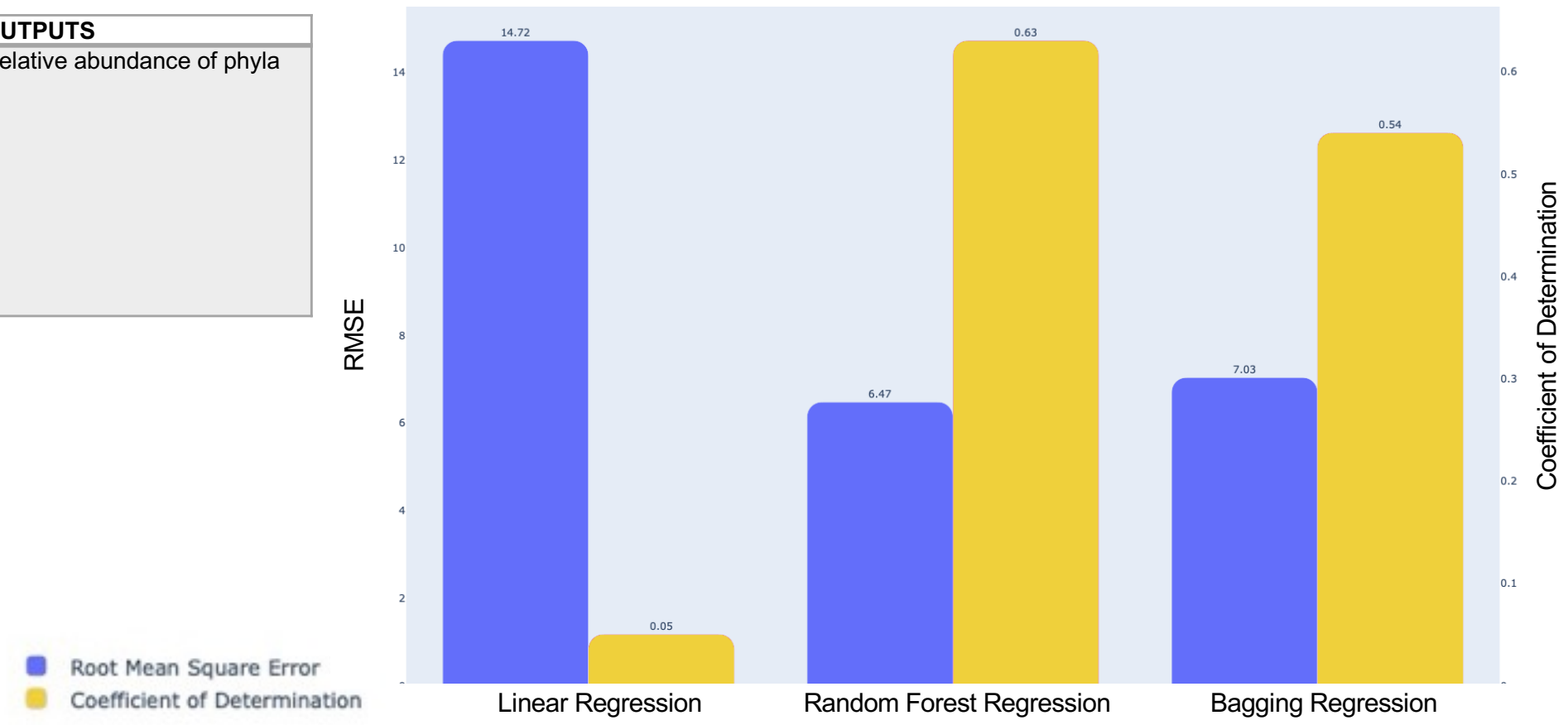

### c) SHAP values of input parameters on top five influential phyla

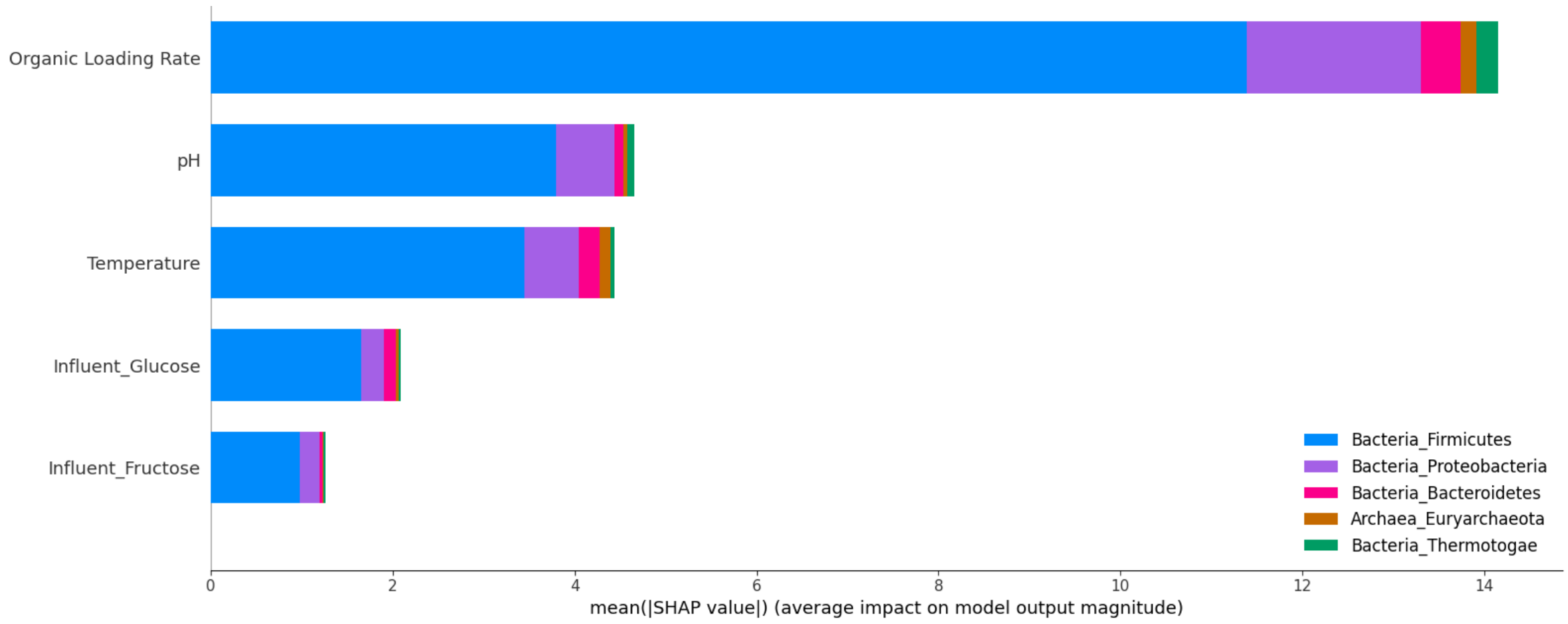

**Supplementary Figure 3.4.** Predictive machine learning models for estimation of phylum relative abundance including detailed feedstock chemical composition. **(a)** Model inputs and outputs. **(b)** Model performance metrics. **(c)** Impact of the top five most influential factors on the top phyla measured by SHAP values.

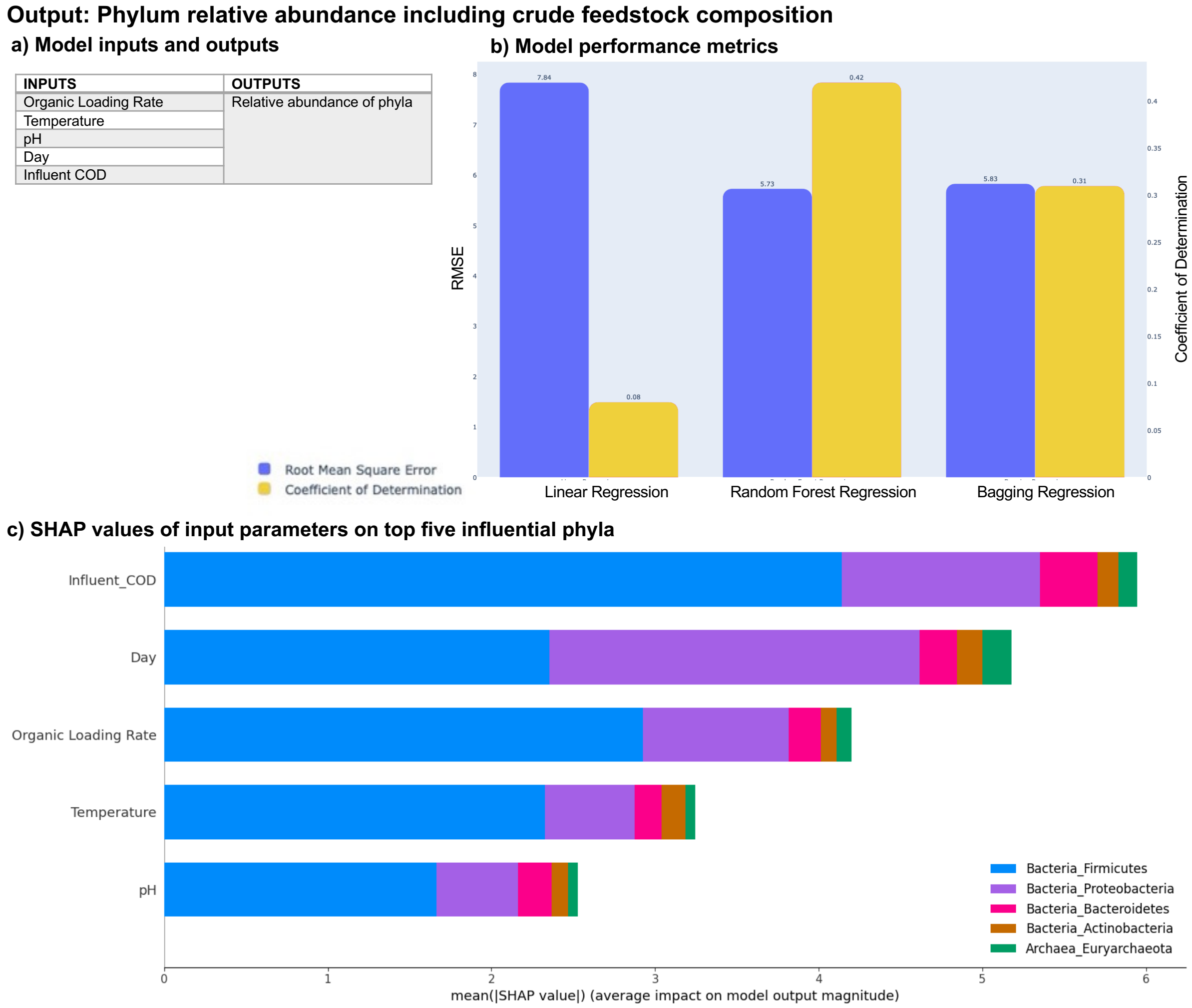

**Supplementary Figure 3.5.** Predictive machine learning models for estimation of phylum relative abundance including crude feedstock characterisation. **(a)** Model inputs and outputs. **(b)** Model performance metrics. **(c)** Impact of the top five most influential factors on the top phyla measured by SHAP values.

## 3.4. Mycoprotein fermentation wastewater additional detailed chemical composition

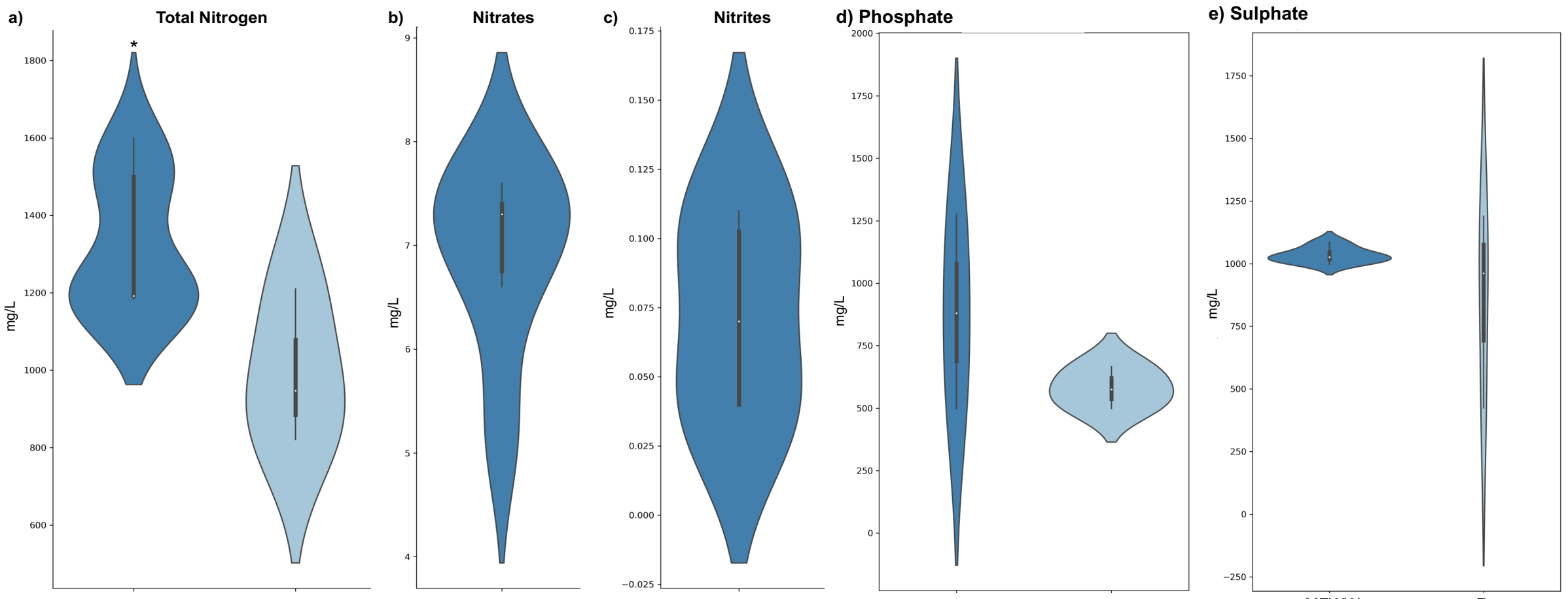

**Supplementary Figure 3.6.** Mycoprotein fermentation wastewater additional detailed chemical composition. **(a)** Total nitrogen, mg/L. **(b)** Nitrates, mg/L. **(c)** Nitrites, mg/L. **(d)** Phosphate, mg/L. **(e)** Sulphate, mg/L.

# 4. Supplementary database 1

Database containing all data relating to:

### 4.1. Literature Study Details

Relative abundance (%) of species reported to be involved in AD and experimental conditions including: Study reference, feedstock source, feedstock type, inoculum source, inoculum source reactor type, inoculum source location, experimental reactor scale, experimental reactor type, experimental reactor total volume (L). From 197 papers

### 4.2. Literature Growth Conditions

Contains all information and notes from literature review, including: microorganism name, classification, NCBI taxonomy number, pH range (optimum), temperature preference, substrate, notes, and reference. ** indicates genera has been reported in literature, but not that particular species

### 4.3. Experimental factors

Experimental factors measured within anaerobic digestion experimental studies for mesophilic (CSTR1.1 to CSTR1.3), thermophilic (CSTR2.1 to CSTR2.3), single-stage (CSTR3.1 to CSTR3.4), first-stage (CSTR4.1 to CSTR4.4), second-stage (CSTR5.1 to CSTR5.4), including: Biogas (L/g COD/d), cumulative biogas (L), specific daily biogas (L/g COD/d), COD removal (%), Organic loading rate (g COD/L/d), Temperature (C), influent COD (g/L), influent pH, Influent volatile fatty acid content (acetic, propionic, isobutyric, butyric, isovaleric, valeric acid, total VFA, g/L), Influent sugar and sugar alcohol (melibiose, maltitol, glucose, mannitol, arabitol, total sugar, g/L), reactor pH, reactor volatile fatty acid, reactor sugar and sugar alcohol

### 4.4. Taxonomy

Relative abundance of taxonomic classification of sequences obtained from mesophilic (CSTR1.1 to CSTR1.3), thermophilic (CSTR2.1 to CSTR2.3) single-stage (CSTR3.1 to CSTR3.4), first-stage (CSTR4.1 to CSTR4.4), second-stage (CSTR5.1 to CSTR5.4). Taxa are reported to the lowest common ancestor and details are given of taxonomy from the kingdom, phylum, class, order, family and genus level.

### 4.5. Nucleotide sequence metadata

The correspondence between sequencing sample ID, taxonomy and metadata.

### 4.6. ENA Accession Numbers

The correspondence between sample ID, accession numbers and taxonomic classification.

### 4.7. Bibliography

References

# 5. Supplementary References